\documentclass[a4paper,11pt]{article}
\usepackage{jheppub}
\usepackage[utf8]{inputenc}

\usepackage{graphicx} 

\usepackage{amsxtra}
\usepackage{float}
\usepackage[dvipsnames]{xcolor}
\usepackage{tikz-cd}
\usepackage{macros}

\usepackage{dcolumn}
\newcolumntype{L}{D{.}{.}{1.1}}

\tikzset{gaugebl/.style={circle,draw=black,fill=black,inner sep=2.5pt}}
\tikzset{hasse/.style={circle, fill,inner sep=2pt}}

\tikzset{hasse/.style={circle, fill,inner sep=2pt}}

\usepackage{todonotes}

\usepackage{multicol}
\usepackage{multirow}

\preprint{Imperial/TP/25/AH/02}

\title{
3d Mirrors and Phase Diagrams of Abelian Gauge Theories
}

\author[a]{Julius F. Grimminger,}
\author[b]{ Amihay Hanany,}
\author[b]{ Deshuo Liu$\,$}

\affiliation[a]{Mathematical Institute, University of Oxford,\\
Andrew Wiles Building, Woodstock Road, Oxford, OX2 6GG, UK}
\affiliation[b]{Theoretical Physics Group, The Blackett Laboratory, Imperial College London,\\ Prince Consort Road
London, SW7 2AZ, UK}

\emailAdd{julius.grimminger@maths.ox.ac.uk}
\emailAdd{a.hanany@imperial.ac.uk}
\emailAdd{deshuo.liu21@imperial.ac.uk}

\abstract{
This paper presents new developments in the study of 3d mirror symmetry and the phase structure of Abelian gauge theories.
Previous works identified 3d mirrors for a specific class of theories, termed ``simple" Abelian theories.
This work extends this framework by proposing 3d mirrors for ``non-simple" Abelian theories with both discrete and continuous gauge group factors.
The proposal is supported by evidence from an exact operator map between the Higgs/Coulomb branch of one theory and %
the %
Coulomb/Higgs branch of its 3d mirror
. Further support is provided by explicit Hilbert series computations.
An algorithm for computing the Hasse (phase) diagram of the Higgs branch of both simple and non-simple Abelian theories is introduced, uncovering a recently discovered family of isolated singularities among the elementary slices. 
A bottom-up algorithm for computing the Coulomb branch Hasse diagram of these theories is also introduced, and the two algorithms are tested against each other via 3d mirror symmetry.
}

\begin{document}

\maketitle

\section{Introduction}

Three dimensional gauge theories are free in the UV and flow to interesting fixed points in the IR. Many flows may end up at the same IR fixed point, and the corresponding UV theories are called IR dual. One theory can have many IR duals. Three dimensional theories with 8 supercharges, i.e.\ $3d$ $\mathcal{N}=4$ theories, distinguish between two types of IR dualities reviewed below. One is called 3d mirror symmetry \cite{Intriligator:1996ex}, and the other isn't.

The R-symmetry of a $3d$ $\mathcal{N}=4$ theory is
\begin{equation}
    \mathrm{Spin}(4)_R=\mathrm{SU}(2)_C\times\mathrm{SU}(2)_H\;,
\end{equation}
which has a $\mathbb{Z}_2$ outer-automorphism. In a $3d$ $\mathcal{N}=4$ gauge theory the two $\mathrm{SU}(2)$ factors are distinguished by the fields which transform under them. Scalars in the hypermultiplets transform non-trivially under $\mathrm{SU}(2)_H$ and trivially under $\mathrm{SU}(2)_C$, while scalars in the vector multiplets transform non-trivially under $\mathrm{SU}(2)_C$ and trivially under $\mathrm{SU}(2)_H$. Vacuum expectation values (VEVs) of hypermutliplet scalars parametrise the Higgs branch \cite{Hitchin:1986ea,Argyres:1996eh,Antoniadis:1996ra}, while VEVs of vector multiplet scalars -- or rather, their corresponding dressed monopole operators -- parametrise the Coulomb branch \cite{Seiberg:1996nz,Gaiotto:2008ak,Cremonesi:2013lqa,Bullimore:2015lsa} in the moduli space of vacua. Both branches are so called symplectic singularities \cite{beauville2000symplectic}. 

An IR duality between two UV theories is called 3d mirror symmetry, if the two $\mathrm{SU}(2)$ factors are exchanged in the IR (i.e.\ the IR superconformal field theories (SCFTs) are related by acting with the $\mathbb{Z}_2$ automorphism). This implies among other things that the Coulomb branch of one theory is the Higgs branch of the other, and vice versa. Two theories related by 3d mirror symmetry are called a 3d mirror pair, or 3d mirrors. In a Hanany-Witten brane system \cite{hanany_type_1997} 3d mirror symmetry is realised as S-duality.

In this paper we focus on 3d $\mathcal{N}=4$ Abelian gauge theories. Consider a theory with gauge group $\mathrm{U}(1)^r$ and $n$ hypermultiplets transforming under the gauge group. The (possibly zero) integer charges of the hypermultiplets under each of the U(1)s can be collected into a rectangular $(n\times r)$ matrix, henceforth called the charge matrix. The Higgs branch of such a theory is a very special type of symplectic singularity, a so called hypertoric variety \cite{Goto:1991pm,Bielawski2000TheGA,Konno2000COHOMOLOGYRO,2002math......3096H,proudfoot2007surveyhypertoricgeometrytopology}, distinguished by the property that the rank of its global symmetry (as a hyperk\"ahler variety) is equal to its quaternionic dimension. The Coulomb branch of such a theory, however, is not always a hypertoric variety. A diagnostic for this is the \emph{Smith Normal Form} (SNF) \cite{smith_i_1861} of the charge matrix. If the invariant factors in the SNF are all equal to $1$, the Coulomb branch is a hypertoric variety. We refer to such a theory as \emph{simple}. If some of the invariant factors in the SNF are greater than $1$, the Coulomb branch is not a hypertoric variety in general.\footnote{In special cases it may be.} We refer to such a theory as \emph{non-simple}. In this case there is a non-trivial electric 1-form symmetry $\Gamma$ determined by the invariant factors \cite{Apruzzi:2021mlh,Bhardwaj:2022dyt,Bhardwaj:2023kri}. Gauging this 1-form symmetry leads to a simple theory whose Coulomb branch is a $\Gamma$-cover of the Coulomb branch of the original non-simple theory. The Higgs branch of both theories is the same. The non-simple theory is obtained by gauging a $\Gamma$ subgroup of the Coulomb branch symmetry of the simple theory.

The 3d mirror of a simple theory is another simple theory \cite{deBoer:1996ck}. The 3d mirror of a non-simple theory, however, can neither be a simple theory, nor a non-simple theory, as both have hypertoric Higgs branches. Since a non-simple theory is obtained by gauging a discrete Abelian subgroup of the Coulomb branch symmetry of a simple theory, the 3d mirror of a non-simple theory should be obtained by gauging a discrete Abelian subgroup of the Higgs branch symmetry of a simple theory. While gauging a discrete Abelian subgroup of the Coulomb branch symmetry of a simple theory does not modify the gauge group but only the charge matrix, gauging a discrete Abelian subgroup of the Higgs branch symmetry of a simple theory adds said discrete Abelian group to the gauge group. Therefore we must consider Abelian theories whose gauge group consists of both $\mathrm{U}(1)$ factors and $\mathbb{Z}_l$ factors. We give a prescription to find a 3d mirror of any such theory.\\

Any symplectic singularity $X$, which may be a Coulomb or Higgs branch of a given theory, is naturally stratified into a finite set of so called symplectic leaves $\mathcal{L}_i$ \cite{kaledin2006symplectic}
\begin{equation}
    X=\bigsqcup_i\mathcal{L}_i\;.
\end{equation}
These leaves are partially ordered by inclusion of closures; we say $\mathcal{L}_i<\mathcal{L}_j$, if $\overline{\mathcal{L}_i}\subset\overline{\mathcal{L}_j}$. Furthermore, between two leaves $\mathcal{L}_i<\mathcal{L}_j$ there exists a \emph{transverse slice} $\mathcal{T}_i^j$ \cite{kaledin2006symplectic}. A partially ordered set can be graphically depicted in a so called Hasse diagram, a graph with nodes and edges. Every leaf $\mathcal{L}_i$ corresponds to a node in the Hasse diagram. If $\mathcal{L}_i<\mathcal{L}_j$, then the node corresponding to $\mathcal{L}_i$ is drawn below the node corresponding to $\mathcal{L}_j$. If there is no $k$, such that $\mathcal{L}_i<\mathcal{L}_k<\mathcal{L}_j$, then an edge is drawn between the nodes corresponding to $\mathcal{L}_i$ and $\mathcal{L}_j$. In this case $\mathcal{L}_i$ is called a \emph{minimal degeneration} of $\mathcal{L}_j$. We usually label each edge with the corresponding transverse slice, called an elementary slice.

As shown in \cite{Bourget:2019aer} the stratification of the Coulomb or Higgs branch into symplectic leaves has a very physical interpretation. On a given leaf every point triggers a Higgs mechanism leading to the same set of massless states. Going from one leaf to another changes the set of massless states. So the Hasse diagram of symplectic leaves is interpreted as the Hasse diagram of partial Higgsings, and can therefore be called a phase diagram. The transverse slice between two leaves corresponds to the moduli one has to tune in order to move from one phase to another. An elementary slice corresponds to a minimal Higgsing.

For quiver gauge theories many techniques have been developed to compute the Hasse diagrams of their Coulomb \cite{Cabrera:2016vvv,Cabrera:2017njm,Cabrera:2018ann,Cabrera:2019izd,Cabrera:2019dob,Hanany:2019tji,Bourget:2020asf,Bourget:2020mez,Grimminger:2020dmg,Bourget:2020gzi,Bourget:2022ehw,Bourget:2022tmw,Bourget:2023dkj,Bourget:2024mgn,Lawrie:2024wan} and Higgs \cite{Bennett:2024loi} branches, based on the underlying graph structure of the quiver.

We provide a novel algorithm to compute the Higgs and Coulomb branch Hasse diagram for any abelian $3d$ $\mathcal{N}=4$ theory, and provide checks via 3d mirror symmetry.

\paragraph{Organisation of the paper.} In Section \ref{sec_discrete} Abelian theories with discrete gauge groups are discussed. In Section \ref{sec_continuous} Abelian theories with continuous gauge groups are discussed. In Section \ref{sec_discretecontinous} Abelian theories with gauge groups with both discrete and continuous factors are discussed. In Section \ref{sec_mirror} 3d mirror symmetry for general Abelian theories is proposed. Moreover, the algorithms to compute both the Higgs and Coulomb branch phase diagrams are provided.

\paragraph{Mathematica code.} A $\mathbf{\mathsf{mathematica}}$ code accompanying the paper, available at \url{https://github.com/DeShuo-Liu/Higgs-Decomposition-for-Abelian-Theories}, implements the algorithms to compute the Higgs and Coulomb branch phase diagrams of abelian theories.

\paragraph{Note added:} During the completion of this paper two other works \cite{Bourget:2024asp,KNBalasubramanian:2024jzm} appeared on overlapping topics.

\section{Abelian theories with discrete gauge groups}
\label{sec_discrete}

In this section we consider 3d $\mathcal{N}=4$ theories which are discrete gaugings of $n$ free hypermultiplets by an Abelian group $\prod_{a=1}^{r}\mathbb{Z}_{l_a}$ (with $r\leq n$), where $l_{a+1}\vert l_a$.
The theory is specified by a diagonal matrix\footnote{The reason we define a diagonal matrix becomes clear in the discussion of 3d mirror symmetry in Section \ref{sec_mirror}.} 
\begin{equation}
    \gamma=\text{diag}(l_1,l_2,\dots,l_r)\;,
    \label{eq_gamma}
\end{equation} and by an $n\times r$ full rank (to avoid freely acting discrete groups) matrix $\mathbf{b}$ with integer entries 
\begin{equation}
    \mathbf{b}_{n\times r}=\begin{pmatrix}
        b_{11} & b_{12} & \dots & b_{1r}\\
        b_{21} & b_{22} & \dots & b_{2r}\\
        \vdots & \vdots & & \vdots\\
        b_{n1} & b_{n2} & \dots & b_{nr}
    \end{pmatrix}\;,
\label{eq_bmatrix}
\end{equation}
where $b_{ia}$ determine the action on the $i$-th hypermutliplet under the $\mathbb{Z}_{l_a}$ factor in the gauge group, as specified in Equation (\ref{eq_bmatrixactionembedding}).

As the representations of $\Z_{l_a}$ are complex, it is useful to choose a 3d $\mathcal{N}=2$ description of the theory, in which each hypermultiplet with an action vector $b_i:=\begin{pmatrix}
    b_{i1} & b_{i2} & \dots & b_{ir}
\end{pmatrix}$ consists of a chiral multiplet of action vector $b_i$ and a chiral multiplet of action vector $-b_i$. This fixes a complex structure on the moduli space. The chiral multiplets span a $\C^{2n}$ affine space with coordinates $(X_1,\dots,X_n,\tilde{X}_1,\dots,\tilde{X}_n)$, where $X_i$ is the coordinate for the chiral multiplet with action vector $b_i$ and $\tilde{X}_i$ is the coordinate for the chiral multiplet with action vector $-b_i$. This space carries a holomorphic symplectic form $\omega_\C=\sum_{i=1}^n dX_i\wedge d\tilde{X}_i$ which is invariant under the gauge group.
The discrete group $\prod_{a=1}^r\Z_{l_a}$ acts on the coordinates as:
\begin{equation}
    (X_i,\tilde{X}_i)\to(\prod_{a=1}^r\omega_{l_a}^{b_{ia}}X_i,\prod_{a=1}^r\omega_{l_a}^{-b_{ia}}\tilde{X}_i)\;,
\label{eq_bmatrixactionembedding}
\end{equation}
where $\omega_{l_a}$ is the $l_a$-th primary root of unity.

\paragraph{$\mathbf{b}$ induces an embedding matrix.}\sloppy
The matrix $\mathbf{b}$ can also be viewed as an embedding matrix for the following reason.
The flavour symmetry of $n$ free hypers is $\mathrm{Sp}(n)$ with maximal torus $\mathrm{U}(1)^n$.
The $i$-th hyper has charge $1$ under the $i$-th $\mathrm{U}(1)$, and charge $0$ under the others, i.e. $\urm(1)^n$ acts on the coordinates as:
\begin{equation}
    (X_i,\tilde{X}_i)\to(e^{i\phi_i}X_i,e^{-i\phi_i}\tilde{X}_i)\;,
\end{equation}
where $\phi_i\in(0,2\pi]$.
The matrix $\mathbf{b}$ induces an embedding
\begin{equation}
    \mathbf{b}:\quad\prod_{a=1}^r\mathbb{Z}_{l_a}\rightarrow\mathrm{U}(1)^n\;,
\end{equation}
such that the discrete group $\prod_{a=1}^r\Z_{l_a}$ acts on the coordinates as in Equation (\ref{eq_bmatrixactionembedding}).

The charge lattice $\Z^n$ of $\urm(1)^n$ determines the normalisation of the $\Z_{l_a}$ charges. 

\paragraph{Higgs branch.}\sloppy The Higgs branch of this theory can be constructed as a discrete symplectic quotient as follows. The embedding matrix $\mathbf{b}$ defines the quotient $\C^{2n}/\prod_{a=1}^{r}\mathbb{Z}_{l_a}$ via the identifications of the coordinates related by action (\ref{eq_bmatrixactionembedding}).

\paragraph{Coulomb branch.} The Coulomb branch of this theory is trivial.

\paragraph{Equivalence relations.} 
The definition of the discrete Abelian gauge theory involves specifying the gauge group factors using Equation (\ref{eq_gamma}) and the embedding matrix as in Equation (\ref{eq_bmatrix}).
There are several equivalence relations which lead to the same moduli space.
One can change the number of factors in the gauge group as is demonstrated in the following example.
Consider a single hypermultiplet transforming under the gauge group $\mathbb{Z}_2\times\mathbb{Z}_3$ with a charge matrix $\mathbf{b}=(1\;1)$ which violates the condition $r\leq n$ and the condition of $l_1\vert l_2$.
This is equivalent to a $\mathbb{Z}_6$ gauge theory with a charge matrix $\mathbf{b}=(2+3)=(5)\sim(1)$. With this choice, the two conditions are satisfied.
Having this example in mind, we can now formulate the following three equivalence relations.
These can be combined to generate the most general form of equivalence.

\begin{enumerate}
    \item{\textbf{Column rescaling}} 
    
The theory $\mathcal{A}$ defined by $\gamma$, Equation (\ref{eq_gamma}), and $\mathbf{b}$, Equation (\ref{eq_bmatrix}), is equivalent to a theory $\tilde{\mathcal{A}}$ defined by $\tilde{\gamma},\tilde{\mathbf{b}}$ where
\begin{equation}
\begin{split}
    \tilde{l}_a&=l_a\\
    \tilde{b}_{ai}&=p b_{ai}\ \text{mod}\ l_a\;,\ \text{for fixed $a$, $\forall i$},
\end{split}
\label{eq_primeshift}
\end{equation}
where $p\in\Z\backslash\{0\}$ and $\text{gcd}(p,l_a)=1$. 
The general column rescaling relation is generated by applying Equation (\ref{eq_primeshift}) to each column. This equivalence relation was also discussed in \cite{Davey:2010px} for complex Abelian orbifolds $\C^n/\Gamma$.

    \item{\textbf{Gauge group factor rescaling}} 
    
The theory $\mathcal{A}$ defined by $\gamma$ and $\mathbf{b}$ has the same moduli space as theory $\tilde{\mathcal{A}}$ defined by $\tilde{\gamma},\tilde{\mathbf{b}}$ where
\begin{equation}
\begin{split}
    \tilde{l}_a&=\alpha l_a\\
    \tilde{b}_{ai}&=\alpha b_{ai}\;, \text{for fixed $a$, $\forall i$}
\end{split}
\label{eq_Zlrescale}
\end{equation}
with $\alpha\in\mathbb{Q}^+$ (requiring that $\tilde{\gamma}$ and $\tilde{\mathbf{b}}$ have integer entries). We can thus reduce the $l_a$ to the lowest possible integers. The general rescaling relation is generated by applying Equation (\ref{eq_Zlrescale}) to each $\Z_{l_a}$ factor. 


    \item{\textbf{Parabolic transformation}}

The theory $\mathcal{A}$ defined by $\gamma$ and $\mathbf{b}$ is invariant under a right action of a parabolic subgroup of $\text{SL}^{\pm}(n,\Z)$ specified by a matrix $P$:
\begin{equation}
    \mathbf{b}\to\tilde{\mathbf{b}}=\mathbf{b}\cdot P.
\label{eq_Zlparabolic}
\end{equation}
The group $\text{SL}^{\pm}(n,\Z)$ is the subgroup of $\text{GL}(n,\Z)$ with determinant $\pm1$.
After the transformation, we get a theory $\tilde{\mathcal{A}}$ with the embedding matrix $\tilde{\mathbf{b}}$, whose Higgs/Coulomb branch of theory $\mathcal{A}$ is the same as the Higgs/Coulomb branch of theory $\tilde{\mathcal{A}}$, respectively.
If some of the $l$s are equal, the discrete group can be written as $\Z_{l_1}^{r_1}\times\dots\times\Z_{l_s}^{r_s}$ where $\sum_i r_i=r$, then the elements of the parabolic subgroup are:
\begin{equation}
    P=\begin{pmatrix}
        S_1&*&*&*\\
        &S_2&*&*\\
        &&\ddots&*\\
        &&&S_n
    \end{pmatrix},
\end{equation}
where $S_i$ is an element of $\text{SL}^{\pm}(r_i,\Z)$.
The parabolic subgroup is $\text{SL}^{\pm}(n,\Z)$ itself if $l_i=l_j$ for all $i\neq j$, and is the Borel subgroup if $l_i\neq l_j$ for all $i\neq j$.
An immediate result of Equation (\ref{eq_Zlrescale}) and Equation (\ref{eq_Zlparabolic}) is that, if $l_{a_1}$ and $l_{a_2}$ are coprime for some $a_1,a_2$, then their discrete gauge factors can be combined to $\mathbb{Z}_{l_{a_1}\times l_{a_2}}$.

The group $\text{SL}^{\pm}(n,\Z)$ can be treated as an $\text{SL}(n,\Z)$ extended by the charge conjugation $\{\pm1\}$ on odd number of hypers.
Since all the theories discussed in this paper are closed under charge conjugation, we will only consider the action of $\text{SL}(n,\Z)$ and its subgroups in the remainder of the paper.
\end{enumerate}

\paragraph{Smith decomposition.}\label{sec_smithdeco} Let $M$ be a $n\times r$ integer matrix have rank $r'\leq r$. One can find two unique matrices, $R\in\mathrm{SL}(n,\mathbb{Z})$ and $S\in\mathrm{SL}(r,\mathbb{Z})$, such that
\begin{equation}
    R\cdot M\cdot S=\begin{pmatrix}
        k_1&&&&0&\dots&0\\
        &k_2&&&\vdots&&\vdots\\
        &&\ddots&&\vdots&&\vdots\\
        &&&k_{r'}&0&\dots&0\\
        0&\dots&\dots&0&0&&\\
        \vdots&&&\vdots&&\ddots&\\
        0&\dots&\dots&0&&&0
    \end{pmatrix}_{n\times r}
\label{eq_snfdecom}
\end{equation}
gives the Smith normal form (SNF) of $M$, where $k_{a+1}|k_a$.\footnote{Note that we use the reverse of the commonly used ordering.}

Let us now consider the case where the rank of $M$ is $r$. The Smith decomposition gives
\begin{equation}
    R\cdot M\cdot S=\begin{pmatrix}
        k_1&&&\\
        &k_2&&\\
        &&\ddots&\\
        &&&k_{r}\\
        0&\dots&\dots&0\\
        \vdots&&&\vdots\\
        0&\dots&\dots&0
    \end{pmatrix}_{n\times r}=\begin{pmatrix}
        1&&&\\
        &1&&\\
        &&\ddots&\\
        &&&1\\
        0&\dots&\dots&0\\
        \vdots&&&\vdots\\
        0&\dots&\dots&0
    \end{pmatrix}_{n\times r}\cdot \begin{pmatrix}
        k_1&&&\\
        &k_2&&\\
        &&\ddots&\\
        &&&k_{r}
    \end{pmatrix}\;.
\end{equation}
We call the matrix
\begin{equation}
    \lambda_{r\times r}=\begin{pmatrix}
        k_1&&&\\
        &k_2&&\\
        &&\ddots&\\
        &&&k_{r}
    \end{pmatrix}
\label{eq_rsnf}
\end{equation}
as the reduced Smith normal form (RSNF) of $M$. The RSNF is called trivial if all $k_a=1$. If $M$ has trivial RSNF we call it \emph{simple}, if not we call it \emph{non-simple}.
We can define a \emph{simplification} $M_0$ of a non-simple $M$ via
\begin{equation}
    M\cdot S=M_0\cdot \lambda\;,
\label{eq_rsnfdeco}
\end{equation}
where $M_0$ has a trivial RSNF.

\paragraph{Example.} Let
\begin{equation}
    M=\begin{pmatrix}
        2 & 3\\
        2 & 3\\
        4 & 3
    \end{pmatrix}\;.
\end{equation}
The Smith decomposition looks as follows:
\begin{equation}
    R\cdot M\cdot S=\begin{pmatrix}
        7 & 0 & -5\\
        3 & 0 & -2\\
        -1 & 1 & 0
    \end{pmatrix}\cdot\begin{pmatrix}
        2 & 3\\
        2 & 3\\
        4 & 3
    \end{pmatrix}\cdot\begin{pmatrix}
        -3 & 1\\
        -2 & 1
    \end{pmatrix}=\begin{pmatrix}
        6 & 0\\
        0 & 1\\
        0 & 0
    \end{pmatrix}\;,
\end{equation}
from which we read
\begin{equation}
    \lambda=\begin{pmatrix}
        6 & 0\\
        0 & 1
    \end{pmatrix}\;.
\end{equation}
We have
\begin{equation}
    M\cdot S=\begin{pmatrix}
        7 & 0 & -5\\
        3 & 0 & -2\\
        -1 & 1 & 0
    \end{pmatrix}^{-1}\cdot\begin{pmatrix}
        1 & 0\\
        0 & 1\\
        0 & 0
    \end{pmatrix}\cdot\begin{pmatrix}
        6 & 0\\
        0 & 1
    \end{pmatrix}=\begin{pmatrix}
        -2 & 5\\
        -2 & 5\\
        -3 & 7
    \end{pmatrix}\cdot\begin{pmatrix}
        6 & 0\\
        0 & 1
    \end{pmatrix}=M_0\cdot\lambda\;.
\end{equation}

\paragraph{Canonical form.}
By the three equivalence relations, one can always bring any theory into a theory admitting a canonical form, such that $l_{a+1}\vert l_{a}$ and $\mathbf{b}$ is full rank and simple.

Consider a $\Z_{\alpha_1}\times\dots\times\Z_{\alpha_{r'}}$ theory $\tilde{\mathcal{A}}$ with $n$ hypers and an $n\times r'$ embedding matrix $\tilde{\mathbf{b}}$ of rank $r\leq r'$.
The $\alpha$s are not required to obey the dividing relation $\alpha_{a+1}\vert\alpha_a$, and $r'$ can be any integer number larger than $r$. 
To find a $\Z_{l_1}\times\dots\times\Z_{l_r}$ theory $\mathcal{A}$ with $n$ hypers in the canonical form which is equivalent to $\tilde{\mathcal{A}}$, one can apply the following steps:

\begin{enumerate}
    \item Using Equation (\ref{eq_Zlrescale}), one can rescale the gauge group to $\Z_\alpha^{r'}$, where $\alpha=\alpha_1\cdots\alpha_{r'}$. The new embedding matrix $\tilde{\mathbf{c}}=\tilde{\mathbf{b}}\cdot \eta$, where $\eta=\text{diag}(\frac{\alpha}{\alpha_1},\dots,\frac{\alpha}{\alpha_{r'}})$;
    \item Using Equation (\ref{eq_Zlparabolic}), one can find a decomposition:
    \begin{equation}
        \tilde{\mathbf{c}}_{n\times r'}\cdot S= \left(\tilde{\mathbf{c}}'_{n\times r'} \quad\alpha\tilde{\mathbf{c}}''_{n\times(r'-r)}\right)\;,
    \end{equation}
    where $\tilde{\mathbf{c}}'$ has rank $r$, and
    \begin{equation}
        \tilde{\mathbf{c}}'=\mathbf{c}\cdot\eta'\;,
    \end{equation} where $\mathbf{c}$ is simple and $\eta'=\text{diag}(\beta_1,\dots,\beta_r)$ is the RSNF; 
    \item Using Equation (\ref{eq_primeshift}) and (\ref{eq_Zlrescale}), one can find a canonical theory $\mathcal{A}$ with gauge group $\Z_{l_1}\times\dots\times\Z_{l_r}$, where $l_a=\frac{\alpha}{\text{gcd}(\alpha,\beta_{r-a+1})}$, and rank $r$ embedding matrix $\mathbf{b}=\mathbf{c}$.
    
\end{enumerate}



Below we discuss the structure of the Higgs branches with examples.

\paragraph{The $h_{n,k}$ singularity.}
\label{sec_hnk}

\begin{figure}[H]
    \centering
    \begin{subfigure}[t]{0.4\textwidth}
      \centering
      \begin{tikzpicture}
      \node[flavour,label=right:{$n$}] (1) at (1,0) {};
            \node[gauge,label=left:{$\Z_k$}] (0) at (0,0) {};
            \draw (0)--(1);
      \end{tikzpicture}
      \caption{}
      \label{subfig_hnkleft}
      \end{subfigure}
     \begin{subfigure}[t]{0.4\textwidth}
    \centering
    \begin{tikzpicture}
    \node[gauge,label=below:{1}] (0) at (0,0) {};
    \node[gauge,label=below:{1}] (1) at (1,0) {};
    \node (2) at (2,0) {$\cdots$};
    \node[gauge,label=below:{1}] (3) at (3,0) {};
    \node[gauge,label=below:{1}] (4) at (4,0) {};
    \node[flavour,label=above:{1}] (0a) at (0,1) {};
    \draw (0a)--(0)--(1)--(2)--(3) (3.6,0.2)--(3.4,0)--(3.6,-0.2);
    \draw[transform canvas={yshift=-2pt}] (3)--(4);
    \draw[transform canvas={yshift=0pt}] (3)--(4);
    \draw[transform canvas={yshift=2pt}] (3)--(4);
     \node at (3.5,.5) {$k$};
     \end{tikzpicture}
     \caption{}
     \label{subfig_hnkright}
     \end{subfigure}
    \caption{A 3d mirror pair. \subref{SQED1...1k}: The quiver description of $\Z_k$ theory with embedding vector $\mathbf{b}=(1,\dots,1)^\mathsf{T}$; \subref{hbarnk}: The quiver description of the mirror theory taken from \cite{Bourget:2021siw}. There are $n$ $\urm(1)$ gauge nodes in the quiver.}
    \label{fig_1...1khnk}
\end{figure}
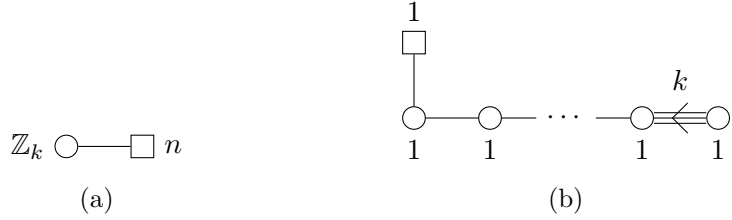

Consider a discrete gauge theory with gauge group $\Z_k$ and $n$ hypers specified by the embedding vector $\mathbf{b}=(1,\dots,1)^\mathsf{T}$.
The Coulomb branch of this theory is trivial.
The Higgs branch of this theory is a $2n$ complex dimensional isolated singularity, denoted $h_{n,k}$ in \cite{Bourget:2021siw}, but constructed there as a Coulomb branch.
The mirror theory in Figure \ref{fig_1...1khnk} is a $\urm(1)^n$ theory which can be describe by a charge matrix as defined in Section \ref{sec_continuous}:
\begin{equation}
    \mathbf{q}^\vee=\begin{pmatrix}
    1&&&&\\
    -1&1&&&\\
    &-1&\ddots&&\\
    &&\ddots&1&\\
    &&&-1&k\\
\end{pmatrix}_{n\times n}.
\label{eq_U(1)nqmirror}
\end{equation}

Combined together, we get the 3d mirror pair (recalling that the Higgs branch for a quiver with non simply laced edges is well defined provided all gauge group factors are Abelian).
The Hasse diagrams reads:
\begin{figure}[H]
    \centering
    \begin{subfigure}[t]{0.4\textwidth}
 \centering
    \begin{tikzpicture}
    \node[hasse] (0) at (0,0) {};
    \end{tikzpicture}
    \caption{}
    \label{subfig_hnkCoulomb}
   \end{subfigure}
    \begin{subfigure}[t]{0.4\textwidth}
 \centering
    \begin{tikzpicture}
    \node[hasse] (0) at (0,0) {};
        \node[hasse] (1) at (0,1.5) {};
        \node at (-0.6,1.7) {$\mathrm{id}$};
        \node at (-0.6,0.2) {$\Z_k$};
        \draw (0)--(1);
        \node at (0.7,0.7) {$h_{n,k}$};
    \end{tikzpicture}
    \caption{}
    \label{subfig_hnkHiggs}
   \end{subfigure}
    \caption{Hasse diagrams for $\Z_k$ gauge theory with embedding vector $\mathbf{b}=(1,\dots,1)^\mathsf{T}$. \subref{subfig_hnkCoulomb}: The Coulomb branch. \subref{subfig_hnkHiggs}: The Higgs branch.}
    \label{fig_HCofhnk}
\end{figure}
The special case $h_{n,2}$ is identified with $c_n$, the closure of the minimal nilpotent orbit of $Sp(n)$.

\paragraph{$h_{n,\gamma,\mathbf{b}}$ singularity.}
We label the Higgs branch of the $\Z_{l_1}\times\dots\times\Z_{l_r}$ gauge theory with embedding matrix $\mathbf{b}$ as $h_{n,\gamma,\mathbf{b}}$, where $\gamma=\text{diag}(l_1,\dots,l_{n-r})$.

\paragraph{Higgsing of $\Z_l$.}
For a single gauge group $\Z_l$ with $\mathbf{b}=(b_1,\dots,b_r)^\mathsf{T}$. The Hasse diagram can be determined by finding the subgroups of $\Z_l$ that fix a subspace of $\mathbb{H}^n$. If $b_i$ divides $l$, then the $n_i$ coordinates with charge $b_i$ span the fixed subspace under the $\Z_{b_i}$ subgroup, it contributes to the transverse slice $\mathbb{H}^{n_i}/\Z_{l/b_i}$ with action induced from the action of $\Z_k$. The leaf closure is $\mathbb{H}^{n-n_i}/\Z_{b_i}$. This is analogous to the Higgs mechanism for continuous gauge groups: the discrete gauge group $\Z_l$ can be broken into subgroup $\Z_{b_i}$ if there are hypers transforming trivially under $\Z_{b_i}$.
The Higgsing pattern for general discrete theory is discussed in Section \ref{sec_nonsimpleAbelianHiggsing}.

\paragraph{Example.}
Take $h_{n_1+n_2+n_3,6,\mathbf{b}}=\mathbb{H}^{n_1+n_2+n_3}/\Z_6$ with $\mathbf{b}=(1^{n_1},2^{n_2},3^{n_3})^\mathsf{T}$ with $n_j>1$.
As $\Z_6$ has two subgroups, the Hasse diagram reads:
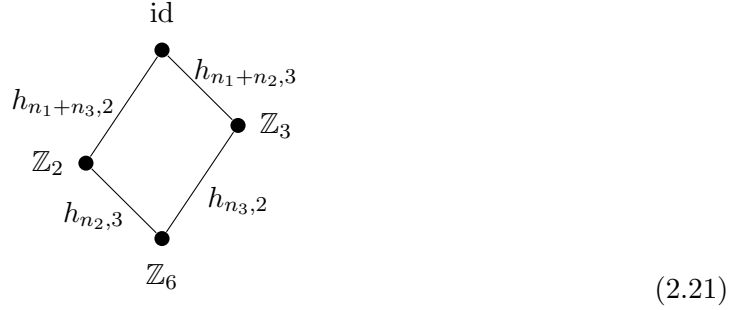
\begin{equation}
        \begin{tikzpicture}
        \node[hasse] (0) at (0,0) {};
        \node[hasse] (1) at (-1,1) {};
        \node[hasse] (2) at (1,1.5) {};
        \node[hasse] (3) at (0,2.5) {};
        \draw (0)--(1)--(3)--(2)--(0);
        \node at (0,3) {$\mathrm{id}$};
        \node at (0,-0.5) {$\Z_6$};
        \node at (-1.5,1) {$\Z_2$};
        \node at (1.5,1.5) {$\Z_3$};
        \node at (-0.9,0.3) {$h_{n_2,3}$};
        \node at (1,0.5) {$h_{n_3,2}$};
        \node at (-1.3,1.8) {$h_{n_1+n_3,2}$};
        \node at (1.1,2.2) {$h_{n_1+n_2,3}$};
    \end{tikzpicture}
\end{equation}

\section{Abelian theories with continuous gauge groups}
\label{sec_continuous}

In this section we consider Abelian 3d $\mathcal{N}=4$ theories with gauge group U$(1)^r$ and $n$ hypermultiplets (with $r\leq n$) whose charges under the gauge group are summarised in a $n\times r$ matrix with integer entries, the charge matrix
\begin{equation}
    \mathbf{q}_{n\times r}=\begin{pmatrix}
        q_{11} & q_{12} & \dots & q_{1r}\\
        q_{21} & q_{22} & \dots & q_{2r}\\
        \vdots & \vdots & & \vdots\\
        q_{n1} & q_{n2} & \dots & q_{nr}
    \end{pmatrix}\;.
\end{equation}
The entry $q_{ia}$ determines the charge of the $i$-th hypermultiplet under the $a$-th U$(1)$ factor in the gauge group. We label this theory as $\mathcal{A}$.

It is useful to choose a 3d $\mathcal{N}=2$ description of the theory, in which each hypermultiplet of charge $q_i:=\begin{pmatrix}
    q_{i1} & q_{i2} & \dots & q_{ir}
\end{pmatrix}$ consists of a chiral multiplet of charge $q_i$ and a chiral multiplet of charge $-q_i$. This fixes a complex structure on the moduli space. The chiral multiplets span a $\C^{2n}$ affine space with coordinates $(X_1,\dots,X_n,\tilde{X}_1,\dots,\tilde{X}_n)$, where $X_i$ is the coordinate for the chiral multiplet with charge $q_i$ and $\tilde{X}_i$ is the coordinate for the chiral multiplet with charge $-q_i$. This space carries a holomorphic symplectic form $\omega_\C=\sum_{i=1}^n dX_i\wedge d\tilde{X}_i$.

\paragraph{Charge matrix induces an embedding.} The charge matrix $\mathbf{q}$ is the embedding matrix of the gauge $\urm(1)^r$ inside the natural $\urm(1)^n$ symmetry of $n$ free hypers:
\begin{equation}
    \mathbf{q}:\quad\urm(1)^r\rightarrow\mathrm{U}(1)^n\;,
\end{equation}
which determines the charges of the hypers under the gauge group, such that $\urm(1)^r$ act on the coordinate as:
\begin{equation}
    (X_i,\tilde{X}_i)\to(\prod_{a=1}^re^{q_{ia}\phi_a}X_i,\prod_{a=1}^re^{-q_{ia}\phi_a}\tilde{X}_i)\;,
\label{eq_u1matrixactionembedding}
\end{equation}
for $q_{ia}\in(0,2\pi]$. The charge lattice $\Z^n$ of $\urm(1)^n$ determines the normalisation of the $\urm(1)^r$ charges.

\paragraph{Flavour matrix induces an embedding.} The flavour matrix $\mathbf{b}$\footnote{The reason to choose the same symbol $\mathbf{b}$ as the charge matrix of discrete gauge theory becomes clear in Section \ref{sec_discretecontinous}.} is the embedding matrix of the flavour $\urm(1)^{n-r}$ (as the maximal torus of the flavour symmetry) inside the natural $\urm(1)^n$ symmetry of $n$ free hypers:
\begin{equation}
    \mathbf{b}:\quad\urm(1)^{n-r}\rightarrow\mathrm{U}(1)^n\;,
\end{equation}
which determines the charges of the hypers under the flavour group. Since the flavour charge is defined module the gauge charge, there is a flexibility in the choice of $\mathbf{b}$.

The combined matrix $(\mathbf{q},\mathbf{b})$ determines the embedding of $\urm(1)^r\times\urm(1)^{n-r}$ into the natural $\urm(1)^n$:
\begin{equation}
    (\mathbf{q},\mathbf{b}):\quad\urm(1)^{r}\times\urm(1)^{n-r}\rightarrow\mathrm{U}(1)^n\;.
\end{equation}
The Abelian theory $\mathcal{A}$ is simple iff the combined matrix $(\mathbf{q},\mathbf{b})\in\text{SL}(n,\Z)$, which implies the charge lattice $\Z^n$ of $\urm(1)^n$ is preserved.

\paragraph{Higgs branch.}\sloppy The Higgs branch of this theory can be constructed as a hyperk\"ahler quotient \cite{Hitchin:1986ea} as follows. The charge matrix $\mathbf{q}$ defines a $(\C^\times)^r$ action on $\C^{2n}$, as the complexified $\urm(1)^r$ action, given by
\begin{equation}
    (X_i,\tilde{X}_i)\mapsto (\prod_{a=1}^r t_a^{q_{ia}}X_i,\prod_{a=1}^r t_a^{-q_{ia}}\tilde{X}_i)\;,
\end{equation}
where $t_a\in\C^{\times}$ is an element of the $a$-th complexified U$(1)$ factor in the gauge group.
There is a complex moment map
\begin{equation}
    \left(\mu_\C\right)_a=\sum_{i=1}^n q_{ia} X_i \tilde{X}_i\;.
    \label{eq_complexmm}
\end{equation}
$\mu_\C=0$ is the F-term equation. The hyperk\"ahler quotient
\begin{equation}
    \mu_\C^{-1}(0)//\left(\C^{\times}\right)^r\;,
\end{equation}
a so called hypertoric variety, is the Higgs branch of our theory.

If $\mathbf{q}$ contains rows where all entries are zero, these correspond to free hypers. We may remove these rows from $\mathbf{q}$ without affecting the Coulomb branch, while removing the $\C^2$ factors corresponding to free hypers from the Higgs branch.

\paragraph{Coulomb branch.} The Coulomb branch is best described as a space of dressed monopole operators. In our 3d $\mathcal{N}=2$ description for each U$(1)$ factor in the gauge group there is an $\mathcal{N}=2$ vector multiplet and a neutral chiral multiplet. For each magnetic charge $m\in\mathbb{Z}^r$ there is a bare monopole operator with conformal dimension
\begin{equation}
    \Delta(m)=\frac{1}{2}\sum_{i=1}^n\left|\sum_{a=1}^r q_{ia}m_a\right|\;.
\end{equation}
Such a bare monopole operator can be dressed with any combination of the neutral chiral multiplets. These dressed monopole operators describe the holomorphic functions on the Coulomb branch. The counting of the monopoles is given by the monopole formula in \cite{Cremonesi:2013lqa} and the explicit Coulomb branch ring is constructed in \cite{Bullimore:2015lsa}.

\paragraph{Gauge group basis change.} The moduli space of an Abelian theory defined by a charge matrix $\mathbf{q}$ is invariant under right action on $\mathbf{q}$ with an element $S\in\text{SL}(r,\Z)$:
\begin{equation}
    \mathbf{q}\mapsto\mathbf{q}\cdot S=\tilde{\mathbf{q}}\;.
\end{equation}
This operation corresponds to choosing a new basis of $\urm(1)^r$ by taking linear combinations of the charges.

If the rank of $\mathbf{q}$ is $r'<r$, then one can find an $S$ such that $r-r'$ columns of $\tilde{\mathbf{q}}$ are zero. We may delete these columns from $\tilde{\mathbf{q}}$ to produce an $n\times r'$ charge matrix $\tilde{\mathbf{q}}'$. The Higgs branch of the theories defined by $\tilde{\mathbf{q}}$ and $\tilde{\mathbf{q}}'$ is the same, while the Coulomb branch of $\tilde{\mathbf{q}}$ is the Coulomb branch of $\tilde{\mathbf{q}}'$ with an extra $\left(\C\times\C^\times\right)^{(r-r')}$ factor. 

\paragraph{Simple theory.}
Given any theory $\mathcal{A}$ with charge matrix $\mathbf{q}$ of rank $r$.
The simplification of charge matrix
\begin{equation}
    \mathbf{q}\cdot S=\mathbf{q}_0\cdot \lambda\;,
\label{eq_rsnfdeco}
\end{equation}
where the $\lambda=\text{diag}(k_1,\dots,k_r)$ is the RSNF and $\mathbf{q}_0$ is simple, defines a simple Abelian theory $\mathcal{A}_0$ with charge matrix $\mathbf{q}_0$.
Since $\lambda$ acts as a rescaling of the $\urm(1)$ charges, the theory defined by $\mathbf{q}$ and by $\mathbf{q}_0$ share the same Higgs branch. The Coulomb branch of $\mathcal{A}$ is a $\Z_{k_1}\times\Z_{k_2}\times\dots\times\mathbb{Z}_{r}$ quotient of the Coulomb branch of $\mathcal{A}_0$, with the embedding specified by $\mathbf{q}_0$.
Further, the RSNF $\lambda$ determines the $\Z_{k_1}\times\Z_{k_2}\times\dots\times\mathbb{Z}_{k_{r'}}$ electric 1-form symmetry of the theory $\mathcal{A}$ \cite{Apruzzi:2021mlh,Bhardwaj:2022dyt,Bhardwaj:2023kri}.

\subsection{Phase diagrams of $\urm(1)$ gauge theories}
\label{sec_rank2}
In this section we consider the 3d $\mathcal{N}=4$ theory with gauge group U$(1)$ and $n$ hypermultiplets with positive integer charges $\mathbf{q}=(q_1,q_2,\dots,q_n)^\mathsf{T}$.
Set $Q=\sum_{i=1}^nq_i$.

A 3d $\mathcal{N}=2$ description may be chosen, in which each hypermultiplet of charge $q_i$ consists of a chiral multiplet of charge $q_i$ and a chiral multiplet of charge $-q_i$. This fixes a complex structure on the moduli space.

The Coulomb branch of the $\urm(1)$ theory is the $A_{Q-1}$ singularity. By turning on individual complex masses for each of the hypermultiplets the Coulomb branch can be deformed into
$n$ singularities of type $A_{q_i-1}$ for $i=1,...,n$.


The phase structure of the Higgs branch is governed by the Higgs mechanism, with each phase corresponding to a possible Higgsing into a residual subgroup. The phase structure of the Higgs branch can be represented by a Hasse diagram which consists of a set of nodes and lines connecting them. The nodes in the Hasse diagram are associated with the unbroken gauge group of the corresponding phase. The Hasse diagram is equipped with a partial order determined by the Higgsing pattern. It is discussed in \cite{Bourget:2019aer} that the Hasse diagram defined above coincides with the Hasse diagram of the symplectic leaves and traverse slices of the corresponding symplectic singularity.
In particular, the \textbf{bottom} leaf is associated with the unbroken gauge group $\urm(1)$, and the \textbf{top} leaf is associated with the minimal residual gauge group, which is a $\Z_{k_1}\times\dots\times\Z_{k_r}$ group determined from the RSNF.
Each leaf in-between represents a possible residual gauge group $\urm(1)$ can be broken into.
As there is no non-trivial continuous subgroup of $\urm(1)$, the residual gauge groups form a set of discrete subgroups of $\urm(1)$. See an example in Equation (\ref{eq_2233}). The slice between two leaves represents the symplectic singularity that forms the transition between the two phases. This symplectic singularity is the Higgs branch of the transverse theory with gauge group $N_{H_2}(H_1)/H_1$, which is the normaliser of the unbroken gauge group $H_1$ within the unbroken gauge group $H_2$ of the lower leaf quotient by itself. The hypermultiplets of this traverse theory are those transforming trivially under $H_1$. If the Higgsing from $H_2$ to $H_1$ is minimal, the slice is minimal degeneration.

In the language of the geometric invariant theory (GIT) \cite{Mumford1965GeometricIT}, the stratification of the Higgsing phases is the \emph{Luna Stratification} \cite{Luna1973,Schwarz1980}, which can be computed using \emph{Luna's Slice Theorem} by finding the stablisers of sub vector spaces of $\C^{2n}$ with non-trivial image in $\mu_{\C}(0)$.

Below we discuss the structure of the Higgs branch using the Hasse diagram.
There are several special cases to cover, as follows.

\subsubsection{Coprime charges}
\label{sec_coprime}
Consider a charge vector $\mathbf{q}=(q_1,q_2,\dots,q_n)^\mathsf{T}$ with $\mathrm{gcd}(q_i,q_j)=1\;\forall\;i\neq j$. Set $Q=\sum_iq_i$.
The Coulomb branch is $A_{Q-1}=\mathbb{C}^2/\mathbb{Z}_{Q}$, which can be mass deformed into $n$ $\mathbb{C}^2/\mathbb{Z}_{q_i}$ singularities.
The Higgs branches of these theories have recently been described by Namikawa \cite{2023arXiv230913877N}, where he proved that they are isolated symplectic singularities with trivial fundamental group.
Henceforth, this isolated singularity is denoted by $\bar{h}_{\mathbf{q}}$, where the explanation for this notation comes from the following special case.

\paragraph{The $\bar{h}_{n,k}$ singularity.}
\label{sec_hbarnk}

Consider the charge matrix $\mathbf{q}=(1^n,k)^\mathsf{T}$. The Coulomb branch of this theory is a $A_{n+k-1}$ singularity. The Higgs branch of this theory is a $2n$ complex dimensional isolated singularity denoted as $\bar{h}_{n,k}$ in \cite{Bourget:2021siw}, constructed as a Coulomb branch.
Combined together with the current example, we get the 3d mirror pair.

\begin{figure}[H]
    \centering
    \begin{subfigure}[t]{0.4\textwidth}
      \centering
      \begin{tikzpicture}
      \node[flavour,label=below:{$1$}] (1) at (1,0) {};
      \node[flavour,label=below:{$n$}] (2) at (-1,0) {};
            \node[gauge,label=below:{$1$}] (0) at (0,0) {};
            \draw (0)--(1);
            \draw[transform canvas={yshift=-2pt}] (0)--(1);
            \draw[transform canvas={yshift=2pt}] (0)--(1);
            \draw (0)--(2);
            \draw (0.4,0.2)--(0.6,0)--(0.4,-0.2);
            \node at (.5,.5) {$k$};
      \end{tikzpicture}
      \caption{}
      \label{SQED1...1k}
      \end{subfigure}
     \begin{subfigure}[t]{0.4\textwidth}
    \centering
    \begin{tikzpicture}
    \node[gauge,label=below:{1}] (0) at (0,0) {};
    \node[gauge,label=below:{1}] (1) at (1,0) {};
    \node (2) at (2,0) {$\cdots$};
    \node[gauge,label=below:{1}] (3) at (3,0) {};
    \node[gauge,label=below:{1}] (4) at (4,0) {};
    \node[flavour,label=above:{1}] (5) at (4,1) {};
    \node[flavour,label=above:{1}] (0a) at (0,1) {};
    \draw (0a)--(0)--(1)--(2)--(3) (3.6,0.2)--(3.4,0)--(3.6,-0.2);
    \draw[transform canvas={yshift=-2pt}] (3)--(4);
    \draw[transform canvas={yshift=0pt}] (3)--(4);
    \draw[transform canvas={yshift=2pt}] (3)--(4);
    \draw (4)--(5);
     \node at (3.5,.5) {$k$};
     \end{tikzpicture}
     \caption{}
     \label{hbarnk}
     \end{subfigure}
    \caption{A 3d mirror pair. \subref{SQED1...1k}: The quiver description of $\urm(1)$ theory with charge matrix $\mathbf{q}=(1,\dots,1,k)^\mathsf{T}$; \subref{hbarnk}: The quiver description of the mirror theory taken from \cite{Bourget:2021siw}, whose charge matrix $\mathbf{q}^\vee$ is in (\ref{eq_chargehnk}). There are $n$ $\urm(1)$ gauge nodes in the quiver.}
    \label{fig_1}
\end{figure}
The charge matrix of the mirror theory in Figure \ref{fig_1} is:
\begin{equation}
    \mathbf{q}^\vee=\begin{pmatrix}
    1&&&&\\
    -1&1&&&\\
    &-1&\ddots&&\\
    &&\ddots&1&\\
    &&&-1&k\\
    &&&&-1
\end{pmatrix}_{(n+1)\times n}.
\label{eq_chargehnk}
\end{equation}

\paragraph{The $\bar{h}_{\mathbf{q}}$ singularity}
As the singularity $\bar{h}_{n,k}$ has a charge vector $\mathbf{q}$, it is natural to generalize to any charge matrix and denote the Higgs branch of an Abelian theory with a continuous gauge group and a general charge matrix $\mathbf{q}$ by $\bar{h}_\mathbf{q}$.

\paragraph{Higgsing}

Let us turn back to the Higgsing of $\urm(1)$ gauge theory with coprime charges. The meson $M_i=X_i\tilde{X}_i$ can obtain a VEV.
The residual gauge group is a discrete subgroup of U(1) acting on these charges.
As for this example all integer charges are coprime, there are no non-trivial residual subgroups, hence there are two leaves in the Hasse diagram of the Higgs branch, representing the no Higgsing phase and the complete Higgsing phase.
The diagrams for the Coulomb branch and for the Higgs branch for a coprime charge matrix $\mathbf{q}$ are given in Figure \ref{fig_HCofU(1)coprime}.

\begin{figure}[h]
    \centering
    \begin{subfigure}[t]{0.4\textwidth}
 \centering
    \begin{tikzpicture}
    \node[hasse] (0) at (0,0) {};
        \node[hasse] (1) at (0,1.5) {};
        \draw (0)--(1);
        \node at (0.7,0.7) {$A_{Q-1}$};
    \end{tikzpicture}
    \caption{}
    \label{subfig_U(1)coprimeCoulomb}
   \end{subfigure}
    \begin{subfigure}[t]{0.4\textwidth}
 \centering
    \begin{tikzpicture}
    \node[hasse] (0) at (0,0) {};
        \node[hasse] (1) at (0,1.5) {};
        \node at (-0.6,1.7) {$\mathrm{id}$};
        \node at (-0.6,0.2) {$\urm(1)$};
        \draw (0)--(1);
        \node at (0.7,0.7) {$\bar{h}_\mathbf{q}$};
    \end{tikzpicture}
    \caption{}
    \label{subfig_U(1)coprimeHiggs}
   \end{subfigure}
    \caption{Hasse diagrams for $\urm(1)$ gauge theory with coprime charges $\mathbf{q}=(q_1,q_2,\dots,q_n)^\mathsf{T}$. \subref{subfig_U(1)coprimeCoulomb}: The Coulomb branch. \subref{subfig_U(1)qNHiggs}: The Higgs branch.}
    \label{fig_HCofU(1)coprime}
\end{figure}

The 3d mirror of the theory with a general charge matrix is a U$(1)^{n-1}$ gauge theory with a quiver description that has generalised $(q_i,q_{i+1})$ edges, where $q_i$ and $q_{i+1}$ denote charges of the hypermultiplet under U$(1)\times$U$(1)$ connected by the edge. This is discussed in Section \ref{sec_mirror}.



\subsubsection{Coprime charges with multiplicities}
As an extension to the coprime charge cases, consider a charge vector $\mathbf{q}=(q_1^{n_1},q_2^{n_2},\dots,q_s^{n_s})^\mathsf{T}$ with $n_i\in\mathbb{N}\backslash\{0\}$ and $\sum_i n_i=n$.
The charges $q_i$ further satisfy the condition $\mathrm{gcd}(q_i,q_j)=1\;\forall\;i\neq j$.
For this set of examples, we will encounter gauge theories with discrete Abelian gauge groups, otherwise known as orbifolds. They are somewhat special as in addition they are symplectic singularities.
Before discussing the Higgs mechanism, we will first look at this class of theories with a single charge $q$.

\paragraph{Equal charges.}
Consider the case where the $n$ hypermultiplets are of equal charge namely $q_i=q\in\mathbb{Z}^+$.
This theory has a moduli space\footnote{Note that by moduli spaces we mean algebraic varieties, not stacks.} which is simple to describe.
The Coulomb branch is $A_{nq-1}=\mathbb{C}^2/\mathbb{Z}_{nq}$, which can be mass deformed into $n$ singularities of type $\mathbb{C}^2/\mathbb{Z}_q$. The Higgs branch is $a_{n-1}$, which can be fully resolved with a generic FI parameter. There is a $\mathbb{Z}_{q}$ electric 1-form symmetry (and sometimes a 2-group symmetry, depending on $n$) \cite{Bhardwaj:2022dyt,Mekareeya:2022spm,Bhardwaj:2023zix,Nawata:2023rdx}. The 3d mirror of this theory is obtained by gauging a $\mathbb{Z}_q$ subgroup of the Higgs branch flavour symmetry of the 3d mirror of U$(1)$ with $n$ charge 1 hypers. 

\begin{figure}[h]
    \centering
    \begin{subfigure}[t]{0.4\textwidth}
 \centering
    \begin{tikzpicture}
        \node[gauge,label=below:{$1$}] (0) at (0,0) {};
            \node[flavour,label=below:{$n$}] (1) at (1.5,0) {};
            \node[] (2) at (0.75,0.5) {$q$};
            \draw[transform canvas={yshift=2pt}] (0)--(1);
            \draw[transform canvas={yshift=-2pt}] (0)--(1);
            \draw[transform canvas={yshift=0pt}] (0)--(1);
            \draw (0.7,0.2)--(0.9,0)--(0.7,-0.2);
    \end{tikzpicture}
    \caption{}
    \label{fig_U(1)Nqleft}
   \end{subfigure}
       \begin{subfigure}[t]{0.4\textwidth}
 \centering
    \begin{tikzpicture}
        \node[gauge,label=below:{$1$}] (0) at (0,0) {};
        \node[flavour,label=above:{$1$}] (0a) at (0,1) {};
        \node[gauge,label=below:{$1$}] (1) at (1,0) {};
        \node[gauge,label=below:{$1$}] (2) at (3,0) {};
        \node[gauge,label=below:{$1$}] (3) at (4,0) {};
        \node[gauge,label=above:{$\Z_q$}] (4) at (4,1) {};
        \node (m) at (2,0) {$\dots$};
        \draw (0)--(1)--(m)--(2)--(3)--(4) (0a)--(0);
    \end{tikzpicture}
    \caption{}
    \label{fig_U(1)Nqright}
   \end{subfigure}
   \caption{A 3d mirror pair. \subref{fig_U(1)Nqleft}: A non-simply laced quiver description of $\urm(1)$ gauge theory with $n$ charge $q$ hypermultiplets. The $q$-laced edge represents a hypermultiplet with charge $q$ under the $\urm(1)$ gauge group and transforms in the fundamental representation under the SU$(n)$ flavor group. \subref{fig_U(1)Nqright}: Quiver description of the 3d mirror with gauge group $\urm(1)^{N-1}\times\Z_q$. The edge between a discrete gauge group and a continuous gauge group represents a hypermultiplet with charge 1 under both $\urm(1)$ and $\Z_q$. For $q=1$ the discrete gauge node should be treated as a flavor node.}
    \label{fig_U(1)Nq}
\end{figure}
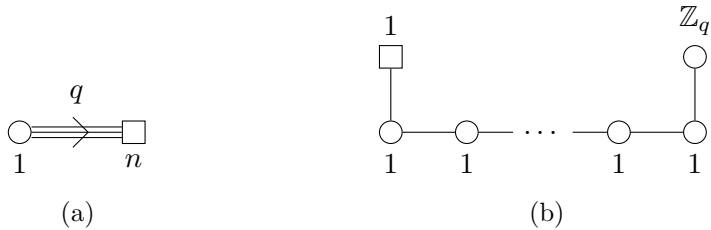

The mirror pair admits a quiver description in Figure \ref{fig_U(1)Nq}, the computation of the mirror theory is discussed in Section \ref{ex_sqedqn}.

Two quiver extensions are used here, as well as in Figure \ref{fig_1}:
(i) A non-simply laced edge can be used for a charge $q>1$, provided the gauge group is U(1). (ii) A gauge node can label a finite cyclic group with its defining representation at its side of the edge. 

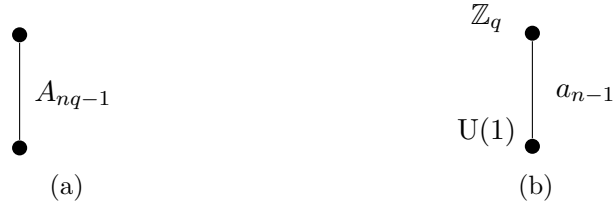
\begin{figure}[h]
    \centering
    \begin{subfigure}[t]{0.4\textwidth}
 \centering
    \begin{tikzpicture}
    \node[hasse] (0) at (0,0) {};
        \node[hasse] (1) at (0,1.5) {};
        \draw (0)--(1);
        \node at (0.7,0.7) {$A_{nq-1}$};
    \end{tikzpicture}
    \caption{}
    \label{subfig_U(1)qNCoulomb}
   \end{subfigure}
    \begin{subfigure}[t]{0.4\textwidth}
 \centering
    \begin{tikzpicture}
    \node[hasse] (0) at (0,0) {};
        \node[hasse] (1) at (0,1.5) {};
        \node at (-0.6,1.7) {$\Z_q$};
        \node at (-0.6,0.2) {$\urm(1)$};
        \draw (0)--(1);
        \node at (0.7,0.7) {$a_{n-1}$};
    \end{tikzpicture}
    \caption{}
    \label{subfig_U(1)qNHiggs}
   \end{subfigure}
    \caption{Hasse diagrams for $\urm(1)$ theory with $N$ hyper multiplets of charge $q$. \subref{subfig_U(1)qNCoulomb}: The Coulomb branch; \subref{subfig_U(1)qNHiggs}: The Higgs branch.}
    \label{fig_HCofU(1)Nq}
\end{figure}

The Hasse diagram of the Coulomb and Higgs branch is given in Figure \ref{fig_HCofU(1)Nq}.

\paragraph{Higgsing.}
Let us turn back to the $\urm(1)$ gauge theory with multiple coprime charges. For each $a\in\{1,\dots,s\}$ with $q_a>1$, there is a rank-$1$ $n_a\times n_a$ meson matrix $M_a=X_a\tilde{X}_a$ which can admit a VEV, Higgsing the theory to a $\mathbb{Z}_{q_a}$ gauge theory with $n-n_a$ hypermultiplets of embedding $\mathbf{b}_a=(q_1^{n_1},\dots,q_{a-1}^{n_{a-1}},q_{a+1}^{n_{a+1}},\dots,q_{s}^{n_{s}})^\mathsf{T}$. Giving VEV to $M_a$ describes a symplectic leaf $\mathcal{L}_a$. The transverse slice from $\mathcal{L}_a$ to the maximal leaf is the Higgs branch of the residual theory $\mathcal{T}_a=h_{n-n_a,q_a,\mathbf{b}_a}$. The transverse slice from the bottom leaf to $\mathcal{L}_a$ (the closure) is $\overline{\mathcal{L}_a}=a_{n_{a-1}}$, which is the Higgs branch of the transverse theory defined as a $\urm(1)$ theory with charges $(q_a^{n_a})$. 

Below we compute the Higgs branch Hasse diagram for several examples.

\begin{enumerate}

\item{$\mathbf{q}=(2^2,3^3)^\mathsf{T}$}

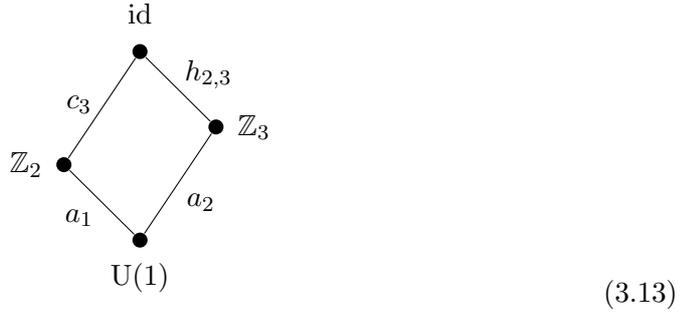
\begin{equation}
    \begin{tikzpicture}
        \node[hasse] (0) at (0,0) {};
        \node[hasse] (1) at (-1,1) {};
        \node[hasse] (2) at (1,1.5) {};
        \node[hasse] (3) at (0,2.5) {};
        \draw (0)--(1)--(3)--(2)--(0);
        \node at (0,3) {$\mathrm{id}$};
        \node at (0,-0.5) {$\urm(1)$};
        \node at (-1.5,1) {$\Z_2$};
        \node at (1.5,1.5) {$\Z_3$};
        \node at (-0.8,0.3) {$a_1$};
        \node at (0.8,0.5) {$a_2$};
        \node at (-0.8,1.8) {$c_3$};
        \node at (0.9,2.2) {$h_{2,3}$};
    \end{tikzpicture}
\label{eq_2233}
\end{equation}

\item{$\mathbf{q}=(4^2,6^3)^\mathsf{T}$}

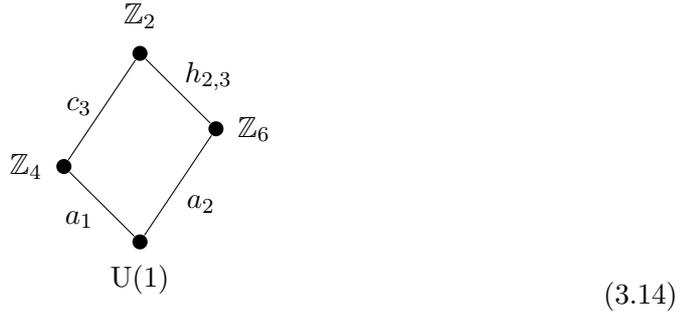
\begin{equation}
    \begin{tikzpicture}
        \node[hasse] (0) at (0,0) {};
        \node[hasse] (1) at (-1,1) {};
        \node[hasse] (2) at (1,1.5) {};
        \node[hasse] (3) at (0,2.5) {};
        \draw (0)--(1)--(3)--(2)--(0);
        \node at (0,3) {$\Z_2$};
        \node at (0,-0.5) {$\urm(1)$};
        \node at (-1.5,1) {$\Z_4$};
        \node at (1.5,1.5) {$\Z_6$};
        \node at (-0.8,0.3) {$a_1$};
        \node at (0.8,0.5) {$a_2$};
        \node at (-0.8,1.8) {$c_3$};
        \node at (0.9,2.2) {$h_{2,3}$};
    \end{tikzpicture}
\label{eq_4466}
\end{equation}

\item{$\mathbf{q}=(2^3,3^2)^\mathsf{T}$}

\begin{equation}
    \begin{tikzpicture}
        \node[hasse] (0) at (0,0) {};
        \node[hasse] (1) at (1,1) {};
        \node[hasse] (2) at (-1,1.5) {};
        \node[hasse] (3) at (0,2.5) {};
        \draw (0)--(1)--(3)--(2)--(0);
        \node at (0,3) {$\mathrm{id}$};
        \node at (0,-0.5) {$\urm(1)$};
        \node at (1.5,1) {$\Z_3$};
        \node at (-1.5,1.5) {$\Z_2$};
        \node at (0.8,0.3) {$a_1$};
        \node at (-0.8,0.5) {$a_2$};
        \node at (0.8,2) {$h_{3,3}$};
        \node at (-0.9,2.2) {$c_2$};
    \end{tikzpicture}
\end{equation}

\item{$\mathbf{q}=(1,2^2,3^3)^\mathsf{T}$}

\begin{equation}
    \begin{tikzpicture}
        \node[hasse] (0) at (0,0) {};
        \node[hasse] (1) at (-1,1) {};
        \node[hasse] (2) at (1,1.5) {};
        \node[hasse] (3) at (0,2.5) {};
        \draw (0)--(1)--(3)--(2)--(0);
        \node at (0,3) {$\mathrm{id}$};
        \node at (0,-0.5) {$\urm(1)$};
        \node at (-1.5,1) {$\Z_2$};
        \node at (1.5,1.5) {$\Z_3$};
        \node at (-0.8,0.3) {$a_1$};
        \node at (0.8,0.5) {$a_2$};
        \node at (-0.8,1.8) {$c_4$};
        \node at (0.9,2.2) {$h_{3,3}$};
    \end{tikzpicture}
\end{equation}

\item{$\mathbf{q}=(1^{n_1},3^{n_2},7^{n_3})^\mathsf{T}$ with $n_j>1$}

\begin{equation}
    \begin{tikzpicture}
        \node[hasse] (0) at (0,0) {};
        \node[hasse] (1) at (-1,1) {};
        \node[hasse] (2) at (1,1.5) {};
        \node[hasse] (3) at (0,2.5) {};
        \draw (0)--(1)--(3)--(2)--(0);
        \node at (0,3) {$\mathrm{id}$};
        \node at (0,-0.5) {$\urm(1)$};
        \node at (-1.5,1) {$\Z_3$};
        \node at (1.5,1.5) {$\Z_7$};
        \node at (-0.9,0.3) {$a_{n_2-1}$};
        \node at (1,0.5) {$a_{n_3-1}$};
        \node at (-1.4,2) {$h_{n_1+n_3,3}$};
        \node at (1.8,2.2) {$h_{n_1+n_2,7,(1^{n_1},3^{n_2})^\mathsf{T}}$};
    \end{tikzpicture}
\end{equation}

\end{enumerate}

\subsubsection{General charges}

Next consider a general charge vector $\mathbf{q}=(q_1,q_2,\dots,q_n)^\mathsf{T}$.
Set $\bold g = \{\text{gcd}(q_i,q_j)>1\}_{i<j}$.
Let us investigate the Higgsing pattern in detail. The theory at the origin can be Higgsed into discrete Abelian theories. The possible residual groups are the cyclic groups $\Z_g$ for $g\in\bold g$. 
The leaves in-between the bottom and the top leaves are bijective to the set $\mathbf{g}$.

Given $g\in\bold g$ define $\mathbf{p}_g=\{q\in\mathbf{q}\vert g \text{ divides }q\}$ and $\bar{\mathbf{p}}_g=\{q\in\bold q|q\notin\bold p_g\}$.
The largest possible residual gauge group is $\Z_g$, where $g\in\mathbf{g}$ does not divide any other element in $\mathbf{g}$.
Naturally, $\mathbf{p}_g$ is uncharged under $\Z_g$.
The residual gauge theory is a $\Z_g$ gauge theory with embedding vector $\bold c_{g}=\bar{\mathbf{p}}_g \mod g$. In the Hasse diagram, this gauge group is represented by a \textbf{next-to bottom} leaf. The transverse slice between the bottom leaf and the next-to bottom leaf $\Z_g$ is the Higgs branch of $\urm(1)$ theory with charge matrix $\mathbf{p}_g$. Here, the slice is a minimal degeneration, which means no intermediate theory is in between $\urm(1)$ and $\Z_g$. 

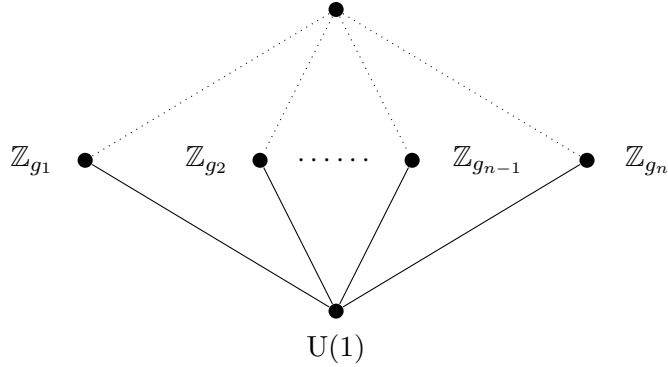
\begin{figure}[H]
\centering
    \begin{tikzpicture}
        \node[hasse] (0) at (0,0) {};
        \node[hasse] (1) at (-3.3,2) {};
        \node[hasse] (2) at (3.3,2) {};
        \node[hasse] (4) at (-1,2) {};
        \node[hasse] (5) at (1,2) {};
        \node[hasse] (3) at (0,4) {};
        \draw (0)--(1) (0)--(2) (0)--(4) (0)--(5);
        \draw[dotted] (3)--(1) (3)--(2) (3)--(4) (3)--(5);
        \node at (0,-0.5) {$\urm(1)$};
        \node at (-4,2) {$\Z_{g_1}$};
        \node at (-1.7,2) {$\Z_{g_2}$};
        \node at (0,2) {$\cdots\cdots$};
        \node at (2,2) {$\Z_{g_{n-1}}$};
        \node at (4.1,2) {$\Z_{g_n}$};
    \end{tikzpicture}
\caption{The Hasse diagram of $\urm(1)$ gauge theory with a general charge vector $\bold q$. The leaves are denoted by the largest residual gauge groups. The full lines denote minimal degenerations. The dotted lines denote slices which are not necessarily minimal degenerations.}
\label{fig_hassegeneral}
\end{figure}
The residual theory of $\Z_{g}$ with embedding vector ${\mathbf{b}}_{g}$ can be Higgsed into $\Z_{g^\prime}$ with embedding vector ${\mathbf{b}}_{g^\prime}$, iff $g^\prime$ divides $g$, where $g^\prime\in\mathbf{g} $. The transverse slice between $\Z_{g}$ and $\Z_{g^\prime}$ is a minimal degeneration iff there is no $g^{\prime\prime}\in\bold g$ which divides $g$ and can be divided by $g^\prime$.

\begin{figure}[H]
\centering
    \begin{tikzpicture}
        \node[hasse] (1) at (-3,0) {};
        \node[hasse] (2) at (3,0) {};
        \node[hasse] (4) at (-1,0) {};
        \node[hasse] (5) at (1,0) {};
        \node[hasse] (3) at (0,2) {};
        \draw[dotted] (3)--(1) (3)--(2) (3)--(4) (3)--(5);
        \node at (-3,-0.5) {$\Z_{g_{1}}$};
        \node at (-1,-0.5) {$\Z_{g_{2}}$};
        \node at (0,0) {$\cdots\cdots$};
        \node at (1.2,-0.5) {$\Z_{g_{{n-1}}}$};
        \node at (3.2,-0.5) {$\Z_{g_{n}}$};
        \node at (0,2.5) {$\Z_{g'}$};
    \end{tikzpicture}
\caption{Sub-diagram of the Hasse diagram of $\urm(1)$ theory. Here $g'$ divides $g_{i}$.}
\label{fig_hassegeneralsub}
\end{figure}
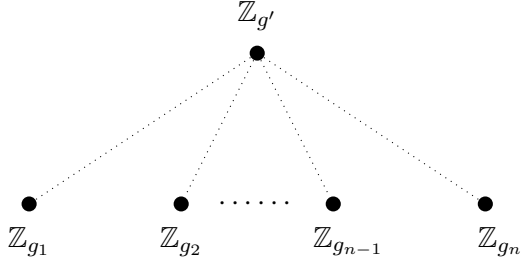

Below we compute the Higgs branch Hasse diagram for several examples.

\begin{enumerate}
     
\item{$\mathbf{q}=(1^{n_1},3^{n_2},6^{n_3})^\mathsf{T}$ with $n_j>1$}

Here the set of gcd is $\mathbf{g}=\{3,6\}$, which gives two possible residual gauge groups $\Z_6$ and $\Z_3$. Because $6$ is divided by $3$, the residual gauge group $\Z_6$ can be broken into $\Z_3$. This gives the partial order in the Hasse diagram. $\mathbf{p}_6=\{6^{n_3}\}$ and $\mathbf{p}_3=\{3^{n_2},6^{n_3}\}$ give the charges of the transverse theories from the bottom leaf to $\Z_6$ and $\Z_3$ respectively. $\bar{\mathbf{p}}_6=\{1^{n_1},3^{n_2}\}$ and $\bar{\mathbf{p}}_3=\{1^{n_1}\}$ give the charges of the residual theories with gauge group $\Z_6$ and $\Z_3$ respectively.

\begin{equation}
    \begin{tikzpicture}
        \node[hasse] (0) at (0,0) {};
        \node[hasse] (1) at (0,1.5) {};
        \node[hasse] (2) at (0,3) {};
        \node[hasse] (2) at (0,4.5) {};
        \draw (0)--(1)--(2)--(3);
        \node at (-0.6,0) {$\urm(1)$};
        \node at (-0.6,1.5) {$\Z_6$};
        \node at (-0.6,3) {$\Z_3$};
        \node at (-0.6,4.5) {$\mathrm{id}$};
        \node at (1,0.8) {$a_{n_3-1}$};
        \node at (1,2.3) {$h_{n_2,2}$};
        \node at (1,3.8) {$h_{n_1,3}$};
    \end{tikzpicture}
\end{equation}

\item{$\mathbf{q}=(11,22,33,5,7)^\mathsf{T}$}

Here the set of gcd $\mathbf{g}=\{11\}$, which gives the residual gauge group $\Z_{11}$. $\mathbf{p}_{11}=\{11,22,33\}$ gives the charges of the transverse theories from the bottom leaf to $\Z_{11}$. $\bar{\mathbf{p}}_{11}=\{5,7\}$ gives the charges of the residual theories with gauge group $\Z_{11}$. The isolated singularity $\bar{h}_{(1,2,3)^\mathsf{T}}$ is the Higgs branch of $\urm(1)$ gauge theory with charge vector $(1,2,3)^\mathsf{T}$, as defined in Section \ref{sec_coprime}.

\begin{equation}
    \begin{tikzpicture}
        \node[hasse] (0) at (0,0) {};
        \node[hasse] (1) at (0,1.5) {};
        \node[hasse] (2) at (0,3) {};
        \draw (0)--(1)--(2);
        \node at (-0.6,0) {$\urm(1)$};
        \node at (-0.6,1.5) {$\Z_{11}$};
        \node at (-0.6,3) {$\mathrm{id}$};
        \node at (1.2,0.8) {$\bar{h}_{(1,2,3)^\mathsf{T}}$};
        \node at (1.4,2.3) {$h_{2,11,(5,7)^\mathsf{T}}$};
    \end{tikzpicture}
\end{equation}

\end{enumerate}

\subsection{Phase diagrams of $\urm(1)^2$ theories}

In this section we consider the 3d $\mathcal{N}=4$ theory with rank $2$ gauge group $\urm(1)^2$ and $n$ hypermultiplets, with charge matrix $\mathbf{q}=\big(\begin{smallmatrix}
  q_{11} & q_{21} & \dots & q_{n1}\\
  q_{12} & q_{22} & \dots & q_{n2}
\end{smallmatrix}\big)^\mathsf{T}$. Additional requirements on the charge matrix are:
\begin{enumerate}
    \item The rank of $\mathbf{q}$ is $2$, i.e.\ no free vectors;
    \item No rows contain only zero elements, i.e. no free hypers;
    \item It cannot be factorised into the block diagonal form, i.e. it is not a product of independent theories.
\end{enumerate}

The Coulomb branch structure is discussed using 3d mirror symmetry in Section \ref{sec_mirror}. In the following, we discuss the Higgsing patterns.

The possible subgroups of $\urm(1)^2$ are $\urm(1)\times\Z_k$, $\urm(1)$, $\Z_{k_1}\times\Z_{k_2}$, $\Z_k$, where $k_2\vert k_1$. 
To allow the breaking pattern to a certain subgroup, this subgroup must be the stabliser of a sub vector space of the vector space spanned by the hypers.
Let us discuss each subgroup in detail.


\begin{itemize}

    \item
For the $\Z_k\times\urm(1)$ residual group, considering the moment map, the complex dimension of the stabilised sub vector space need to be larger or equal to $4$, i.e at least $2$ hypers need to transform trivially under this group. The stabilisation conditions put constraints on the charge matrix. To be stabilised by the $\urm(1)$, it requires the existence of an $(1+s)\times2$ submatrix of the charge matrix whose rank is $1$. To be further stabilised by the $\Z_k$, it requires that the elements of the $(1+s)\times2$ submatrix have $k$ as a common divisor, i.e. this submatrix can be transformed into the form:
\begin{equation}
\begin{pmatrix}
  &a_{11} k_1 \\
  0_{(1+s)\times1}&\vdots\\
  &a_{(1+s)1} k_1
\end{pmatrix}.
\end{equation}
If $k$ is the gcd, then this subgroup labels a bottom leaf. The Higgs branch of the residual theory $\Z_k\times\urm(1)$ are discussed in Section \ref{sec_discretecontinous}.

\item For $\urm(1)$ residual group, it can be treated as special case of $\Z_k\times\urm(1)$ with $k=1$, i.e, no requirement on common divisor. Further if their common gcd is $1$, then this subgroup labels a next-to-bottom leaf.

\item For $\Z_{k_1}\times\Z_{k_2}$ residual group, considering the moment map, the complex dimension of the stabilised sub vector space need to be larger or equal to $6$, i.e, at least $3$ hypers transform trivially under this group. To be stabilised by the $\Z_{k_1}\times\Z_{k_2}$, it requires the existence of a $(2+s)\times2$ submatrix which can be transformed by SL$(2,\Z)$ into the form:
\begin{equation}
\begin{pmatrix}
  a_{11} k_1 & a_{12}k_2 &\\
  \vdots & \vdots\\
  a_{(2+s)1} k_1 & a_{(2+s)2} k_2&
\end{pmatrix},
\end{equation}
where $a_{ij}\in\Z$.
If the determinants of the $2\times2$ submatrices within the $(2+s)\times2$ submatrix have gcd $k_1k_2$, i.e. the RSNF is diag$(k_1,k_2)$, then this subgroup labels a next-to-bottom leaf.

\item For $\Z_k$ residual group, it can be treated as special case of $\Z_{k_1}\times\Z_{k_2}$ with $(k_1,k_2)=(k,1)$. If the determinants of the $2\times2$ submatrices within the $(2+s)\times2$ submatrix have common gcd $k$, then this subgroup labels a next-to-bottom leaf.

\end{itemize}

The Higgs branch is an isolated singularity, if for any $(2+s)\times2$ submatrix, all the determinants of the $2\times2$ submatrices are non-zero and the gcd of any pair of these determinants is $1$. This isolated singularity aligns with the construction in \cite{2023arXiv230913877N}. This is a generalisation to the singularity $\bar{h}_\mathbf{q}$ introduced in the previous section, hence we retain the same label.

Below we compute the Higgs branch Hasse diagram for several examples.

\begin{enumerate}

\item{$\mathbf{q}=\big(\begin{smallmatrix}
  1 & 0 & 2 & 2\\
  0 & 1 & 2 & 2
\end{smallmatrix}\big)^\mathsf{T}$}

\begin{equation}
    \begin{tikzpicture}
        \node[hasse] (0) at (0,0) {};
        \node[hasse] (1) at (0,1.5) {};
        \node[hasse] (2) at (0,3) {};
        \draw (0)--(1)--(2);
        \node at (-0.7,0) {$\urm(1)^2$};
        \node at (-1,1.5) {$\urm(1)\times\Z_2$};
        \node at (-0.6,3) {$\mathrm{id}$};
        \node at (0.8,0.8) {$a_1$};
        \node at (0.8,2.3) {$A_3$};
    \end{tikzpicture}
\end{equation}

\item{$\mathbf{q}=\big(\begin{smallmatrix}
  1 & 0 & 3 & 3\\
  0 & 1 & 3 & 3
\end{smallmatrix}\big)^\mathsf{T}$}

\begin{equation}
    \begin{tikzpicture}
        \node[hasse] (0) at (0,0) {};
        \node[hasse] (1) at (0,1.5) {};
        \node[hasse] (2) at (0,3) {};
        \draw (0)--(1)--(2);
        \node at (-0.7,0) {$\urm(1)^2$};
        \node at (-1,1.5) {$\urm(1)\times\Z_3$};
        \node at (-0.6,3) {$\mathrm{id}$};
        \node at (0.8,0.8) {$a_1$};
        \node at (0.8,2.3) {$A_5$};
    \end{tikzpicture}
\end{equation}

\item{$\mathbf{q}=\big(\begin{smallmatrix}
  1 & 1 & 0 & 2 & 2\\
  0 & 0 & 1 & 2 & 2
\end{smallmatrix}\big)^\mathsf{T}$}

\begin{equation}
    \begin{tikzpicture}
        \node[hasse] (0) at (0,0) {};
        \node[hasse] (1) at (-1.5,1.5) {};
        \node[hasse] (3) at (1.5,1.5) {};
        \node[hasse] (2) at (0,3) {};
        \node[hasse] (4) at (0,4.5) {};
        \draw (0)--(1)--(2) (0)--(3)--(2) (2)--(4);
        \node at (0,-0.5) {$\urm(1)^2$};
        \node at (-2.5,1.5) {$\urm(1)\times\Z_2$};
        \node at (2.2,1.5) {$\urm(1)$};
        \node at (-0.5,3) {$\Z_2$};
        \node at (0,5) {$\mathrm{id}$};
        \node at (-1,0.6) {$a_1$};
        \node at (1,0.6) {$a_1$};
        \node at (-1.2,2.4) {$a_1$};
        \node at (1.2,2.4) {$a_1$};
        \node at (0.4,3.8) {$A_1$};
    \end{tikzpicture}
\end{equation}

\item{$\mathbf{q}=\big(\begin{smallmatrix}
   1 & 0 & 2 & 2 & 2\\
   0 & 1 & 2 & 2 & 2
\end{smallmatrix}\big)^\mathsf{T}$}

\begin{equation}
    \begin{tikzpicture}
        \node[hasse] (0) at (0,0) {};
        \node[hasse] (1) at (0,1.5) {};
        \node[hasse] (2) at (0,3) {};
        \draw (0)--(1)--(2);
        \node at (-0.7,0) {$\urm(1)^2$};
        \node at (-1,1.5) {$\urm(1)\times\Z_2$};
        \node at (-0.6,3) {$\mathrm{id}$};
        \node at (0.8,0.8) {$a_2$};
        \node at (0.8,2.3) {$A_3$};
    \end{tikzpicture}
\end{equation}

\item{$\mathbf{q}=\big(\begin{smallmatrix}
  1 & 0 & 2 & 2 & 3 & 3\\
  0 & 1 & 2 & 2 & 3 & 3
\end{smallmatrix}\big)^\mathsf{T}$}

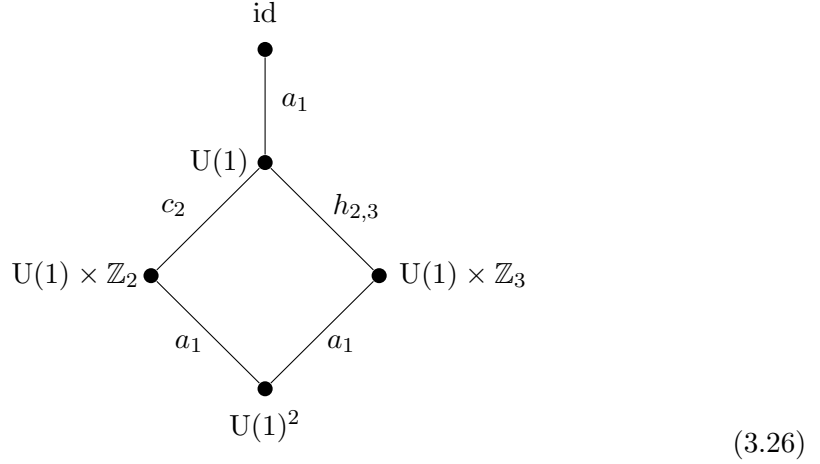
\begin{equation}
    \begin{tikzpicture}
        \node[hasse] (0) at (0,0) {};
        \node[hasse] (1) at (-1.5,1.5) {};
        \node[hasse] (3) at (1.5,1.5) {};
        \node[hasse] (2) at (0,3) {};
        \node[hasse] (4) at (0,4.5) {};
        \draw (0)--(1)--(2) (0)--(3)--(2) (2)--(4);
        \node at (0,-0.5) {$\urm(1)^2$};
        \node at (-2.5,1.5) {$\urm(1)\times\Z_2$};
        \node at (2.6,1.5) {$\urm(1)\times\Z_3$};
        \node at (-0.6,3) {$\urm(1)$};
        \node at (0,5) {$\mathrm{id}$};
        \node at (-1,0.6) {$a_1$};
        \node at (1,0.6) {$a_1$};
        \node at (-1.2,2.4) {$c_2$};
        \node at (1.2,2.4) {$h_{2,3 }$};
        \node at (0.4,3.8) {$a_1$};
    \end{tikzpicture}
\end{equation}

\end{enumerate}

\subsection{Phase diagrams of $\urm(1)^r$ theories}
\label{sec_rankr}

In this section we consider a 3d $\mathcal{N}=4$ theory with gauge group $\urm(1)^r$ and $n$ hypermultiplets, with charge matrix $\mathbf{q}$ which is an $n\times r$ rectangular matrix. Additional requirements on the charge matrix are:
\begin{enumerate}
\item The matrix $\mathbf{q}$ is rank $r$, i.e.\ no free vectors;
    \item No rows contains only zero elements, i.e. no free hypers;
    \item It cannot be factorised into the block diagonal form, i.e. it is not a product of independent theories.
\end{enumerate}



The Coulomb branch structure is discussed using 3d mirror symmetry in Section \ref{sec_mirror}.
Below we focus on the Higgs branch structure.



In the following, we discuss the Higgsing patterns. The most general form of the residual gauge group is $\urm(1)^{r-s}\times\Z_{k_1}\times\dots\times\Z_{k_s}$, where $k_i\in \Z^+$ and $k_{i+1}\vert k_i$.
This breaking pattern, 
requires the existence of an $(s+s')\times r$ submatrix with rank no larger than $s$ and it can be transformed by an SL$(r,\Z)$ matrix into the form:
\begin{equation}
\begin{pmatrix}
  &a_{11} k_1 & \dots & a_{1s}k_s\\
  0_{(s+s')\times (r-s)}&\vdots & \ddots & \vdots\\
  &a_{(s+s')1} k_1 & \dots & a_{(s+s')s} k_s
\end{pmatrix},
\end{equation}
where $a_{ij}\in\Z$.
This implies that there are at least $s+1$ hypers transforming trivially under the residual group.
It labels a bottom leaf iff the rank is $s$ and $k_i$ is the greatest common divisor for the $i$-th row, i.e. the left part of the matrix takes the form of the RSNF $\lambda=\text{diag}(k_1,\dots,k_s)$. If no such submatrix can be found for any value of $s$ and $k_i$, unless for $s=r$ and $k_i=1$, the Higgs branch is an isolated singularity.

The Higgs branch is an isolated singularity, if for any $(r+s)\times r$ submatrix of the simplified charge matrix $\mathbf{q}_0$, all the determinants of the $r\times r$ submatrices are non-zero and the gcd of any pair of these determinants is $1$. This agrees with the condition given in \cite{2023arXiv230913877N}.
As in the $\urm(1)^2$ case, we label the singularity as $\bar{h}_\mathbf{q}$.

\section{Abelian theories with discrete and continuous gauge group factors}
\label{sec_discretecontinous}

In this section, we discuss Abelian theories with both discrete and continuous factors in the gauge group.
In Section \ref{sec_continuous}, a theory $\mathcal{A}$ is defined by a charge matrix $\mathbf{q}$ of the $\urm(1)^r$ gauge group.
If the RSNF of $\mathbf{q}$ is non-trivial and takes the form $\lambda=\text{diag}(k_1,\dots,k_r)$, where $k_{i+1}\vert k_i$, then theory $\mathcal{A}$ can be obtained by gauging a $\Z_{k_1}\times\dots\times\Z_{k_r}$ subgroup of the topological $\urm(1)^r$ symmetry of the simple Abelian theory $\mathcal{A}_0$ with a charge matrix $\mathbf{q}_0$ defined in Equation (\ref{eq_rsnfdeco}), with the natural embedding induced by $\mathrm{Id}_r$:
\begin{equation}
\mathrm{Id}_r:\Z_{k_1}\times\dots\times\Z_{k_r}\rightarrow\mathrm{U}(1)^r\;,    
\end{equation}
such that it acts on the monopole operator $v_\mathbf{m}$ with charge $\mathbf{m}=(m_1,\dots,m_r)$ as:
\begin{equation}
    v_\mathbf{m}\to \prod_{a=1}^r\omega_{k_a} v_\mathbf{m}.
\end{equation}

The discrete gauging gives rise to an electric 1-from symmetry \cite{Gaiotto:2014kfa}, which is a symmetry of Wilson line operators.
Under the discrete gauging, the Higgs branch of theory $\mathcal{A}$ is the same as the Higgs branch of theory $\mathcal{A}_0$, and the Coulomb branch of theory $\mathcal{A}$ is a $\Z_{k_1}\times\dots\times\Z_{k_r}$ quotient of the Coulomb branch of theory $\mathcal{A}_0$.

Another family of Abelian theories can be constructed by gauging a discrete subgroup of the $\urm(1)^{n-r}$ flavour symmetry.

Without loss of the generality, the discrete subgroup can be chosen to be $\Z_{l_1}\times\dots\times\Z_{l_{n-r}}$, satisfying $l_{i+1}\vert l_i$, where the embedding of the discrete group inside the natural $\urm(1)^n$ is specified by a full rank matrix $\mathbf{b}$ of dimensions $n\times(n-r)$:
\begin{equation}
\mathbf{b}:\Z_{l_1}\times\dots\times\Z_{l_{n-r}}\rightarrow\mathrm{U}(1)^n\;.    
\end{equation}
Equivalently, if we choose $\mathbf{b}$ to be the flavour matrix determing the embedding of the flavour $\urm(1)^{n-r}$ inside $\urm(1)^n$, then the discrete group has natural embedding inside the flavour symmetry induced by $\mathrm{Id}_{n-r}$:
\begin{equation}
\mathrm{Id}_{n-r}:\Z_{l_1}\times\dots\times\Z_{l_{n-r}}\rightarrow\mathrm{U}(1)^{n-r}\;.    
\end{equation}

The resulting gauge theory $\mathcal{A}'$ has a gauge group made of continuous and discrete factors, $\urm(1)^r\times\Z_{l_1}\times\dots\times\Z_{l_{n-r}}$.
If $r=0$, theory $\mathcal{A}'$ reduces to a discrete gauge theory which is discussed in detail in Section \ref{sec_discrete}.
The discrete gauging gives rise to a magnetic 1-from symmetry, which is a symmetry of vortex line operators. 
Under the discrete gauging, the Coulomb branch of theory $\mathcal{A}'$ is the same as the Coulomb branch of theory $\mathcal{A}$, and the Higgs branch of theory $\mathcal{A}'$ becomes a $\Z_{l_1}\times\dots\times\Z_{l_{n-r}}$ quotient of the Higgs branch of theory $\mathcal{A}$ with the embedding given by the matrix $\mathbf{b}$.
This extension leads to the following family of symplectic singularities.

\paragraph{$\bar{h}_{\mathbf{q},\gamma,\mathbf{b}}$ singularity.}
\label{sec_barhnquotientk}
Consider a simple $\urm(1)^r\times \Z_{l_1}\times\dots\times\Z_{l_{n-r}}$ theory with charge matrix $\mathbf{q}$ and embedding matrix $\mathbf{b}$. The Higgs branch is the quotient $\bar{h}_\mathbf{q}/\Z_{l_1}\times\dots\times\Z_{l_{n-r}}$ with embedding $\mathbf{b}$. We label it as $\bar{h}_{\mathbf{q},\gamma,\mathbf{b}}$, where $\gamma=\text{diag}(l_1,\dots,l_{n-r})$.


\paragraph{General Abelian theory.} Next, the general Abelian theory is discussed, with a clear distinction made between whether a $\urm(1)$ is electric or magnetic to avoid ambiguity.
The general Abelian theory can be acquired by steps of gauging from the theory of $n$ free hypers with electric flavour symmetry $\urm(1)^n_E$. The combined matrix $(\mathbf{q}_0,\mathbf{b})$ gives an embedding of $\urm(1)^r_E\times\urm(1)^{n-r}_E$ inside $\urm(1)^n_E$. By gauging the $\urm(1)^r_E$ symmetry, one get a simple $\urm(1)^r$ theory, with matrix $\mathbf{q}_0$ as the gauge charge matrix, and $\mathbf{b}$ as the flavour charge matrix. After the gauging, there is a new $\urm(1)^r_M$ magnetic flavour symmetry (or topological symmetry) which is dual to the gauged $\urm(1)^r_E$. 
From the simple Abelian theory, a gauging of $\Z_{k_1}\times\dots\times\Z_{k_r}$ subgroup of the $\urm(1)^r_M$ with natural embedding specified by $\text{Id}_r$ gives a non-simple $\urm(1)^r$ theory with electric RSNF $\lambda$ and charge matrix $\mathbf{q}_0\cdot\lambda$. 
On the other hand, a gauging of $\Z_{l_1}\times\dots\times\Z_{l_{n-r}}$ subgroup of the $\urm(1)^{n-r}_E$ with the natural embedding specified by $\text{Id}_{n-r}$ gives a $\urm(1)^r\times\Z_{l_1}\times\dots\times\Z_{l_{n-r}}$ theory with magnetic RSNF $\gamma$, charge matrix $\mathbf{q}_0$ and embedding matrix $\mathbf{b}$.
Together, it gives the general Abelian theory.

The general Abelian theory with gauge group $\urm(1)^{r}\times{\Z_{l_1}}\times\dots\times{\Z_{l_{n-r}}}$ can be defined by a generalised charge matrix tuple $[\lambda,(\mathbf{q}_0,\mathbf{b}),\gamma]$. In this tuple:
\begin{itemize}
    \item $\lambda=\text{diag}(k_1,\dots,k_r)$ with $k_{i+1}\vert k_i$ is the electric Smith normal form, which is the RSNF of the electric charge matrix $\mathbf{q}$, representing the gauging of the $\Z_{k_1}\times\dots\times\Z_{k_r}$ subgroup of the magnetic flavour (or topological) $\urm(1)^r$, giving the electric $1$-form symmetry;
    \item $\mathbf{q}_0$ is the simplified charge matrix of electric gauge $\urm(1)^r$ of size $n\times r$, such that $\mathbf{q}$ can be transformed into the form $\mathbf{q}_0\cdot\lambda$ under SL$(r,\Z)$;
    \item $\gamma=\text{diag}(l_1,\dots,l_{n-r})$ with $l_{i+1}\vert l_i$ is the magentic Smith normal form, representing the gauging of $\Z_{l_1}\times\dots\times\Z_{l_{n-r}}$ subgroup of the electric flavour $\urm(1)^{n-r}$, giving the magnetic $1$-form symmetry;
    \item $\mathbf{b}$ is the embedding matrix of $\Z_{l_1}\times\dots\times\Z_{l_{n-r}}$ gauge group inside the electric $\urm(1)^{n}$ of size $n\times (n-r)$, which gives the representation of the hypers under the discrete gauge group. and it fixes a splitting in the short exact sequence (\ref{eq_sequence}).
\end{itemize}

This class of Abelian theories form a closed set under 3d mirror symmetry. The details of the 3d mirror symmetry and the Higgsing patterns are discuss in Section \ref{sec_mirror}.

\paragraph{Equivalent form.} Consider a general Abelian theory $\mathcal{A}$. The matrices $\mathbf{q}$ and $\mathbf{b}$ satisfy the following equivalence relation.
Combine $\mathbf{q}$ and $\mathbf{b}$ into an $n\times n$ matrix, an element of SL$(n,\Z)$.
Under a right action of a parabolic subgroup of SL$(n,\Z)$ specified by a matrix $P$:
\begin{equation}
    (\mathbf{q},\mathbf{b})\to(\mathbf{q},\mathbf{b})\cdot P=(\tilde{\mathbf{q}},\tilde{\mathbf{b}}),
\end{equation}
we get a theory $\tilde{\mathcal{A}}$ with the same gauge group and the combined charge matrix $(\tilde{\mathbf{q}},\tilde{\mathbf{b}})$. The Higgs/Coulomb branch of theory $\mathcal{A}$ is the same as the Higgs/Coulomb branch of $\tilde{\mathcal{A}}$, respectively.

The matrix $P$ takes a specific form,
\begin{equation}
    P=\begin{pmatrix}
        S_{r\times r}&M_{r\times(n-r)}\\
        0_{(n-r)\times r}&B_{(n-r)\times(n-r)}
    \end{pmatrix},
    \label{eq_parabolic}
\end{equation}
where $S$ is an element of SL$(r,\Z)$, $M$ is an $r\times(n-r)$ integer matrix, and $B$ is an element of a parabolic subgroup of SL$(n-r,\Z)$.
If some of the $l$s are equal, the discrete group can be written as $\Z_{l_1}^{r_1}\times\dots\times\Z_{l_s}^{r_s}$, with $\sum_i r_i=n-r$, then $B$ takes a specific form,
\begin{equation}
    B=\begin{pmatrix}
        S_1&*&*&*\\
        &S_2&*&*\\
        &&\ddots&*\\
        &&&S_n
    \end{pmatrix},
\end{equation}
where $S_i$ is an element of SL$(r_i,\Z)$.

\section{3d mirror symmetry}
\label{sec_mirror}

\subsection{Simple Abelian theories}
\label{sec_sat}

$3$d mirror symmetry for simple Abelian theories, i.e. Abelian theories with trivial $1$-form symmetries, were first discussed in \cite{deBoer:1996ck,deBoer:1996mp}.
It goes under the name of \emph{Gale duality} for hypertoric variety in the math literature, see for example \cite{2002math......3096H,Braden:2008kny}.
Following the description in \cite{Ballin:2023rmt}, for an Abelian theory $\mathcal{A}$ of gauge group $\urm(1)^r$ with a simple electric charge matrix $\mathbf{q}$ of size $n\times r$, there is a short exact sequence:
\begin{equation}
\begin{tikzcd}
0 \arrow{r} & \Z^r 
\arrow{r}{\mathbf{q}} &
\Z^n \arrow{r}{\mathbf{\tau}}
&\Z^{n-r} \arrow{r} \arrow[bend left]{l}{\mathbf{b}} & 0,
\end{tikzcd}
\label{eq_sequence}
\end{equation}
where $\mathbf{\tau}$ is an $(n-r)\times n$ integer matrix, and $\mathbf{b}$ is an $n\times (n-r)$ integer matrix called splitting. They satisfy the relations:
\begin{equation}
    \tau\cdot\mathbf{q}=0_{(n-r)\times r}
    \label{eq_tq}
\end{equation}
\begin{equation}
\mathbf{\tau}\cdot\mathbf{b}=\text{Id}_{n-r}.
\label{eq_tb}
\end{equation}

The matrix $\mathbf{b}$ is the electric flavour charge matrix of theory $\mathcal{A}$.
The combined matrix $(\mathbf{q},\mathbf{b})$ is an element of SL$(n,\Z)$. 

3d mirror symmetry can be expressed as the inverse-transpose of the combined matrix $(\mathbf{q},\mathbf{b})$:
\begin{equation}
    (\mathbf{q},\mathbf{b}) \xleftrightarrow{\text{mirror}} (\mathbf{b},\mathbf{q})^{-1,\mathsf{T}}=(\tau^\mathsf{T},\mathbf{b}^\vee)=(\mathbf{q}^\vee,\mathbf{b}^\vee),
    \label{eq_inversemirrorsimple}
\end{equation}
where $\mathbf{q}^\vee$ is an $n\times (n-r)$ matrix and $\mathbf{b}^\vee$ is an $n\times r$ matrix.
The mirror theory is the Abelian theory $\mathcal{A}^\vee$ of gauge group $\urm(1)^{n-r}$ with the electric charge matrix $\mathbf{q}^\vee$ and the electric flavour charge matrix $\mathbf{b}^\vee$. Due to taking the inverse in Equation (\ref{eq_inversemirrorsimple}), the electric charges $\mathbf{q}^\vee$ and $\mathbf{b}^\vee$ for theory $\mathcal{A}^\vee$ are magnetic charges for theory $\mathcal{A}$.

A similar short exact sequence exists for the mirror theory:
\begin{equation}
\begin{tikzcd}
0 \arrow{r} & \Z^{n-r} 
\arrow{r}{\mathbf{q}^\vee} &
\Z^n \arrow{r}{\tau^\vee}
&\Z^{r} \arrow{r} \arrow[bend left]{l}{\mathbf{b}^\vee} & 0,
\end{tikzcd}
\label{eq_mirrorsequence}
\end{equation}
with the important identifications:
\begin{equation}
    \begin{split}
        \mathbf{q}^\vee=\mathbf{\tau}^{\mathsf{T}},\\
        \tau^\vee=\mathbf{q}^{\mathsf{T}}.
    \end{split}
\label{eq_qveetvee}
\end{equation}

Equations (\ref{eq_sequence}) and (\ref{eq_mirrorsequence}) can be combined together, so that the mirror pair is encoded by the short exact sequence with splitting $\mathbf{b}$ and co-splitting $\mathbf{b}^{\vee,\mathsf{T}}$ as:
\begin{equation}
\begin{tikzcd}
0 \arrow{r} & \Z^r 
\arrow{r}{\mathbf{q}} &
\Z^n \arrow{r}{\mathbf{q}^{\vee,\mathsf{T}}} \arrow[bend left]{l}{\mathbf{b}^{\vee,\mathsf{T}}}
&\Z^{n-r} \arrow{r} \arrow[bend left]{l}{\mathbf{b}} & 0.
\end{tikzcd}
\label{eq_sescosplitting}
\end{equation}
In this short exact sequence, arrows going into $\Z^n$ represent electric (respectively magnetic) charges of theory $\mathcal{A}$ (respectively of theory $\mathcal{A}^\vee$), while arrows coming out of $\Z^n$ represent magnetic (respectively electric) charges of theory $\mathcal{A}$ (respectively of theory $\mathcal{A}^\vee$).

The matrices $\tau$ and $\mathbf{b}$ have the following dual physical interpretations for theory $\mathcal{A}$ and for the mirror theory $\mathcal{A}^\vee$.
\begin{itemize}
    \item 
From Equation (\ref{eq_tq}) $\tau^\mathsf{T}$ spans the magnetic charge lattice $\Z^{n-r}_M$ of theory $\mathcal{A}$ which is orthogonal to the electric charge lattice $\Z^r_E$ of theory $\mathcal{A}$ spanned by the electric charge matrix $\mathbf{q}$.
    \item 
From Equation (\ref{eq_tb}) $\mathbf{b}$ spans the electric flavor charge lattice $\Z^{n-r}_E$ of theory $\mathcal{A}$ which is dual to the magnetic charge lattice $\Z^{n-r}_M$ of theory $\mathcal{A}$ spanned by $\tau^\mathsf{T}$.
    \item 
The choice of $\mathbf{q}$ and $\mathbf{b}$ splits the electric charge lattice $\Z^n_E$ of theory $\mathcal{A}$ into $\Z^r_E\times\Z^{n-r}_E$, where the first factor is a lattice of gauge charges, while the second factor is a lattice of flavor charges. The combined matrix $(\mathbf{q},\mathbf{b})$ is an element of SL$(n,\Z)$. 
    \item 
From Equation (\ref{eq_tq}) $\tau^\mathsf{T}$ spans the electric charge lattice $\Z^{n-r}_E$ of theory $\mathcal{A}^\vee$ which is orthogonal to the magnetic charge lattice $\Z^r_M$ of theory $\mathcal{A}^\vee$ spanned by $\mathbf{q}$.
    \item 
From Equation (\ref{eq_tb}) $\mathbf{b}$ spans the magnetic charge lattice $\Z^{n-r}_M$ of theory $\mathcal{A}^\vee$ which is dual to the electric charge lattice $\Z^{n-r}_E$ of theory $\mathcal{A}^\vee$ spanned by $\tau^\mathsf{T}$.
\end{itemize}
Similar interpretations apply for the matrices $\tau^\vee$ and $\mathbf{b}^\vee$.

To summarize the relations under 3d mirror symmetry, label the Higgs and Coulomb branch of the theory $\mathcal{A}$ as $(\Hcal,\Ccal)$ and the Coulomb and Higgs branch of the mirror theory $\mathcal{A}^\vee$ as $(\Ccal^\vee, \Hcal^\vee)$. The Coulomb and Higgs branch exchanges under 3d mirror symmetry, i.e. $(\Hcal,\Ccal)=(\Ccal^\vee, \Hcal^\vee)$. Moreover, the dictionary of $3$d mirror symmetry reads:

\begin{align}
\Z^n_E &\leftrightarrow \Z^n_M \notag\\
    (\Hcal,\Ccal) &\leftrightarrow (\Ccal^\vee,\Hcal^\vee) \notag\\
    \left(\surm(2)_\Hcal,\surm(2)_\Ccal\right) &\leftrightarrow \left(\surm(2)_{\Ccal^\vee},\surm(2)_{\Hcal^\vee}\right) \notag\\
    (G_{\Hcal},G_{\Ccal}) &\leftrightarrow (G_{\Ccal^\vee},G_{\Hcal^\vee})\\
    (\text{vortex line},\text{Wilson line}) &\leftrightarrow (\text{Wilson line}^\vee,\text{vortex line}^\vee)  \notag\\
    (\text{magnetic},\text{electric}) \text{$1$-form symmetry} &\leftrightarrow (\text{electric},\text{magnetic}) \text{$1$-form symmetry}^\vee \notag\\
    &\dots \notag
\end{align}

\subsection{Non-simple Abelian theories}
Consider a non-simple Abelian theory $\mathcal{A}$ of gauge group $\urm(1)^r$ and a charge matrix $\mathbf{q}$ of size $n\times r$, where $\mathbf{q}$ has a non-trivial RSNF $\lambda=\text{diag}(k_1,\dots,k_r)$ with $k_{i+1}\vert k_i$.
$\mathbf{q}$ can be uniquely transformed into the form $\mathbf{q}_0\cdot\lambda$ under SL$(r,\Z)$, where $\mathbf{q}_0$ is simple.
Define a simple Abelian theory $\mathcal{A}_0$ with a charge matrix $\mathbf{q}_0$ and a splitting $\mathbf{b}$, which is given below equation (\ref{eq_sequence}). Using the equation (\ref{eq_inversemirrorsimple}), one can find its mirror theory $\mathcal{A}_0^\vee$ with charge matrix $\mathbf{q}_0^\vee$ and splitting $\mathbf{b}^\vee$.
The Abelian theory $\mathcal{A}$ is computed from the Abelian theory $\mathcal{A}_0$, by a $\Z_{k_1}\times\dots\times\Z_{k_r}$ gauging of the topological $\urm(1)^r$ symmetry:
\newcommand{\widesim}[2][1.5]{
  \mathrel{\overset{#2}{\scalebox{#1}[1]{$\sim$}}}
}
\begin{equation}
(\mathbf{q}_0,\mathbf{b})\quad \xrightarrow[\text{ gauging}]{\Z_{k_1}\times\dots\times\Z_{k_r}} \quad(\mathbf{q}_0\cdot \lambda,\mathbf{b})\quad \widesim[3]{\text{SL}(r,\Z)} \quad (\mathbf{q},\mathbf{b}). 
\end{equation}

The mirror theory $\mathcal{A}^\vee$ is a $\Z_{k_1}\times\dots\times\Z_{k_r}$ gauging of the flavour symmetry of the simple Abelian theory $\mathcal{A}_0^\vee$. The splitting $\mathbf{b}^\vee$ is the embedding matrix which gives the representations of the hypers under the $\Z_{k_1}\times\dots\times\Z_{k_r}$ gauge group. This gauging also results in a magnetic $1$-form symmetry.

\subsection{General Abelian theories}

For an Abelian theory $\mathcal{A}$ defined by a generalised charge matrix tuple $[\lambda,(\mathbf{q}_0,\mathbf{b}),\gamma]$ introduced in Section \ref{sec_discretecontinous}, the $3$d mirror symmetry acts by:
\begin{equation}
    [\lambda,(\mathbf{q}_0,\mathbf{b}),\gamma] \xleftrightarrow{\text{mirror}} [\gamma,(\mathbf{b},\mathbf{q}_0)^{-1,\mathsf{T}},\lambda]=[\lambda^\vee,(\mathbf{q}_0^\vee,\mathbf{b}^\vee),\gamma^\vee].
\label{eq_generalmirror}
\end{equation}
The mirror theory $\mathcal{A}^\vee$ is defined by the mirror tuple $[\lambda^\vee,(\mathbf{q}_0^\vee,\mathbf{b}^\vee),\gamma^\vee]$, which contains:
\begin{itemize}
    \item $\lambda^\vee=\gamma$ is the electric Smith normal form, which is the RSNF of the electric charge matrix $\mathbf{q}^\vee$, representing the gauging of $\Z_{l_1}\times\dots\times\Z_{l_{n-r}}$ subgroup of the magnetic flavour (or topological) $\urm(1)^{n-r}$, giving the electric $1$-form symmetry; 
    \item $\mathbf{q}_0^\vee$ is the simplified charge matrix under electric gauge $\urm(1)^{n-r}$, such that $\mathbf{q}^\vee$ can be transformed into the form $\mathbf{q}_0^\vee\cdot\lambda^\vee$ under SL$(n\!-\!r,\Z)$;
    \item $\gamma^\vee=\lambda$ is the magentic Smith normal form, representing the gauging of $\Z_{k_1}\times\dots\times\Z_{k_r}$ subgroup of the electric flavour $\urm(1)^r$, giving the magnetic $1$-form symmetry;
    \item $\mathbf{b}^\vee$ is the embedding matrix of the $\Z_{k_1}\times\dots\times\Z_{k_r}$ gauge group inside electric $\urm(1)^n$, which gives the representation of the hypers under the discrete gauge group. It fixes a splitting in the sequence (\ref{eq_mirrorsequence}).
\end{itemize}


An important consistency check of this 3d mirror relation is given in the subsection below.

\subsection{Monopole formula and Molien-Weyl integral}

To check the $3$d mirror relation, we can compare the refined Hilbert series of the Higgs and Coulomb branch.
For a simple Abelian theory defined by $(\mathbf{q},\mathbf{b})$ as in Section \ref{sec_sat}, the monopole formula determines the Hilbert series of the Coulomb branch,
\begin{equation}
    \text{HS}_\mathcal{C}=\frac{1}{(1-t^2)^r}\sum_{(m_1,\dots,m_r)\in\Z^r}{\left(\prod_{a=1}^r z_a^{m_a}\right) t^{2\Delta(m_1,\dots,m_r)}},
\label{eq_monosimple}
\end{equation}
where $z_a$ is the fugacity for the $a$-th topological $\urm(1)$, $t$ is the fugacity for the R-symmetry $\urm(1)_\mathcal{C}\subset\surm(2)_\mathcal{C}$, and the conformal dimension:
\begin{equation}
    \Delta(m_1,\dots,m_r)=\frac{1}{2}\sum_{i=1}^n\left\vert\sum_{a=1}^rq_{ia} m_a\right\vert. 
\end{equation}
The sum in Equation (\ref{eq_monosimple}) runs over the magnetic lattice of U$(1)^r$, which is $\Z^r$.

The Weyl integral determines the Hilbert series of the Higgs branch,
\begin{equation}
    \text{HS}_\mathcal{H}=\oint_{\vert x_a\vert=1} \left(\prod_{a=1}^r\frac{dx_a}{x_a}\right)\text{PE}\left(\sum_{i=1}^n\left( \prod_{a=1}^r x_a^{q_{ia}}\right)\left( \prod_{j=1}^{n-r} z_j^{b_{ij}}\right)t+c.c-rt^2\right),
\end{equation}
where $x_a$ is the fugacity for the $a$-th gauge $\urm(1)$, $z_i$ is the fugacity for the $i$-th flavour $\urm(1)$, and $t$ is the fugacity for the R-symmetry $\urm(1)_\mathcal{H}\subset\surm(2)_\mathcal{H}$. Note that both the $\text{HS}_\mathcal{C}$ and $\text{HS}_\mathcal{H}$ are invariant under SL$(r,\Z)$ transformation on $\mathbf{q}$.

For a non-simple Abelian theory defined by $[\lambda,(\mathbf{q}_0,\mathbf{b}),\gamma]$, the monopole formula takes the same form as in (\ref{eq_monosimple}).
Also, as shown in \cite{Bourget:2020bxh,hanany_actions_2024}, the monopole formula of the theory with charge matrix $\mathbf{q}_0\cdot\lambda$ can be computed from the monopole formula of the simple Abelian theory with charge matrix $\mathbf{q}_0$ by taking the Molien sum over $\Z_{k_1}\times\dots\times\Z_{k_r}$, up to a redefinition of the topological fugacity:
\begin{equation}
    \text{HS}_\mathcal{C}=\frac{1}{k_1\cdots k_r}\frac{1}{(1-t^2)^r}\sum_{(s_1,\dots,s_r)\in\Z_{k_1}\times\dots\Z_{k_r}}\sum_{(m_1,\dots,m_r)\in\Z^r}{\left(\prod_{a=1}^r (\omega_{k_a}^{s_a}z_a)^{m_a}\right) t^{2\Delta_0(m_1,\dots,m_r)}},
\end{equation}
where $\omega_k$ is the primary $k$-th root of unity, and $\Delta_0(m_1,\dots,m_r)$ is the conformal dimension of the simple Abelian theory with charge matrix $\mathbf{q}_0$.

On the Higgs branch, the Weyl integral can also be computed from the simple theory with $(\mathbf{q}_0,\mathbf{b})$ by averaging over $\Z_{l_1}\times\dots\times\Z_{l_{n-r}}$ using the Molien sum:
\begin{align}
    \text{HS}_\mathcal{H}=\frac{1}{l_1\cdots l_{n-r}}\sum_{(s_1,\dots,s_{n-r})\in\Z_{l_1}\times\dots\times\Z_{l_{n-r}}}\oint_{\vert x_a\vert=1} \left(\prod_{a=1}^r\frac{dx_a}{x_a}\right)\notag\\
    \times\text{PE}\left(\sum_{i=1}^n\left( \prod_{a=1}^r x_a^{q_{ia}}\right)\left( \prod_{j=1}^{n-r} (\omega_{l_j}^{s_j} z_j)^{b_{ij}}\right)t+c.c-rt^2\right).
\end{align}

\subsubsection{Simple Abelian theories}
In the following, we show the Hilbert series of the Higgs branch of the theory $\mathcal{A}$ agrees with the monopole formula of the Coulomb branch of the mirror theory $\mathcal{A}^\vee$.
Instead of using the Weyl integral, we can directly count the gauge invariant operators on the Higgs branch.
Denote the scalars in the $i$-th hypermultiplet as $X_i$ and $\tilde{X}_i$, which have charges $q_{ia}$ and $-q_{ia}$ under the $a$-th $\urm(1)$ gauge group, respectively. The most general GIOs $Y$ takes the form:
\begin{equation}
    Y=\prod_{i=1}^n X_i^{d_i} \tilde{X}_i^{\tilde{d}_i},
\label{eq_gio}
\end{equation}
where invariance of $Y$ under the $\urm(1)^r$ gauge symmetry implies constraints on $d_i,\tilde{d}_i\in(\Z^{\geq0})^n$,
\begin{equation}
    \sum_{i=1}^n q_{ia}(d_i-\tilde{d}_i)=0,
    \label{eq_cos1}
\end{equation}
for $a=1,\dots,r$. To proceed with the proof we rewrite $Y$ as:
\begin{equation}
    Y=\prod_{i=1}^n M_i B,
\end{equation}
where $M_i$ is a meson-like operator:
\begin{equation}
    M_i=\begin{cases}
        X_i^{d_i} \tilde{X}_i^{d_i},\quad d_i\leq\tilde{d}_i;\\
        X_i^{\tilde{d}_i} \tilde{X}_i^{\tilde{d}_i},\quad d_i>\tilde{d}_i,
    \end{cases}
\end{equation}
and $B$ is a baryon-like operator:
\begin{equation}
    B=\prod_{i=1}^nB_i,
\end{equation}
where
\begin{equation}
    B_i=\begin{cases}
        \tilde{X}_i^{\tilde{d}_i-d_i},\quad d_i\leq\tilde{d}_i;\\
        X_i^{d_i-\tilde{d}_i},\quad d_i>\tilde{d}_i.
    \end{cases}
\end{equation}
The Hilbert series of the Higgs branch is the product of the Hilbert series of the meson-like operators and the Hilbert series of the baryon-like operators with respect to the F-term relations.

The meson-like operators are generated by $n$ degree $2$ independent operators of the form: $X_i\tilde{X}_i$. Hence, the Hilbert series of the meson-like operators is $\frac{1}{(1-t^2)^n}$. The contribution of the F-term relations for $r$ $\urm(1)$s gauge fields is $(1-t^2)^r$, since there are $r$ independent relations on degree $2$ meson-like operators. Write $h_i=d_i-\tilde{d}_i$, the Hilbert series of the baryon-like operators can be written as a sum over the $\Z^n$ lattice with basis $h_i$ and with $r$ constraints in (\ref{eq_cos1}):
\begin{equation}
    {\sum_{h_i\in\Z^{n}}}_{\vert_{\sum_{i=1}^n q_{ia} h_i=0}}\prod_{j=1}^{n-r} z_j^{b_{ij}h_i}t^{\sum_{i=1}^n\vert h_i \vert}.
\end{equation} One can always find a new basis $m_j$ of the resulting $\Z^{n-r}$ lattice, such that $h_i=\sum_{j=1}^{n-r} \tau_{ji} m_j$, where $\sum_{i=1}^n\tau_{ji}q_{ia}=0$ and can be uniquely fixed by imposing that $\sum_{i=1}^n \tau_{ji} b_{ij'}=\text{Id}_{jj'}$. Hence, the Hilbert series of the baryon-like operators can be rewritten as:
\begin{equation}
    \sum_{m_j\in\Z^{n-r}}\prod_{j=1}^{n-r}z_j^{m_j}t^{\sum_{i=1}^n\left\vert \sum_{j=1}^{n-r}\tau_{ji} m_j \right\vert}.
\end{equation} The Hilbert series of the Higgs branch is the product of the above contributions, which reads:
\begin{equation}
    \HS_\Hcal=\frac{1}{(1-t^2)^{n-r}}\sum_{m_j\in\Z^{n-r}}z_j^{m_j}t^{\sum_{i=1}^n\left\vert \sum_{j=1}^{n-r} \tau_{ji} m_j \right\vert}=\HS_{\Ccal^\vee},
\end{equation}
which is exactly the monopole formula of the $\urm(1)^{n-r}$ theory with the charge matrix $\mathbf{\tau}^\mathsf{T}$. One can see $\tau$ satisfies the short exact sequence of $\mathcal{A}$ \ref{eq_sequence} and $\mathbf{\tau}^\mathsf{T}$ is the charge matrix of the mirror theory $\mathcal{A}^\vee$ as shown in Equation (\ref{eq_qveetvee}).
It is straightforward to apply a similar argument to the Higgs branch $\Hcal^\vee$ of the mirror theory $\mathcal{A}^\vee$ and to find agreement between $\HS_{\Hcal^\vee}$ and $\HS_{\Ccal}$.

The operator map and their contribution to the Hilbert series are summarised below:
\renewcommand{\arraystretch}{1.5}
\begin{table}[!h]
        \centering
\begin{tabular}{|p{0.25\textwidth}|p{0.34\textwidth}|p{0.29\textwidth}|}
\hline 
  $\Hcal$ & Hilbert Series & $\Ccal^\vee$ \\
\hline 
  F-term relations & $(1-t^2)^r$ & \multirow{2}{*}{Casimirs} \\
\cline{1-2} 
 Meson-like operators & $\frac{1}{(1-t^2)^n}$ & \\
 \hline
 Baryon-like operators & $\sum_{m_j\in\Z^{n-r}}z_j^{m_j}t^{\sum_{i=1}^n\left\vert \sum_{j=1}^{n-r}\tau_{ji} m_j \right\vert}$ & Bare monopole operators \\
\hline 
\end{tabular}
\end{table}

\subsubsection{General Abelian theories}
Now consider the non-simple theory $\mathcal{A}$ defined by $[\lambda,(\mathbf{q}_0,\mathbf{b}),\gamma]$.
On the Higgs branch $\Hcal$, the GIOs $Y$ take the same form as in Equation (\ref{eq_gio}), satisfying the constraints (\ref{eq_cos1}).
Invariance under the discrete factors $\Z_{l_1}\times\dots\times\Z_{l_{n-r}}$ gives additional constraints:
\begin{equation}
    \sum_{i=1}^n b_{ij} h_i=0\text{ mod }l_j,
    \label{eq_cos2}
\end{equation}
for $j=1,\dots,n-r$. By taking $h_i=\sum_{j=1}^{n-r} \tau_{ji} m_j$, the constraint (\ref{eq_cos1}) can be satisfied by requiring $\sum_{i=1}^n\tau_{ji}q_{ia}=0$ and the constraint (\ref{eq_cos2}) can be satisfied by requiring $\sum_{i=1}^n\tau_{ji} b_{ij'}=\gamma_{jj'}=l_j\delta_{jj'}$. The Hilbert series reads:
\begin{equation}
    \HS_\Hcal=\frac{1}{(1-t^2)^{n-r}}\sum_{m_j\in\Z^{n-r}}z_j^{l_jm_j}t^{\sum_{i=1}^n\left\vert \sum_{j=1}^{n-r} l_j \tau_{ji} m_j \right\vert}=\HS_{\Ccal^\vee}.
\end{equation}
After a redefinition of the fugacity $z_j=z_j^{l_j}$, we get exactly the monopole formula of the Coulomb branch $\Ccal^\vee$ of the mirror theory $\mathcal{A}^\vee$ defined by the mirror tuple $[\lambda^\vee,(\mathbf{q}_0^\vee,\mathbf{b}^\vee),\gamma^\vee]$.
It is straightforward to apply a similar argument to the Higgs branch $\Hcal^\vee$ of the mirror theory $\mathcal{A}^\vee$ and to find agreement between $\HS_{\Hcal^\vee}$ and $\HS_{\Ccal}$.

\subsection{Higgsing algorithm revisited}
\label{sec_revisit}

In Section \ref{sec_rankr}, the algorithm for computing the Hasse diagram for simple Abelian theories is given.
In this section we review this algorithm and generalise it to a non-simple Abelian theory with continuous and discrete gauge group factors.
The algorithm is done on the Higgs branch. Using 3d mirror symmetry, there is a similar algorithm on the Coulomb branch. We call the algorithm on the Higgs branch the \emph{Higgs decomposition}, and the algorithm on the Coulomb branch the \emph{Coulomb decomposition}. \footnote{An implementation of the algorithms in \texttt{Mathematica} is provided at the address: \url{https://github.com/DeShuo-Liu/Higgs-Decomposition-for-Abelian-Theories}}

\subsubsection{Simple Abelian theories}
\label{sec_simpleAbelianHiggsing}

Consider a simple Abelian theory defined by an $n\times r$ charge matrix $\mathbf{q}$ of rank $r$ and with trivial RSNF. 

\paragraph{Higgs decomposition.}
The Higgsing on the Higgs branch $\Hcal$ can be summarised into following steps:

\begin{enumerate}
    \item Find the set $M_1$ of lagestest $(1+s')\times r$ submatrices of rank $1$ with unique RSNF, where $s'\geq1$. Each element in this set labels a unique Higgsing pattern; 
    \item Find the set $M_s$ of lagestest $(s+s')\times r$ submatrices of rank $s$ with unique RSNF, such that if an element contains a submatrix as an element in $M_{s-s''}$, $s'\geq s''+1$. Each element in this set labels a unique Higgsing pattern; 
    \item For each submatrix $\mathbf{m}\in M_s$, find the SL$(r,\Z)$ element $S$, such that
    \begin{equation}
        \mathbf{m}\cdot S=
        \begin{pmatrix}
            0_{(s+s')\times(r-s)} & \mathbf{p}_{(s+s')\times s}
        \end{pmatrix},
        \end{equation}
    where $\mathbf{p}$ has rank $s$ and 
    \begin{equation}
        \mathbf{p}=\mathbf{p}_0\cdot\lambda_\mathbf{p}
    \end{equation} where $\lambda_\mathbf{p}=\text{diag}(k_1,\dots,k_s)$ with $k_{i+1}\vert k_i$ is the RSNF of both $\mathbf{m}$ and $\mathbf{p}$. Such element $S$ always exist, due to the existence of Smith decomposition;
    \item Act the element $S$ on the charge matrix $\mathbf{q}$. After certain row permutation $T$, one can find the decomposition:
    \begin{equation}T\cdot\mathbf{q}\cdot S=
    \begin{pmatrix}
    \mathbf{q}'_{(n-s-s')\times (r-s)}&\mathbf{b}'_{(n-s-s')\times s}\\
    0_{(s+s')\times (r-s)}&\mathbf{p}_{(s+s')\times s}
    \end{pmatrix},
    \label{eq_HiggsHiggsing}
    \end{equation}
    where $\mathbf{q}'$ is simple;
    \item The residual gauge theory has a $\urm(1)^{r-s}\times\Z_{k_1}\times\dots\times\Z_{k_s}$ gauge group and $n-s-s'$ hypers with a charge matrix $\mathbf{q'}$ and an embedding matrix $\mathbf{b}'$. As discussed in Section \ref{sec_discrete}, this theory can be rearranged into a $\urm(1)^{r-s}\times\Z_{k_1'}\times\dots\times\Z_{k_{n-r-s'}'}$ gauge theory $n-s-s'$ hypers, where $k_{i+1}'\vert k_i'$, with a charge matrix $\mathbf{q'}$ and an $(n-s-s')\times (n-r-s')$ embedding matrix induced from $\mathbf{b}'$;
    \item The transverse theory is a $\urm(1)^s$ gauge theory with $s+s'$ hypers, with a charge matrix $\mathbf{p}_0$.
\end{enumerate}

The Higgs branch is an isolated singularity iff there is only one element in $M_s$, and this element is $\mathbf{q}$ itself. It is equivalent to the condition that the determinants of $r\times r$ submatrices of any $(r+1)\times r$ submatrix are non-zero and coprime to each other.
\\

One can find a mirror charge matrix $\mathbf{q}^\vee$ of (\ref{eq_HiggsHiggsing}) in the form:
    \begin{equation}
    \begin{pmatrix}
    \mathbf{q}'^\vee_{(n-s-s')\times (n-r-s')}&0_{(n-s-s')\times s'}\\
    \mathbf{b}'^\vee_{(s+s')\times (n-r-s')}&\mathbf{p}^\vee_{(s+s')\times s'}
    \end{pmatrix},
    \end{equation}
where $\mathbf{q}'^\vee=\mathbf{q}'^\vee_0\cdot\lambda_{\mathbf{q'^\vee}}$ with $\lambda_{\mathbf{q'^\vee}}=\text{diag}(k'_1,\dots,k'_{n-r-s'})$, and $\mathbf{b}'^{\mathsf{T}}\cdot\mathbf{q}'^\vee+\mathbf{p}^\mathsf{T}\cdot\mathbf{b}'^\vee=0$. Such that, a $\urm(1)^{n-r-s'}$ gauge theory with charge matrix $\mathbf{q}'^\vee$ gives the mirror theory of residual theory, and a $\urm(1)^{s'}$ gauge theory with charge matrix $\mathbf{p}^\vee$ gives the mirror theory of the transverse theory. This gives a ``Higgsing" pattern on the mirror Coulomb branch $\Ccal^\vee$.

\paragraph{Coulomb decomposition.}
By considering the Higgsing of the mirror Higgs branch $\Hcal^\vee$, the similar algorithm on the Coulomb branch $\Ccal$ reads:

\begin{enumerate}
    \item For any fixed $0<s\leq r$, find the set $M_s$ of largest $(n-s-s')\times r$ submatrices of rank $r-s$ and with unique RSNF, with an additional condition that the submatrix of any element in the set is not a element in the set. These sets can always be found by examining the Smith decomposition of all the possible $(n-s-s')\times r$ submatrices of $\mathbf{q}$. Each element in these sets labels a unique ``Higgsing" pattern;
    \item For each submatrix $\mathbf{m}\in M_s$, find the SL$(r,\Z)$ element $S$, such that
    \begin{equation}
        \mathbf{m}\cdot S=
        \begin{pmatrix}
            0_{(n-s-s')\times s} & \mathbf{q}'_{(n-s-s')\times (r-s)}
        \end{pmatrix},
        \end{equation}
    where $\mathbf{q}'$ has rank $r-s$ and 
    \begin{equation}
        \mathbf{q}'=\mathbf{q}'_0\cdot\lambda_{\mathbf{q}'}
    \end{equation}
    where $\lambda_{\mathbf{q}'}=\text{diag}(l_1,\dots,l_{r-s})$ with $l_{i+1}\vert l_i$ is the RSNF of both $\mathbf{m}$ and $\mathbf{q}'$. Such element $S$ always exist, due to the existence of Smith decomposition;
    \item Act the element $S$ on the charge matrix $\mathbf{q}$. After certain row permutation $T$, one can find the decomposition:
    \begin{equation}
    T\cdot\mathbf{q}\cdot S=\begin{pmatrix}
    \mathbf{p}_{(s+s')\times s}&\mathbf{c}_{(s+s')\times (r-s)}\\
    0_{(n-s-s')\times s}&\mathbf{q}'_{(n-s-s')\times (r-s)}
    \end{pmatrix},
    \label{eq_CoulombHiggsing}
    \end{equation}
    where $\mathbf{p}$ is simple;
    \item The ``residual" gauge theory is a $\urm(1)^{r-s}$ gauge theory with $n-s-s'$ hypers, with a charge matrix $\mathbf{q'}$; 
    \item The ``transverse" theory is a $\urm(1)^{s}$ gauge theory with $s+s'$ hypers, with a charge matrix $\mathbf{p}$.
\end{enumerate}

The Coulomb branch is an isolated singularity iff there is only one element in $M_s$, and this element is $\mathbf{q}$ itself.

The Hasse diagram of the mixed branches can be acquired by applying the two algorithms in turns.


\subsubsection{General Abelian theories}
\label{sec_nonsimpleAbelianHiggsing}

Consider a general Abelian theory defined by the tuple $[\lambda,(\mathbf{q}_0,\mathbf{b}),\gamma]$, where the RSNF $\lambda=\text{diag}(k_1,\dots,k_r)$, the charge matrix $\mathbf{q}_0$ has size $n\times r$ and rank $r$, the magentic SNF $\gamma=\text{diag}(l_1,\dots,l_{n-r})$, the embedding matrix $\mathbf{b}$ has size $n\times (n-r)$ and rank $n-r$. 
Denote $\mathbf{q}$ as $\mathbf{\mathbf{q}_0}\cdot\lambda$ under possible SL$(r,\Z)$ transformation.

If the magentic SNF $\gamma$ is trivial, the same Higgsing algorithm applies as simple theory. An interesting observation is that the for any rank $r$ theory $\mathcal{A}$ with $n$ hypers defined by $[\lambda,(\mathbf{q}_0,\mathbf{b}),\gamma]$, it can be treated as a Higgsing phase from a rank $n$ theory $\mathcal{A}'$ with $3n-2r$ hypers with charge matrix:
\begin{equation}
    \begin{pmatrix}
        (\mathbf{q}_0\cdot\lambda)_{n\times r}&&\mathbf{b}_{n\times (n-r)}\\
     \multirow{2}{*}{$0_{2(n-r)\times r}$}   &&\gamma_{(n-r)\times(n-r)}\\
    &&\gamma_{(n-r)\times(n-r)}
    \end{pmatrix}_{(3n-2r)\times n}.
\end{equation}

Hence the Higgsing of $\mathcal{A}$ can be computed from the Higgsing of $\mathcal{A}'$.

Alternatively, one can apply the following steps directly on the combined charge matrix $(\mathbf{q},\mathbf{b})$.
\paragraph{Higgs decomposition.}
The Higgsing pattern on the Higgs branch $\Hcal$ is:
\begin{enumerate}
    \item For any $0<s\leq r$ and $0<s'<n-r$, find a parabolic transformation (\ref{eq_parabolic}) which takes the combined charge matrix $(\mathbf{q},\mathbf{b})$ into the form:
    \begin{equation}
    \begin{pmatrix}
    \mathbf{q}'_{(n-s-s')\times (r-s)}&\mathbf{b'}_{(n-s-s')\times s}&\mathbf{b}''_{(n-s-s')\times (n-r)}\\
    0_{(s+s')\times (r-s)}&\mathbf{p}_{(s+s')\times s}&\mathbf{c}'_{(s+s')\times (n-r)}
    \end{pmatrix},
    \end{equation}
    where $\mathbf{p}$ has rank $s$ and:
    \begin{equation}
     \mathbf{p}=\mathbf{p}_0\cdot\lambda_\mathbf{p},
    \label{eq_tranversematrix}
    \end{equation}
    where $\lambda_\mathbf{p}=\text{diag}(\alpha_1,\dots,\alpha_s)$ with $\alpha_{i+1}\vert \alpha_i$ is the RSNF; And where $\mathbf{c}'$ has the form:
    \begin{equation}
     \mathbf{c}'=\begin{pmatrix}
    c_{11} \beta_1 & \dots & c_{1(n-r)}\beta_{n-r}\\
    \vdots & \ddots & \vdots\\
    c_{(s+s')1} \beta_1 & \dots & c_{(s+s')(n-r)} \beta_{n-r}
    \end{pmatrix},
    \end{equation}
    where $\beta_i\vert l_i$ and $\text{gcd}(c_{ji}, l_i)=1$; Different choices of parabolic transformations with the same $\mathbf{p}$ and $\mathbf{c}'$ give equivalent residual gauge theory;
    \item The residual gauge theory is a $\urm(1)^{r-s}\times\Z_{\alpha_1}\times\dots\times\Z_{\alpha_s}\times\Z_{\beta_1}\times\cdots\times\Z_{\beta_{n-r}}$ gauge theory with $n-s-s'$ hypers, with a charge matrix $\mathbf{q}'$ and an embedding matrix $(\mathbf{b}',\mathbf{b}'')$. This theory can be rearranged into a $\urm(1)^{r-s}\times\Z_{\alpha_1'}\times\dots\times\Z_{\alpha_{n-r-s'}'}$ gauge theory, where $\alpha_{i+1}'\vert\alpha_{i}'$, with a charge matrix $\mathbf{q'}$ and an $(n-s-s')\times(n-r-s')$ embedding matrix induced from $(\mathbf{b}',\mathbf{b}'')$;
    \item The transverse theory is a $\urm(1)^{s}\times\Z_{\frac{l_1}{\beta_1}}\times\dots\times\Z_{\frac{l_{n-r}}{\beta_{n-r}}}$ gauge theory with $s+s'$ hypers, with a charge matrix $\mathbf{p}_0$ and an embedding matrix $\mathbf{c}'$. It must have a non-trivial Higgs branch. This theory can be rearranged into a $\urm(1)^s\times\Z_{\beta_1'}\times\dots\times\Z_{\beta_{s'}'}$ gauge theory, where $\beta_{i+1}'\vert\beta_{i}'$, with a charge matrix $\mathbf{p}_0$ and an $(s+s')\times s'$ embedding matrix induced from $\mathbf{c}'$.
\end{enumerate}


\paragraph{Coulomb decomposition.}
On the Coulomb branch $\Ccal$, the steps to decompose the charge matrix is the same as in the simple theory, but we read the residual theory and the transverse theory differently:
\begin{enumerate}
    \item For any fixed $0<s\leq r$, find the set $M_s$ of largest $(n-s-s')\times r$ submatrices of rank $r-s$ and with unique RSNF, such that the submatrix of any element in the set is not a element in the set. These sets can always be found by examining the Smith decomposition of all the possible $(n-s-s')\times r$ submatrices of $\mathbf{q}$. Each element in these sets labels a unique Higgsing pattern;
    \item For each submatrix $\mathbf{m}\in M_s$, find the SL$(r,\Z)$ element $S$, such that
    \begin{equation}
        \mathbf{m}\cdot S=
        \begin{pmatrix}
            0_{(n-s-s')\times s} & \mathbf{q}'_{(n-s-s')\times (r-s)}
        \end{pmatrix},
        \end{equation}
    where $\mathbf{q}'$ has rank $r-s$ and 
    \begin{equation}
        \mathbf{q}'=\mathbf{q}'_0\cdot\lambda_{\mathbf{q}'}
    \end{equation} where $\lambda_{\mathbf{q}'}=\text{diag}(\alpha_1,\dots,\alpha_{r-s})$ with $\alpha_{i+1}\vert \alpha_i$ is the RSNF of both $\mathbf{m}$ and $\mathbf{q}'$. Such element $S$ always exist, due to the existence of Smith decomposition;
    \item Act the element $S$ on the combined charge matrix $(\mathbf{q},\mathbf{b})$. After certain row permutation $T$, one can find the decomposition:
    \begin{equation}
    T\cdot\mathbf{q}\cdot S=\begin{pmatrix}
    \mathbf{p}_{(s+s')\times s}&\mathbf{c}_{(s+s')\times (r-s)}&\mathbf{c}'_{(s+s')\times (n-r)}\\
    0_{(n-s-s')\times s}&\mathbf{q}'_{(n-s-s')\times (r-s)}&\mathbf{b}'_{(n-s-s')\times (n-r)}
    \end{pmatrix};
    \end{equation}
    \item The residual gauge theory is a $\urm(1)^{r-s}\times\Z_{l_1}\times\dots\times\Z_{l_{n-r}}$ gauge theory with charge matrix $\mathbf{q'}$ and embedding matrix $\mathbf{b'}$;
    \item The transverse theory is a $\urm(1)^{s}$ gauge theory with charge matrix $\mathbf{p}$.
\end{enumerate}


The Hasse diagram of the mixed branches can be acquired by applying the two algorithms in turns.
An example of the application of the algorithms is shown in Section \ref{ex_qqq} and Section \ref{ex_212}.

\subsection{Examples}

In this section, we apply the methods developed in the previous sections to several examples.

\subsubsection{Quiver Notation}
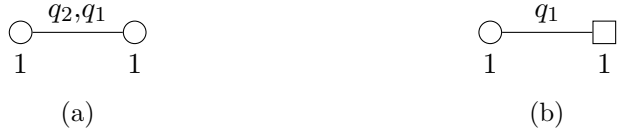
\begin{figure}[H]
    \centering
      \centering
\begin{subfigure}[t]{0.4\textwidth}
      \centering
    \begin{tikzpicture}
\node[gauge,label=below:{$1$}] (0) at (0,0) {};
            \node[gauge,label=below:{$1$}] (1) at (1.5,0) {};
            \draw (0)--(1)node[pos=0.5,above,sloped]{$q_2$,$q_1$};
    \end{tikzpicture}
    \caption{}
    \label{subfig_q2q1}
   \end{subfigure}
      \begin{subfigure}[t]{0.4\textwidth}
      \centering
    \begin{tikzpicture}
\node[gauge,label=below:{$1$}] (0) at (0,0) {};
            \node[flavour,label=below:{$1$}] (1) at (1.5,0) {};
            \draw (0)--(1)node[pos=0.5,above,sloped]{$q_1$};
    \end{tikzpicture}
    \caption{}
    \label{subfig_q1}
   \end{subfigure}
    \caption{\subref{subfig_q2q1}: The edge represents a hypermultiplet with charge $(q_1,-q_2)$, $q_1$ on the right $\urm(1)$ node and $-q_2$ on the left $\urm(1)$ node; \subref{subfig_q1}: The edge represents a hypermultiplet with charge $q_1$ under the gauge $\urm(1)$, while the charge under the flavour node is determined by the rest part of the quiver.}
    \label{fig_quivernotation}
\end{figure}

It is always helpful to express the theory using a quiver notation.
If all hypermultiplets are charged under at most two $\urm(1)$ gauge or flavor groups, such a quiver notation is possible.
Recall that in Section \ref{sec_rank2}, non simply laced edges are used for charges higher than 1.
However, this happens when one side of the edge has either charge 1 or fundamental under a non Abelian flavor symmetry.
We will need a more general notation where a single edge carries both charges higher than 1.
For this purpose, we introduce a new notation to represent the edge with charge $(q_1,-q_2)$ as summarised in Figure \ref{fig_quivernotation}.

\subsubsection{$\text{SQED}$ with $\mathbf{q}=(q_1,q_2)^{\mathsf{T}}$ -- self 3d mirror}

In Section \ref{sec_coprime}, SQED with $n$ hypers of coprime charges is discussed.
Here we consider the case with two hypers to demonstrate the procedure described in Section \ref{sec_sat} for finding the 3d mirror.
Consider $\text{SQED}$ with two hypermultiplets of charge vector $\mathbf{q}=(q_1,q_2)^{\mathsf{T}}$ that satisfies gcd$(q_1,q_2)=1$.
The global symmetry before gauging is $\urm(1)^2$. After gauging one $\urm(1)$ symmetry we are left with one $\urm(1)$ flavor symmetry.
Recall the relations $\tau\cdot\mathbf{q}=0$ and $\tau\cdot\mathbf{b}=1$.
This gives a unique $\tau=(-q_2,q_1)$.
We can choose $\mathbf{b}=(p_1,p_2)^\mathsf{T}$, where $p_1,p_2$ are two integers such that $\text{det}(\mathbf{q},\mathbf{b})=p_2q_1-p_1q_2=1$.
By B\'ezout's identity, the coefficients $p_1,p_2$ always exist and they are not unique, as $p_1+q_1,p_2+q_2$ also satisfy the same relation.
The short exact sequence (\ref{eq_sescosplitting}) reads:
\begin{equation}
\begin{tikzcd}
0 \arrow{r} & \Z 
\arrow{rr}{\mathbf{q}=\big(\begin{smallmatrix}
    q_1\\q_2
\end{smallmatrix}\big)} & &
\Z^2 \arrow{rr}{\mathbf{q}^{\vee,\mathsf{T}}=(-q_2,q_1)} \arrow[bend left]{ll}{\mathbf{b}^{\vee,\mathsf{T}}=(p_2,-p_1)}
& & \Z \arrow{r} \arrow[bend left]{ll}{\mathbf{b}=\big(\begin{smallmatrix}
    p_1\\p_2
\end{smallmatrix}\big)} & 0.
\end{tikzcd}
\end{equation}
Hence, the mirror theory is SQED with two hypermultiplets of charge vector $\mathbf{q}^\vee=(-q_2,q_1)^\mathsf{T}$, with a choice of flavour charge specified by $\mathbf{b}^\vee=(p_2,-p_1)^\mathsf{T}$. Further, it is easy to see that $\mathbf{q}$ and $\mathbf{q}^\vee$ define the same theory, since a hypermultiplet of charge $q$ is the same as a hypermultiplet of charge $-q$, namely, a hypermultiplet is invariant under charge conjugation. Hence, $\urm(1)$ gauge theory with two hypermultiplets of coprime charges is self-mirror. A special case is $q_1=1, q_2=1$ which is well known to be self 3d mirror.

Both Higgs and Couomb branches are the Klein $A_{q_1+q_2-1}$ singularity.



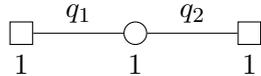
\begin{figure}[H]
    \centering
    \begin{tikzpicture}
            \node[flavour,label=below:{$1$}] (0) at (0,0) {};
            \node[flavour,label=below:{$1$}] (1) at (3,0) {};
            \node[gauge,label=below:{$1$}] (2) at (1.5,0) {};
            \draw (2)--(0)node[pos=0.5,above,sloped]{$q_1$};
            \draw (2)--(1)node[pos=0.5,above,sloped]{$q_2$};
    \end{tikzpicture}
    \caption{Quiver description of SQED with $\mathbf{q}=(q_1,q_2)^\mathsf{T}$.}
    \label{fig:SQEDpq}
\end{figure}

\subsubsection{A two dimensional self-mirror family}

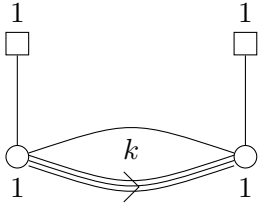
\begin{figure}[H]
    \centering
    \begin{tikzpicture}
            \node[flavour,label=above:{$1$}] (0) at (0,1.5) {};
            \node[flavour,label=above:{$1$}] (1) at (3,1.5) {};
            \node[gauge,label=below:{$1$}] (2) at (0,0) {};
            \node[gauge,label=below:{$1$}] (3) at (3,0) {};
            \draw (2)--(0);
            \draw (3)--(1);
            \draw (2) .. controls (1.5,0.5) .. (3);
            \draw[transform canvas={yshift=2pt}] (2) .. controls (1.5,-0.5) .. (3);
            \draw[transform canvas={yshift=0pt}] (2) .. controls (1.5,-0.5) .. (3);
            \draw[transform canvas={yshift=-2pt}] (2) .. controls (1.5,-0.5) .. (3);
            \node[] () at (1.5,0.1) {$k$};
            \draw (1.4,-0.2)--(1.6,-0.4)--(1.4,-0.6);
    \end{tikzpicture}
    \caption{Quiver of the theory defined by $\mathbf{q}=\begin{pmatrix}
    k&1&0&1\\
    -1&-1&1&0
\end{pmatrix}^\mathsf{T}$.}
    \label{fig_rank2mirror}
\end{figure}

Consider a $\urm(1)^2$ gauge theory with 4 hypermultiplets of charge matrix:
\begin{equation}
\mathbf{q}=\begin{pmatrix}
    k&-1\\
    1&-1\\
    0&1\\
    1&0
\end{pmatrix}.\end{equation}
This theory admits a non-simply laced quiver description as shown in Figure \ref{fig_rank2mirror}. One can find the charge matrix of the mirror theory:
\begin{equation}
    \mathbf{q}^\vee=\begin{pmatrix}
    -1&0\\
    0&1\\
    -1&1\\
    k&-1
\end{pmatrix},\end{equation} by applying (\ref{eq_inversemirrorsimple}). It is easy to see that $\mathbf{q}$ and $\mathbf{q}^\vee$ defines the same theory, hence, this theory is self-mirror.

If $k>1$, all the four branches $\Hcal$, $\Ccal$, $\Hcal^\vee$, $\Ccal^\vee$ are the complex 2-dimensional isolated singularity $\bar{h}_{\mathbf{q}}$.

\paragraph{Hilbert series} One can compute the Hilbert series:
\begin{equation}
    \text{HS}=\frac{N(t;k)}{(1-t^3)(1-t^{k+1})(1-t^{k+2})(1-t^{2k})},
\end{equation}
where $N(t;k)$ is a palindromic polynomial of degree $4k+2$. Below is the expression of $N(t;k)$ for $k\leq4$:
\begin{equation}
\begin{split}
N(t;1)&=1+2t^2+2t^3+2t^4+t^6,\\
N(t;2)&=1 + 2 t^2 + 2 t^3 + 5 t^4 + 4 t^5 + 5 t^6 + 2 t^7 + 2 t^8 + t^{10},\\
N(t;3)&=1 + 2 t^2 + t^3 + 4 t^4 + 5 t^5 + 7 t^6 + 6 t^7 + 7 t^8 + 5 t^9 + 4 t^{10} + t^{11} + 2 t^{12} + t^{14},\\
N(t;4)&=1 + 2 t^2 + t^3 + 3 t^4 + 3 t^5 + 7 t^6 + 7 t^7 + 9 t^8 + 8 t^9+ 
 9 t^{10} + 7 t^{11} + 7 t^{12} \\ &\quad\quad  + 3 t^{13} + 3 t^{14} + t^{15} + 2 t^{16} + t^{18}.\\
\end{split}
\end{equation}

\subsubsection{$\text{SQED}$ with $\mathbf{q}=(q_1,q_2,q_3)^{\mathsf{T}}$}
Consider $\text{SQED}$ with $\mathbf{q}=(q_1,q_2,q_3)^{\mathsf{T}}$, with $\text{gcd}(q_i,q_j)=1$ for $i\neq j$.
Using (\ref{eq_inversemirrorsimple}) one can find the charge matrix of the mirror theory:
\begin{equation}
    \mathbf{q}^\vee=\begin{pmatrix}
    q_2&0\\
    -q_1&q_3\\
    0&-q_2
\end{pmatrix}.
\end{equation} 

In this case, the Coulomb branch $\Ccal$ and the mirror Higgs branch $\Hcal^\vee$ are $A_{q_1+q_2+q_3-1}$ singularity. The Higgs branch $\Hcal$ and the mirror Coulomb branch $\Ccal^\vee$ are $\bar{h}_{(q_1,q_2,q_3)^\mathsf{T}}$ singularity.

One can also find a corank $1$ charge matrix:
\begin{equation}
    \mathbf{q}^\vee=\begin{pmatrix}
    q_2&0&-q_3\\
    -q_1&q_3&0\\
    0&-q_2&q_1
\end{pmatrix},
\end{equation}
giving the same Coulomb and Higgs branch after decoupling a free $\urm(1)$. The corank 1 matrix and the original charge matrix are related by the generalised \emph{Crawley-Boevey move} \cite{Crawley-Boevey1} which applies to a unitary quiver gauge theory. This corank 1 matrix gives rise to an unframed non-simply laced quiver, as shown below.

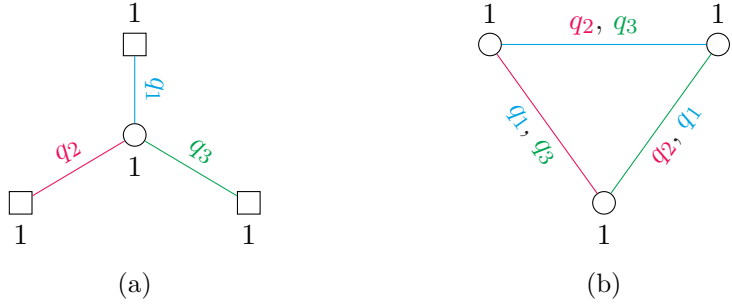
\begin{figure}[H]
    \centering
      \centering
\begin{subfigure}[t]{0.4\textwidth}
      \centering
    \raisebox{-0.0\height}{\begin{tikzpicture}
\node[flavour,label=below:{$1$}] (1) at (0,0) {};
            \node[flavour,label=below:{$1$}] (2) at (3,0) {};
            \node[flavour,label=above:{$1$}] (0) at (1.5,2.1) {};
            \node[gauge,label=below:{$1$}] (3) at (1.5,0.9) {};
            \draw[color=Cerulean!100] (0)--(3)node[pos=0.5,above,sloped]{\textcolor{Cerulean}{$q_1$}};
            \draw[color=OrangeRed!100] (1)--(3)node[pos=0.5,above,sloped]{\textcolor{OrangeRed}{$q_2$}};
            \draw[color=Green!100] (2)--(3)node[pos=0.5,above,sloped]{\textcolor{Green}{$q_3$}};
    \end{tikzpicture}}
    \caption{}
    \label{fig_q1q2q3}
    \end{subfigure}
    \begin{subfigure}[t]{0.4\textwidth}
      \centering
    \begin{tikzpicture}
\node[gauge,label=above:{$1$}] (1) at (0,0) {};
            \node[gauge,label=above:{$1$}] (2) at (3,0) {};
            \node[gauge,label=below:{$1$}] (0) at (1.5,-2.1) {};
            \draw[color=OrangeRed] (0)--(1)node[pos=0.5,below,sloped]{\textcolor{Cerulean}{$q_1$}\textcolor{Black}{,} \textcolor{Green}{$q_3$}};
            \draw[color=Green] (0)--(2)node[pos=0.5,below,sloped]{\textcolor{OrangeRed}{$q_2$}\textcolor{Black}{,} \textcolor{Cerulean}{$q_1$}};
            \draw[color=Cerulean] (1)--(2)node[pos=0.5,above,sloped]{\textcolor{OrangeRed}{$q_2$}\textcolor{Black}{,} \textcolor{Green}{$q_3$}};
    \end{tikzpicture}
    \caption{}
    \label{fig_q1q2q3mirror}
    \end{subfigure}
    \caption{\subref{fig_q1q2q3}: The quiver description of SQED with 3 hypermultiplets of charge vector $\mathbf{q}=(q_1,q_2,q_3)^\mathsf{T}$ satisfying $\text{gcd}(q_i,q_j)=1$ for $i\neq j$; \subref{fig_q1q2q3mirror}: The unframed quiver description of the mirror theory with charge matrix $\mathbf{q}^\vee$. The colors are chosen to illustrate the graph dual which is discussed in Section \ref{ex_startriangle}.}
    \label{fig_q1q2q3both}
\end{figure}

\subsubsection{$\text{SQED}$ with $\mathbf{q}=(1,\dots,1,k)^{\mathsf{T}}$}
Here we revisit the example discussed in Section \ref{sec_hbarnk}: $\text{SQED}$ with $n+1$ hypermultiplets of charge vector $\mathbf{q}=(1,\dots,1,k)^{\mathsf{T}}$, where $k\geq1$.
Using (\ref{eq_inversemirrorsimple}) one can find the charge matrix of the mirror theory: \begin{equation}
    \mathbf{q}^\vee=\begin{pmatrix}
    1&&&&\\
    -1&1&&&\\
    &-1&\ddots&&\\
    &&\ddots&1&\\
    &&&-1&k\\
    &&&&-1
\end{pmatrix}_{(n+1)\times n}.\end{equation}

One can also find a corank $1$ matrix:
\begin{equation}
\begin{pmatrix}
    1&&&&&-k\\
    -1&1&&&&\\
    &-1&\ddots&&&\\
    &&\ddots&1&&\\
    &&&-1&k&\\
    &&&&-1&1
\end{pmatrix}_{(n+1)\times (n+1)},\end{equation} giving the same Coulomb and Higgs branch after decoupling a free $\urm(1)$. This corank $1$ matrix gives rise to an unframed non-simply laced quiver as in Figure \ref{fig_hbarnkstar}. Here we rearrange the SQED quiver into a star shaped graph, and the mirror quiver into a necklace graph. They demonstrate a graph dual pattern as discussed in Section \ref{ex_startriangle}.

\begin{figure}[H]
    \centering
    \begin{subfigure}[t]{0.4\textwidth}
      \centering
      \begin{tikzpicture}
      \node[flavour,label=above:{$1$}] (1) at (0,2) {};
            \node[flavour,label=above:{$1$}] (2) at ({2*cos(18)},{2*sin(18)}) {};
            \node[flavour,label=above:{$1$}] (3) at ({-2*cos(18)},{2*sin(18)}) {};
            \node[flavour,label=below:{$1$}] (4) at ({2*cos(54)},{-2*sin(54)}) {};
            \node[flavour,label=below:{$1$}] (5) at ({-2*cos(54)},{-2*sin(54)}) {};
            \node[gauge,label=below:{$1$}] (0) at (0,0) {};
            \draw[color=OrangeRed!100] (0)--(1)node[pos=0.5,below,sloped]{\textcolor{OrangeRed}{$k$}};
            \draw (0)--(2);
            \draw (0)--(3);
            \draw (0)--(4);
            \draw (0)--(5);
            \path (4) -- (5) node[midway] (dotsr) {$\cdots$};
      \end{tikzpicture}
      \caption{}
      \label{SQED1...1kstar}
      \end{subfigure}
     \begin{subfigure}[t]{0.4\textwidth}
    \centering
    \raisebox{-0\height}{\begin{tikzpicture}
\node[] (1) at (0,0) {$\cdots$};
            \node[gauge,label=below:{$1$}] (2) at ({2*cos(18)},{2-2*sin(18)}) {};
            \node[gauge,label=below:{$1$}] (3) at ({-2*cos(18)},{2-2*sin(18)}) {};
            \node[gauge,label=above:{$1$}] (4) at ({2*cos(54)},{2+2*sin(54)}) {};
            \node[gauge,label=above:{$1$}] (5) at ({-2*cos(54)},{2+2*sin(54)}) {};
            \draw[transform canvas={xshift=-2pt}] (2)--(4);
            \draw[transform canvas={xshift=0pt}] (2)--(4);
            \draw[transform canvas={xshift=2pt}] (2)--(4)node[pos=0.5,above,sloped]{\textcolor{OrangeRed}{$k$}};
            \draw[color=OrangeRed] (4)--(5);
            \draw[] (1)--(2);
            \draw[] (1)--(3);
            \draw[transform canvas={xshift=2pt}] (3)--(5);
            \draw[] (3)--(5);
            \draw[transform canvas={xshift=-2pt}] (3)--(5)node[pos=0.5,above,sloped]{\textcolor{OrangeRed}{$k$}};
            \draw (-1.6,3)--(-1.5,2.6)--(-1.2,2.9);
            \draw (1.6,3)--(1.5,2.6)--(1.2,2.9);
    \end{tikzpicture}}
     \caption{}
     \label{fig_hbarnkstar}
     \end{subfigure}
    \caption{\sloppy\subref{SQED1...1k}: The quiver description of SQED with charge vector $\mathbf{q}=(1,\dots,1,k)^\mathsf{T}$;\quad\quad
    \subref{fig_hbarnkstar}: The unframed quiver description of the mirror theory defined by $\mathbf{q}^\vee$. The colors are chosen to illustrate the graph dual which is discussed in Section \ref{ex_startriangle}.}
    \label{fig_hbarnkstar}
\end{figure}
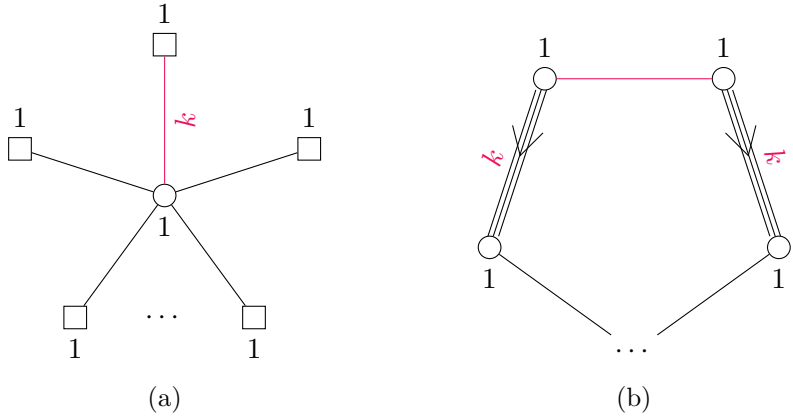

\subsubsection{$\text{SQED}$ and Star-Polygon duality}
\label{ex_startriangle}

Consider $\text{SQED}$ with $n$ hypermultiplets of charge vector $\mathbf{q}=(q_1,\dots,q_n)^{\mathsf{T}}$ such that gcd$(q_i,q_j)=1$ for $i\neq j$.
Using (\ref{eq_inversemirrorsimple}) one can find the charge matrix of the mirror theory: \begin{equation}
    \begin{pmatrix}
    -q_2&&&\\
    q_1&-q_3&&\\
    &q_2&\ddots&\\
    &&\ddots&-q_{n}\\
    &&&q_{n-1}\\
\end{pmatrix}_{n\times n-1}.\end{equation}

One can also find a corank $1$ matrix:
\begin{equation}
\begin{pmatrix}
    -q_2&&&&q_n\\
    q_1&-q_3&&&\\
    &q_2&\ddots&&\\
    &&\ddots&-q_{n}&\\
    &&&q_{n-1}&-q_1\\
\end{pmatrix}_{n\times n},
\end{equation}
giving the same Coulomb and Higgs branch after decoupling a free $\urm(1)$. 
This corank $1$ matrix gives rise to an unframed non-simply laced quiver.
By comparing the quiver description of the SQED thery with the mirror theory, one can find a Star-Polygon duality pattern: the gauge node is dual to a closed surface and the flavour node is dual to an open surface, the edge is dual to an orthogonal edge, the charge of the edge determines the charges of adjacent edges of the dual edge.

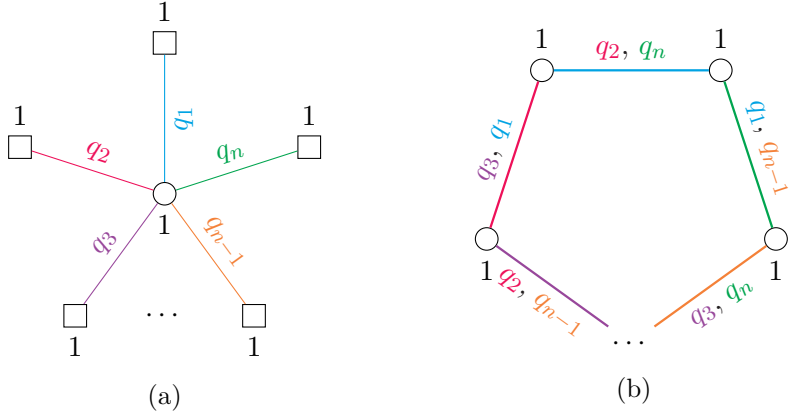
\begin{figure}[H]
    \centering
      \centering
        \begin{subfigure}[t]{0.4\textwidth}
      \centering
    \raisebox{-0.2\height}{\begin{tikzpicture}
        \node[flavour,label=above:{$1$}] (1) at (0,2) {};
            \node[flavour,label=above:{$1$}] (2) at ({2*cos(18)},{2*sin(18)}) {};
            \node[flavour,label=above:{$1$}] (3) at ({-2*cos(18)},{2*sin(18)}) {};
            \node[flavour,label=below:{$1$}] (4) at ({2*cos(54)},{-2*sin(54)}) {};
            \node[flavour,label=below:{$1$}] (5) at ({-2*cos(54)},{-2*sin(54)}) {};
            \node[gauge,label=below:{$1$}] (0) at (0,0) {};
            \draw[color=Cerulean!100] (0)--(1)node[pos=0.5,below,sloped]{\textcolor{Cerulean}{$q_1$}};
            \draw[color=Green!100] (0)--(2)node[pos=0.5,above,sloped]{ \textcolor{Green}{$q_n$}};
            \draw[color=OrangeRed!100] (0)--(3)node[pos=0.5,above,sloped]{ \textcolor{OrangeRed}{$q_2$}};
            \draw[color=Orange!100] (0)--(4)node[pos=0.5,above,sloped]{ \textcolor{Orange}{$q_{n-1}$}};
            \draw[color=Purple!100] (0)--(5)node[pos=0.5,above,sloped]{ \textcolor{Purple}{$q_3$}};
            \path (4) -- (5) node[midway] (dotsr) {$\cdots$};
    \end{tikzpicture}}
    \caption{}
    \label{SQEDn}
    \end{subfigure}
        \begin{subfigure}[t]{0.4\textwidth}
      \centering
    \raisebox{-0.2\height}{\begin{tikzpicture}
\node[] (1) at (0,0) {$\cdots$};
            \node[gauge,label=below:{$1$}] (2) at ({2*cos(18)},{2-2*sin(18)}) {};
            \node[gauge,label=below:{$1$}] (3) at ({-2*cos(18)},{2-2*sin(18)}) {};
            \node[gauge,label=above:{$1$}] (4) at ({2*cos(54)},{2+2*sin(54)}) {};
            \node[gauge,label=above:{$1$}] (5) at ({-2*cos(54)},{2+2*sin(54)}) {};
            \draw[line width=0.3mm, color=Green] (2)--(4)node[pos=0.5,above,sloped]{\textcolor{Cerulean}{$q_1$}\textcolor{Black}{,} \textcolor{Orange}{$q_{n-1}$}};
            \draw[line width=0.3mm,color=Cerulean] (4)--(5)node[pos=0.5,above,sloped]{ \textcolor{OrangeRed}{$q_2$}\textcolor{Black}{,} \textcolor{Green}{$q_n$}};
            \draw[line width=0.3mm,color=Orange] (1)--(2)node[pos=0.5,below,sloped]{ \textcolor{Purple}{$q_3$}\textcolor{Black}{,} \textcolor{Green}{$q_n$}};
            \draw[line width=0.3mm,color=Purple] (1)--(3)node[pos=0.5,below,sloped]{ \textcolor{OrangeRed}{$q_2$}\textcolor{Black}{,} \textcolor{Orange}{$q_{n-1}$}};
            \draw[line width=0.3mm,color=OrangeRed] (3)--(5)node[pos=0.5,above,sloped]{ \textcolor{Purple}{$q_3$}\textcolor{Black}{,} \textcolor{Cerulean}{$q_1$}};
    \end{tikzpicture}}
    \caption{}
    \label{aan}
    \end{subfigure}

     \caption{\subref{SQEDn}: The star-shaped quiver description of SQED with charge vector \\ $\mathbf{q}=(q_1,\cdots,q_n)^\mathsf{T}$. \subref{aan}: The unframed quiver description of the mirror theory defined by $\mathbf{q}^\vee$. One can see the 3d mirror relation gives rise to a Star-Polygon duality.}
    \label{fig_thestarmirror}
\end{figure}

\subsubsection{$\text{SQED}$ with multiple charge $q$ hypers}
\label{ex_sqedqn}

Consider SQED with $n$ charge $q$ hypers as in Figure \ref{fig_U(1)Nq}. the charge vector is given by $\mathbf{q}=(q,\dots,q)^\mathsf{T}$. This theory is non-simple as the RSNF $\lambda=q$. The charge matrix can be decomposed into $\mathbf{q}=\mathbf{q}_0\cdot q$, where $\mathbf{q}_0=(1,\dots,1)^\mathsf{T}$. After choosing a flavour matrix $\mathbf{b}$ as in (\ref{eq_sequence}), the theory can be described by the tuple:

\begin{equation}
   [\lambda,(\mathbf{q}_0,\mathbf{b}),\gamma] = [(q),\left(\begin{pmatrix}
        1\\
        \vdots\\
        \vdots\\
        1
    \end{pmatrix}_{n\times 1},\begin{pmatrix}
        1&\dots&1\\
        &\ddots&\vdots\\
        &&1\\
        0&\cdots&0
    \end{pmatrix}_{n\times n-1}\right),\text{Id}_{n-1}].
\end{equation}

By applying the 3d mirror formula (\ref{eq_generalmirror}), one finds the mirror theory is a $\urm(1)^{n-1}\times\Z_q$ gauge theory defined by the tuple:

\begin{equation}
    [\lambda^\vee,(\mathbf{q}_0^\vee,\mathbf{b}^\vee),\gamma^\vee] = [\text{Id}_{n-1},\left(\begin{pmatrix}
    1&&\\
    -1&\ddots&\\
    &\ddots&1\\
    &&-1\\
\end{pmatrix}_{n\times n-1},\begin{pmatrix}
        0\\
        \vdots\\
        0\\
        1
    \end{pmatrix}_{n\times 1}\right),(q)].
\end{equation}

The Coulomb branch $\Ccal$ and the mirror Higgs branch $\Hcal^\vee$ is the $A_{qn-1}$ singularity. The Higgs branch $\Hcal$ and the mirror Coulomb branch $\Ccal^\vee$ is the $a_{n-1}$ singularity.

\subsubsection{$\text{SQED}$ with multiple charge $q$ hypers and $\Z_q$ quotient}

Consider a $\urm(1)\times\Z_q$ gauge theory with $n$ hypermutiplets described by the tuple:

\begin{equation}
   [\lambda,(\mathbf{q}_0,\mathbf{b}),\gamma] = [(q),\left(\begin{pmatrix}
        1\\
        \vdots\\
        \vdots\\
        1
    \end{pmatrix}_{n\times 1},\begin{pmatrix}
        1&\dots&1\\
        &\ddots&\vdots\\
        &&1\\
        0&\cdots&0
    \end{pmatrix}_{n\times (n-1)}\right),\text{diag}(q,1,\dots,1)_{n-1}].
    \label{eq_sqedzqzq}
\end{equation}
This theory can be obtained by $\Z_q$ gauging of both the topological and flavour symmetry of SQED with $n$ hypers.
By applying the 3d mirror formula (\ref{eq_generalmirror}), one finds the mirror theory is a $\urm(1)^{n-1}\times\Z_q$ gauge theory defined by the tuple:

\begin{equation}
    [\lambda^\vee,(\mathbf{q}_0^\vee,\mathbf{b}^\vee),\gamma^\vee] = [\text{diag}(q,1,\dots,1)_{n-1},\left(\begin{pmatrix}
    1&&\\
    -1&\ddots&\\
    &\ddots&1\\
    &&-1\\
\end{pmatrix}_{n\times (n-1)},\begin{pmatrix}
        0\\
        \vdots\\
        0\\
        1
    \end{pmatrix}_{n\times 1}\right),(q)].
\end{equation}

The theory is self-mirror if $n=2$.

The quiver descriptions of the mirror pair are shown below.
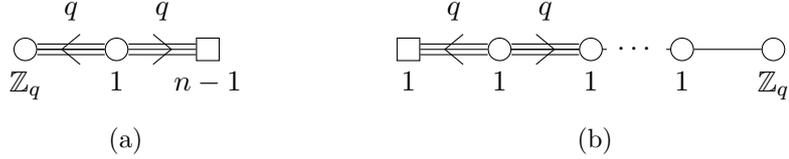
\begin{figure}[H]
    \centering
    \begin{subfigure}[t]{0.4\textwidth}
 \centering
    \begin{tikzpicture}
        \node[gauge,label=below:{$1$}] (0) at (0,0) {};
            \node[flavour,label=below:{$n-1$}] (1) at (1.2,0) {};
            \node[gauge,label=below:{$\Z_  q$}] (3) at (-1.2,0) {};
            \node[] (2) at (0.6,0.5) {$q$};
            \node at (-.6,.5) {$q$};
            \draw[transform canvas={yshift=2pt}] (0)--(1);
            \draw[transform canvas={yshift=-2pt}] (0)--(1);
            \draw[transform canvas={yshift=0pt}] (0)--(1);
            \draw[transform canvas={yshift=2pt}] (0)--(3);
            \draw[transform canvas={yshift=-2pt}] (0)--(3);
            \draw[transform canvas={yshift=0pt}] (0)--(3);
            \draw (0)--(3);
            \draw (0.48,0.2)--(0.72,0)--(0.48,-0.2);
            \draw (-.48,0.2)--(-.72,0)--(-.48,-0.2);
    \end{tikzpicture}
    \caption{}
    \label{subfig_sqedzqzq}
   \end{subfigure}
       \begin{subfigure}[t]{0.4\textwidth}
 \centering
\begin{tikzpicture}
    \node[gauge,label=below:{1}] (0) at (0,0) {};
    \node[gauge,label=below:{1}] (1) at (1.2,0) {};
    \node (2) at (1.8,0) {$\cdots$};
    \node[gauge,label=below:{1}] (3) at (2.4,0) {};
    \node[gauge,label=below:{$\Z_q$}] (4) at (3.6,0) {};
    \node[flavour,label=below:{1}] (0a) at (-1.2,0) {};
    \draw (0)--(1)--(2)--(3);
    \draw[transform canvas={yshift=-2pt}] (0)--(1);
    \draw[transform canvas={yshift=0pt}] (0)--(1);
    \draw[transform canvas={yshift=2pt}] (0)--(1);
    \draw[] (3)--(4);
    \draw[transform canvas={yshift=-2pt}] (0)--(0a);
    \draw[transform canvas={yshift=0pt}] (0)--(0a);
    \draw[transform canvas={yshift=2pt}] (0)--(0a);
    \draw (.48,0.2)--(.72,0)--(.48,-0.2);
    \draw (-.48,0.2)--(-.72,0)--(-.48,-0.2);
     \node at (.6,.5) {$q$};
     \node at (-.6,.5) {$q$};
     \end{tikzpicture}
    \caption{}
    \label{subfig_sqedzqzqmirror}
   \end{subfigure}
    \caption{\subref{subfig_sqedzqzq}: The quiver description of theory defined by (\ref{eq_sqedzqzq}); \subref{subfig_sqedzqzqmirror}: The quiver description of the mirror theory.}
    \label{fig_sqedzqzq}
\end{figure}

The Coulomb branch $\Ccal$ and the mirror Higgs branch $\Hcal^\vee$ is the $A_{nq-1}$ singularity. The Higgs branch $\Hcal$ and the mirror Coulomb branch $\Ccal^\vee$ is the $\bar{h}_{\mathbf{q},q,\mathbf{b}}=a_{n-1}/\Z_q$ singularity. Using mesons $M_{ij}=X_i\tilde{X}_j$ as coordinates, the polynomial ring of $a_{n-1}$ can be expressed as $\C[M_{ij}]/\langle \text{tr}(M)=0,\text{rk}(M)\leq1 \rangle$. The $\Z_q$ acts on the mesons as:
\begin{equation}
\begin{split}
    &M_{ii} \to M_{ii}, \\
    &M_{1i} \to \omega_qM_{1i},\quad\quad\quad\ \  i\neq1, \\
    &M_{i1} \to \omega_q^{-1}M_{i1},\ \ \ \quad\quad i\neq1, \\
    &M_{ij} \to M_{ij},\quad\quad\quad\quad i,j\neq1.
\end{split}
\end{equation}

\subsubsection{Affine $A_2$ quiver with charge $q$ hypers}
\label{ex_qqq}
Consider a $\urm(1)^2$ gauge theory with 3 hypermultiplets of charge matrix:
\begin{equation}
\mathbf{q}=\begin{pmatrix}q&0\\0&q\\q&-q\\ \end{pmatrix},
\label{eq_qqq}
\end{equation}
where $q>1$. This theory is non-simple, with RSNF $\lambda=\text{diag}(q,q)$.
Under an SL$(2,\Z)$ transformation, the matrix $\mathbf{q}$ factorises into the form $\mathbf{q}_0\cdot\lambda$, where:
\begin{equation}
\mathbf{q}_0=\begin{pmatrix}1&0\\0&1\\1&-1\end{pmatrix}.
\label{eq_simpleqqq}
\end{equation}
After choosing a flavour matrix $\mathbf{b}$ as in (\ref{eq_sequence}), the theory can be arranged into the tuple:
\begin{equation}
   [\lambda,(\mathbf{q}_0,\mathbf{b}),\gamma] = [\begin{pmatrix}q&0\\
   0&q\end{pmatrix},\left(\begin{pmatrix}
        1&0\\
        0&1\\
        1&-1
    \end{pmatrix},\begin{pmatrix}
        0\\
        0\\
        1
    \end{pmatrix}\right),(1)].
\end{equation}

By applying the 3d mirror formula (\ref{eq_generalmirror}), one finds the mirror theory as a $\urm(1)\times\Z_q\times\Z_q$ gauge theory defined by the tuple:


\begin{equation}
   [\lambda^\vee,(\mathbf{q}_0^\vee,\mathbf{b}^\vee),\gamma^\vee] = [(1),\left(\begin{pmatrix}
        -1\\
        1\\
        1
    \end{pmatrix},\begin{pmatrix}
        1&0\\
        0&1\\
        0&0\\
    \end{pmatrix}\right),\begin{pmatrix}q&0\\
   0&q\end{pmatrix}].
\end{equation}

Both theories admit a quiver description.

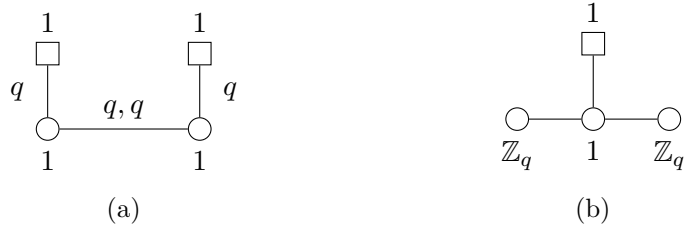
\begin{figure}[H]
    \centering
    \begin{subfigure}[t]{0.4\textwidth}
      \centering
      \begin{tikzpicture}
      \node[flavour,label=above:{$1$}] (2) at (0,1) {};
            \node[gauge,label=below:{$1$}] (0) at (0,0) {};
            \node[gauge,label=below:{$1$}] (1) at (2,0) {};
            \node[flavour,label=above:{$1$}] (3) at (2,1) {};
            \draw[] (0)--(2);
            \draw[] (0)--(1)node[pos=0.5,above,sloped]{{$q,q$}};
            \draw[] (1)--(3);
            \node at (-.4,.5) {$q$};
            \node at (2.4,.5) {$q$};
      \end{tikzpicture}
      \caption{}
      \label{A2ql}
      \end{subfigure}
     \begin{subfigure}[t]{0.4\textwidth}
    \centering
      \begin{tikzpicture}
            \node[gauge,label=below:{$1$}] (0) at (0,0) {};
            \node[gauge,label=below:{$\Z_q$}] (1) at (-1,0) {};
            \node[gauge,label=below:{$\Z_q$}] (2) at (1,0) {};
            \node[flavour,label=above:{$1$}] (3) at (0,1) {};
            \draw[] (0)--(1) (0)--(2) (0)--(3);
      \end{tikzpicture}
     \caption{}
     \label{A2qr}
     \end{subfigure}
    \caption{\subref{A2ql}: The quiver description of theory defined by (\ref{eq_qqq}); \subref{A2qr}: The quiver description of the mirror theory.}
    \label{fig_qqq}
\end{figure}

\paragraph{Hasse diagram}
Below we compute the Hasse diagram of the mirror pair. It is easy to see that the Higgs branch $\Hcal$ and the mirror Coulomb branch $\Ccal^\vee$ is the $A_2$ singularity. For the Coulomb branch $\Ccal$ and the mirror Higgs branch $\Hcal^\vee$, we apply the decomposition algorithm introduced in Section \ref{sec_revisit}.

On the Coulomb $\Ccal$, one can find three non-trivial Coulomb decompositions:

(The colors are chosen to help identify the elements of corresponding submatrices)
\begin{equation}
\begin{pmatrix}\color{OrangeRed}\mathbf{q}'_1&\\&\color{Cerulean}\mathbf{p}_1\\ \end{pmatrix}=\begin{pmatrix}\color{OrangeRed}q&0\\0&\color{Cerulean}q\\q&\color{Cerulean}-q\\ \end{pmatrix},
\end{equation}
where the residual gauge theory is a $\urm(1)$ gauge theory with charge matrix $(\mathbf{q}'_1)$, and the transverse theory is a $\urm(1)^2$ gauge theory with charge matrix $(\mathbf{p}_1)$.
\begin{equation}
\begin{pmatrix}\color{OrangeRed}\mathbf{q}'_2&\\&\color{Cerulean}\mathbf{p}_2\\ \end{pmatrix}=\begin{pmatrix}\color{Cerulean}q&0\\0&\color{OrangeRed}q\\\color{Cerulean}q&-q\\ \end{pmatrix},
\end{equation}
where the residual gauge theory is a $\urm(1)$ gauge theory with charge matrix $(\mathbf{q}'_2)$, and the transverse theory is a $\urm(1)^2$ gauge theory with charge matrix $(\mathbf{p}_2)$.

After an SL$(2,\Z)$ transformation given by $S$:
\begin{equation}
S=\begin{pmatrix}1&1\\0&1\\ \end{pmatrix},
\end{equation}
one finds another decomposition:
\begin{equation}
\begin{pmatrix}&\color{Cerulean}\mathbf{p}_3\\\color{OrangeRed}\mathbf{q}'_3&\\ \end{pmatrix}=\begin{pmatrix}q&\color{Cerulean}q\\0&\color{Cerulean}q\\\color{OrangeRed}q&0\\ \end{pmatrix},
\end{equation}
where the residual gauge theory is a $\urm(1)$ gauge theory with charge matrix $(\mathbf{q}'_3)$, and the transverse theory is a $\urm(1)^2$ gauge theory with charge matrix $(\mathbf{p}_3)$.

On the mirror Higgs branch $\Hcal^\vee$, one can also find three non-trivial Higgs decompositions:

\begin{equation}
\begin{pmatrix}&\color{Orange}\mathbf{b}''^\vee_1&\\\color{Cerulean}\mathbf{p}^\vee_1&&\color{DeepPurple}\mathbf{c}'^\vee_1\\ \end{pmatrix}=
    \left(\begin{pmatrix}
        -1\\
        \color{Cerulean}1\\
        \color{Cerulean}1
    \end{pmatrix},\begin{pmatrix}
        \color{Orange}1&0\\
        0&\color{DeepPurple}1\\
        0&\color{DeepPurple}0\\
    \end{pmatrix}\right),
\end{equation}
where the residual gauge theory is a $\Z_q$ gauge theory with embedding matrix $(\mathbf{b}''^\vee_1)$, and the transverse theory is a $\urm(1)\times\Z_q$ gauge theory with combined charge matrix $(\mathbf{p}^\vee_1,\mathbf{c}'^\vee_1)$.
\begin{equation}
\begin{pmatrix}\color{Cerulean}\mathbf{p}^\vee_2&\color{DeepPurple}\mathbf{c}'^\vee_2&\\&&\color{Orange}\mathbf{b}''^\vee_2\end{pmatrix}=
    \left(\begin{pmatrix}
        \color{Cerulean}-1\\
        1\\
        \color{Cerulean}1
    \end{pmatrix},\begin{pmatrix}
        \color{DeepPurple}1&0\\
        0&\color{Orange}1\\
        \color{DeepPurple}0&0\\
    \end{pmatrix}\right),
\end{equation}
where the residual gauge theory is a $\Z_q$ gauge theory with embedding matrix $(\mathbf{b}''^\vee_2)$, and the transverse theory is a $\urm(1)\times\Z_q$ gauge theory with combined charge matrix $(\mathbf{p}^\vee_2,\mathbf{c}'^\vee_2)$.

And after a parabolic transformation given by $P$:

\begin{equation}
P=
    \begin{pmatrix}
        1&0&-1\\
        0&1&-1\\
        0&0&1
    \end{pmatrix},
\end{equation}
one finds another decomposition:
\begin{equation}
\begin{pmatrix}\color{Cerulean}\mathbf{p}^\vee_3&\color{DeepPurple}\mathbf{c}'^\vee_3&\\&&\color{Orange}\mathbf{b}''^\vee_3\end{pmatrix}=
    \left(\begin{pmatrix}
        \color{Cerulean}-1\\
        \color{Cerulean}1\\
        1
    \end{pmatrix},\begin{pmatrix}
        \color{DeepPurple}1&0\\
        \color{DeepPurple}0&0\\
        0&\color{Orange}-1\\
    \end{pmatrix}\right),
\end{equation}
where the residual gauge theory is a $\Z_q$ gauge theory with embedding matrix $(\mathbf{b}''^\vee_3)$, and the transverse theory is a $\urm(1)\times\Z_q$ gauge theory with combined charge matrix $(\mathbf{p}^\vee_3,\mathbf{c}'^\vee_3)$.

The Hasse diagrams are shown below.
\begin{figure}[H]
    \centering
    \begin{subfigure}[t]{0.4\textwidth}
      \centering
    \begin{tikzpicture}
        \node[hasse] (0) at (0,0) {};
        \node[hasse] (1) at (0,1.5) {};
        \node at (0,-0.5) {$\urm(1)^2$};
        \node at (0,2) {$\Z_q\times\Z_q$};
        \node at (0.5,0.7) {$A_2$};
        \draw[] (0)--(1);
    \end{tikzpicture}
      \caption{}
      \label{qqqleft}
      \end{subfigure}
     \begin{subfigure}[t]{0.4\textwidth}
    \centering
    \begin{tikzpicture}
        \node[hasse] (0) at (0,0) {};
        \node[hasse] (11) at (-2,1.5) {};
        \node[hasse] (12) at (0,1.5) {};
        \node[hasse] (13) at (2,1.5) {};
        \node[hasse] (2) at (0,3) {};
        \draw (0)--(11)--(2) (0)--(12)--(2) (0)--(13)--(2);
        \node at (0,-0.5) {$\urm(1)\times\Z_q\times\Z_q$};
        \node at (-2.5,1.5) {$\Z_q$};
        \node at (2.5,1.5) {$\Z_q$};
        \node at (0.5,1.5) {$\Z_q$};
        \node at (0,3.5) {$\mathrm{id}$};
        \node at (1.4,0.6) {$A_{2q-1}$};
        \node at (1.4,2.4) {$A_{q-1}$};
        \node at (-1.4,0.6) {$A_{2q-1}$};
        \node at (-1.4,2.4) {$A_{q-1}$};
        \node at (-0.5,1) {$A_{2q-1}$};
        \node at (-0.4,2) {$A_{q-1}$};
    \end{tikzpicture}
     \caption{}
     \label{qqqright}
     \end{subfigure}
    \caption{Hasse diagrams of the mirror pair defined in Section \ref{ex_qqq}. \subref{qqqleft}: The Hasse diagram of the Higgs branch $\Hcal$ and the mirror Coulomb branch $\Ccal^\vee$; \subref{qqqright}: The Hasse diagram of the mirror Higgs branch $\Hcal^\vee$ and the Coulomb branch $\Ccal$.}
    \label{fig_qqqHasse}
\end{figure}

\subsubsection{A non-simple $\urm(1)^2$ theory}
\label{ex_212}

Consider a $\urm(1)^2$ theory with $5$ hypermultiplets of charge matrix:
\begin{equation}
\mathbf{q}=\begin{pmatrix}6&0\\6&0\\6&-2\\0&-2\\0&-2\end{pmatrix}.
\label{eq_212}
\end{equation}
This theory is non-simple, with RSNF $\lambda=\text{diag}(6,2)$. Under an  SL$(2,\Z)$ transformation, the matrix $\mathbf{q}$ factorises into the form $\mathbf{q}_0\cdot\lambda$, where:
\begin{equation}
\mathbf{q}_0=\begin{pmatrix}1&0\\1&0\\1&-1\\0&-1\\0&-1\end{pmatrix}.
\label{eq_simple212}
\end{equation}
After choosing a flavour matrix $\mathbf{b}$ as in (\ref{eq_sequence}), the theory can be arranged into the tuple:
\begin{equation}
   [\lambda,(\mathbf{q}_0,\mathbf{b}),\gamma] = [\begin{pmatrix}6&0\\
   0&2\end{pmatrix},\left(\begin{pmatrix}
        1&0\\
        1&0\\
        1&-1\\
        0&-1\\
        0&-1
    \end{pmatrix},\begin{pmatrix}
        0&0&0\\
        1&0&0\\
        0&0&0\\
        0&1&0\\
        0&0&1
    \end{pmatrix}\right),\text{Id}_3].
\end{equation}

By apply the 3d mirror formula (\ref{eq_generalmirror}), one finds the mirror theory as a $\urm(1)^3\times\Z_6\times\Z_2$ gauge theory defined by the tuple:


\begin{equation}
   [\lambda^\vee,(\mathbf{q}_0^\vee,\mathbf{b}^\vee),\gamma^\vee] = [\text{Id}_3,\left(\begin{pmatrix}
        -1&1&1\\
        1&0&0\\
        0&-1&-1\\
        0&1&0\\
        0&0&1
    \end{pmatrix},\begin{pmatrix}
        1&1\\
        0&0\\
        0&-1\\
        0&0\\
        0&0
    \end{pmatrix}\right),\begin{pmatrix}6&0\\
   0&2\end{pmatrix}].
\end{equation}

\paragraph{Hilbert series} One can compute the Hilbert series and find agreement:
\begin{align}
\text{HS}_\mathcal{H}&=\text{HS}_\mathcal{C^\vee}=\frac{(1-t)(1+t+4t^2+9t^3+13t^4+12t^5+13t^6+9t^7+4t^8+t^9+t^{10})}{(1-t^2)^4(1-t^3)^3},\\
\text{HS}_\mathcal{C}&=\text{HS}_\mathcal{H^\vee}=\frac{1 + 2 t^6 +2t^{12}+3t^{18}+2t^{20}+2t^{22}+3t^{24}+2t^{30}+2t^{36}+t^{42}}{(1-t^2)^2(1-t^{18})(1-t^{24})}.
\end{align}

\paragraph{Hasse diagram}
Below we compute the Hasse diagram of the mirror pair.




Consider the Higgs branch $\Hcal$, one can find two non-trivial Higgs decompositions:
\begin{equation}
\begin{pmatrix}\color{Cerulean}\mathbf{p}_1&\\&\color{OrangeRed}\mathbf{q}_1\end{pmatrix}
=\begin{pmatrix}\color{Cerulean}6&0\\\color{Cerulean}6&0\\6&\color{OrangeRed}-2\\0&\color{OrangeRed}-2\\0&\color{OrangeRed}-2\end{pmatrix},
\end{equation}
where the residual gauge theory is a $\urm(1)$ gauge theory with charge matrix $\mathbf{q}_1$, and the transverse theory is a $\urm(1)$ gauge theory with charge matrix $\mathbf{p}_1$.
And another Higgs decomposition is:
\begin{equation}
\begin{pmatrix}\color{OrangeRed}\mathbf{q}_2&\\&\color{Cerulean}\mathbf{p}_2\end{pmatrix}
=\begin{pmatrix}\color{OrangeRed}6&0\\\color{OrangeRed}6&0\\\color{OrangeRed}6&-2\\0&\color{Cerulean}-2\\0&\color{Cerulean}-2\end{pmatrix},
\end{equation}
where the residual gauge theory is a $\urm(1)$ gauge theory with charge matrix $\mathbf{q}_2$, and the transverse theory is a $\urm(1)$ gauge theory with charge matrix $\mathbf{p}_2$.

Now consider mirror the Higgs branch $\Hcal^\vee$, one can find a non-trivial Higgs decompositions:
\begin{equation}
    \left(\begin{pmatrix}\color{OrangeRed}\mathbf{q}^\vee_1&\\&\color{Cerulean}\mathbf{p}^\vee_1\end{pmatrix},\begin{pmatrix}\color{Orange}\mathbf{b}''^\vee_1&\\&\color{DeepPurple}\mathbf{c}'^\vee_1\end{pmatrix}\right)
=\left(\begin{pmatrix}
        \color{OrangeRed}-1&1&1\\
        \color{OrangeRed}1&0&0\\
       0&\color{Cerulean}{-1}&\color{Cerulean}{-1}\\
        0&\color{Cerulean}1&\color{Cerulean}0\\
        0&\color{Cerulean}0&\color{Cerulean}1
    \end{pmatrix},\begin{pmatrix}
        \color{Orange}1&1\\
        \color{Orange}0&0\\
        0&\color{DeepPurple}{-1}\\
        0&\color{DeepPurple}0\\
        0&\color{DeepPurple}0
    \end{pmatrix}\right),
\end{equation}
where the residual gauge theory is a $\urm(1)\times\Z_6$ gauge theory with combined charge matrix $(\mathbf{q}^\vee_1,\mathbf{b}''^\vee_1)$, and the transverse theory is a $\urm(1)^2\times\Z_2$ gauge theory with combined charge matrix $(\mathbf{p}^\vee_1,\mathbf{c}'^\vee_1)$.

And after a parabolic transformation given by $P$:
\begin{equation}
    P=\begin{pmatrix}
    1&0&0&0&0\\
    0&1&-1&0&-1\\
    0&0&1&0&0\\
    0&0&0&1&0\\
    0&0&0&0&1\\
    \end{pmatrix},
\end{equation}
one finds another non-trivial Higgs decomposition:
\begin{equation}
    \left(\begin{pmatrix}\color{Cerulean}\mathbf{p}^\vee_2&\\&\color{OrangeRed}\mathbf{q}^\vee_2\end{pmatrix},\begin{pmatrix}\color{DeepPurple}\mathbf{c}'^\vee_2&\\&\color{Orange}\mathbf{b}''^\vee_2\end{pmatrix}\right)
=\left(\begin{pmatrix}
        \color{Cerulean}-1&\color{Cerulean}1&0\\
       \color{Cerulean} 1&\color{Cerulean}0&0\\
       \color{Cerulean}0&\color{Cerulean}{-1}&0\\
        0&1&\color{OrangeRed}-1\\
        0&0&\color{OrangeRed}1
    \end{pmatrix},\begin{pmatrix}
        \color{DeepPurple}1&0\\
        \color{DeepPurple}0&0\\
        \color{DeepPurple}0&0\\
        0&\color{Orange}-1\\
        0&\color{Orange}0
    \end{pmatrix}\right),
\end{equation}
where the residual gauge theory is a $\urm(1)\times\Z_2$ gauge theory with combined charge matrix $(\mathbf{q}^\vee_2,\mathbf{b}''^\vee_2)$, and the transverse theory is a $\urm(1)^2\times\Z_6$ gauge theory with combined charge matrix $(\mathbf{p}^\vee_2,\mathbf{c}'^\vee_2)$.

The Coulomb branch Higgsing pattern can be computed from above by exchanging the role of the residual theory and the transverse theory.

The Hasse diagrams are shown below.
\begin{figure}[H]
    \centering
    \begin{subfigure}[t]{0.5\textwidth}
      \centering
    \begin{tikzpicture}
        \node[hasse] (0) at (0,0) {};
        \node[hasse] (11) at (-1.5,1.5) {};
        \node[hasse] (12) at (1.5,1.5) {};
        \node[hasse] (2) at (0,4) {};
        \draw (0)--(11)--(2) (0)--(12)--(2);
        \node at (0,-0.5) {$\urm(1)^2$};
        \node at (-2.5,1.5) {$\urm(1)\times\Z_6$};
        \node at (2.5,1.5) {$\urm(1)\times\Z_2$};
        \node at (0,4.5) {$\Z_6\times\Z_2$};
        \node at (1.2,0.6) {$a_1$};
        \node at (1.2,2.7) {$a_2$};
        \node at (-1.2,0.6) {$a_1$};
        \node at (-1.2,2.7) {$a_2$};
    \end{tikzpicture}
      \caption{}
      \label{212left}
      \end{subfigure}
     \begin{subfigure}[t]{0.45\textwidth}
    \centering
    \begin{tikzpicture}
        \node[hasse] (0) at (0,0) {};
        \node[hasse] (11) at (-1.5,1.5) {};
        \node[hasse] (12) at (1.5,1.5) {};
        \node[hasse] (2) at (0,3) {};
        \draw (0)--(11)--(2) (0)--(12)--(2);
        \node at (0,-0.5) {$\urm(1)^3\times\Z_6\times\Z_2$};
        \node at (-2.5,1.5) {$\urm(1)\times\Z_2$};
        \node at (2.5,1.5) {$\urm(1)\times\Z_6$};
        \node at (0,3.5) {$\mathrm{id}$};
        \node at (1.2,0.6) {$A_5$};
        \node at (1.2,2.4) {$A_{11}$};
        \node at (-1.2,0.6) {$A_{17}$};
        \node at (-1.2,2.4) {$A_3$};
    \end{tikzpicture}
     \caption{}
     \label{212right}
     \end{subfigure}
    \caption{Hasse diagrams of the mirror pair defined in Section \ref{ex_212}. \subref{212left}: The Hasse diagram of the Higgs branch $\Hcal$ and the mirror Coulomb branch $\Ccal^\vee$; \subref{212right}: The Hasse diagram of the mirror Higgs branch $\Hcal^\vee$ and the Coulomb branch $\Ccal$.}
    \label{fig_212Hasse}
\end{figure}

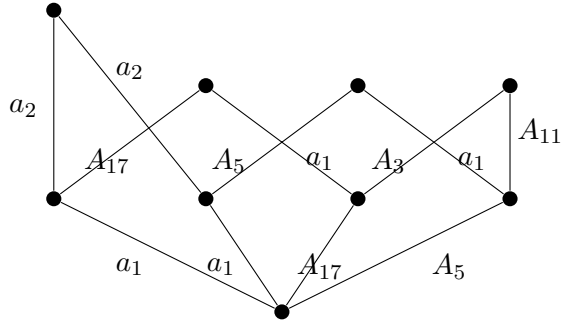
\begin{figure}[H]
    \centering
    \begin{tikzpicture}
        \node[hasse] (0) at (2,0) {};
        \node[hasse] (11l) at (-1,1.5) {};
        \node[hasse] (12l) at (1,1.5) {};
        \node[hasse] (2l) at (-1,4) {};
        \draw (0)--(11l)--(2l) (0)--(12l)--(2l);
        \node[hasse] (11r) at (-1+4,1.5) {};
        \node[hasse] (12r) at (1+4,1.5) {};
        \node[hasse] (2r) at (0+5,3) {};
        \draw (0)--(11r)--(2r) (0)--(12r)--(2r);
        \node[hasse] (1m) at (1,3) {};
        \node[hasse] (2m) at (3,3) {};
        \draw (11l)--(1m)--(11r) (12l)--(2m)--(12r);
        \node at (1.2,0.6) {$a_1$};
        \node at (0,3.2) {$a_2$};
        \node at (0,0.6) {$a_1$};
        \node at (-1.4,2.7) {$a_2$};
        \node at (4.2,0.6) {$A_5$};
        \node at (5.4,2.4) {$A_{11}$};
        \node at (2.5,0.6) {$A_{17}$};
        \node at (3.4,2) {$A_3$};
        \node at (-0.3,2) {$A_{17}$};
        \node at (1.3,2) {$A_5$};
        \node at (2.5,2) {$a_1$};
        \node at (4.5,2) {$a_1$};
    \end{tikzpicture}
    \caption{Hasse diagrams of the mixed branches of the mirror pair defined in Section \ref{ex_212}.}
    \label{fig_212HasseMix}
\end{figure}


\subsubsection{Another non-simple $\urm(1)^2$ theory}
\label{ex_224333}

Consider a $\urm(1)^2$ theory with $3$ hypermultiplets and a charge matrix:
\begin{equation}
\mathbf{q}=\begin{pmatrix}2&3\\2&3\\4&3\end{pmatrix}.
\label{eq_224333}
\end{equation}
This theory is non-simple, with RSNF $\lambda=\text{diag}(6,1)$.
Under certain SL$(2,\Z)$ transformation, the matrix $\mathbf{q}$ factorises into the form $\mathbf{q}_0\cdot\lambda$, where:
\begin{equation}
\mathbf{q}_0=\begin{pmatrix}-2&5\\-2&5\\-3&7\end{pmatrix}.
\label{eq_simple224333}
\end{equation}
After choosing a flavour matrix $\mathbf{b}$ as in (\ref{eq_sequence}), the thoery can be arranged into the tuple:
\begin{equation}
   [\lambda,(\mathbf{q}_0,\mathbf{b}),\gamma] = [\begin{pmatrix}6&0\\
   0&1\end{pmatrix},\left(\begin{pmatrix}
        -2&5\\
        -2&5\\
        -3&7
    \end{pmatrix},\begin{pmatrix}
        0\\
        1\\
        0
    \end{pmatrix}\right),(1)].
\end{equation}

By apply the 3d mirror formula (\ref{eq_generalmirror}), one finds the mirror theory as a $\urm(1)\times\Z_6$ gauge theory defined by the tuple:


\begin{equation}
   [\lambda^\vee,(\mathbf{q}_0^\vee,\mathbf{b}^\vee),\gamma^\vee] = [(1),\left(\begin{pmatrix}
        -1\\
        1\\
        0
    \end{pmatrix},\begin{pmatrix}
        7&3\\
        0&0\\
        -5&-2
    \end{pmatrix}\right),\begin{pmatrix}6&0\\
   0&1\end{pmatrix}].
\end{equation}

\paragraph{Hilbert series} One can work out the Hilbert series and find agreement:
\begin{align}
\text{HS}_\mathcal{H}&=\text{HS}_\mathcal{C^\vee}=\frac{1-t^4}{(1-t^2)^3},\\
\text{HS}_\mathcal{C}&=\text{HS}_\mathcal{H^\vee}=\frac{1 + t^3 + 2 t^6 + 2 t^7 + 2 t^8 + 2 t^9 + t^{12} + t^{15}}{(1-t^2)^2(1-t^3)(1-t^{12})}.
\end{align}


\paragraph{Hasse diagram}
The Hasse diagrams of the mirror pair are shown in Figure \ref{fig_224333Hasse}, using the method discussed in the previous section.


\begin{figure}[H]
    \centering
    \begin{subfigure}[t]{0.4\textwidth}
      \centering
    \begin{tikzpicture}
        \node[hasse] (0) at (0,0) {};
        \node[hasse] (1) at (0,1.5) {};
        \node at (-1.3,1.5) {$\urm(1)\times\Z_6$};
        \node at (-1.3,0) {$\urm(1)\times\urm(1)$};
        \draw (0)--(1);
        \node at (0.7,0.7) {$A_1$};
    \end{tikzpicture}
      \caption{}
      \label{224333left}
      \end{subfigure}
     \begin{subfigure}[t]{0.4\textwidth}
    \centering
    \begin{tikzpicture}
        \node[hasse] (0) at (0,0) {};
        \node[hasse] (1) at (0,1.5) {};
        \node[hasse] (2) at (0,3) {};
        \draw (0)--(1)--(2);
        \node at (-1,0) {$\urm(1)\times\Z_6$};
        \node at (-0.7,1.5) {$\urm(1)$};
        \node at (-0.6,3) {$\mathrm{id}$};
        \node at (0.8,0.8) {$A_5$};
        \node at (0.8,2.3) {$a_1$};
    \end{tikzpicture}
     \caption{}
     \label{224333right}
     \end{subfigure}
    \caption{Hasse diagrams of the mirror pair in Section \ref{ex_224333}. \subref{224333left}: The Hasse diagram of the Higgs branch $\Hcal$ and the mirror Coulomb branch $\Ccal^\vee$; \subref{224333right}: The Hasse diagram of the mirror Higgs branch $\Hcal^\vee$ and the Coulomb branch $\Ccal$.}
    \label{fig_224333Hasse}
\end{figure}

\subsubsection{Yet another non-simple $\urm(1)^2$ theory}
\label{ex_224303}

Consider a $\urm(1)^2$ theory with $3$ hypermultiplets and a charge matrix:
\begin{equation}
\mathbf{q}=\begin{pmatrix}2&3\\2&0\\4&3\end{pmatrix}.
\end{equation}
This theory is non-simple, with RSNF $\lambda=\text{diag}(6,1)$.
Under certain SL$(2,\Z)$ transformation, the matrix $\mathbf{q}$ factorises into the form $\mathbf{q}_0\cdot\lambda$, where:
\begin{equation}
\mathbf{q}_0=\begin{pmatrix}-2&5\\-1&2\\-3&7\end{pmatrix}.
\end{equation}
After choosing a flavour matrix $\mathbf{b}$ as in (\ref{eq_sequence}), the thoery can be arranged into the tuple:
\begin{equation}
   [\lambda,(\mathbf{q}_0,\mathbf{b}),\gamma] = [\begin{pmatrix}6&0\\
   0&1\end{pmatrix},\left(\begin{pmatrix}
        -2&5\\
        -1&2\\
        -3&7
    \end{pmatrix},\begin{pmatrix}
        0\\
        1\\
        0
    \end{pmatrix}\right),(1)].
\end{equation}

By apply the 3d mirror formula (\ref{eq_generalmirror}), one finds the mirror theory is a $\urm(1)\times\Z_6$ gauge theory defined by the tuple:

\begin{equation}
   [\lambda^\vee,(\mathbf{q}_0^\vee,\mathbf{b}^\vee),\gamma^\vee] = [(1),\left(\begin{pmatrix}
        1\\
        1\\
        -1
    \end{pmatrix},\begin{pmatrix}
        7&3\\
        0&0\\
        5&2
    \end{pmatrix}\right),\begin{pmatrix}6&0\\
   0&1\end{pmatrix}].
\end{equation}

\paragraph{Hilbert series} One can work out the Hilbert series and find agreement:
\begin{align}
\text{HS}_\mathcal{H}&=\text{HS}_\mathcal{C^\vee}=\frac{1-t^6}{(1-t^2)(1-t^3)^2 },\\
\text{HS}_\mathcal{C}&=\text{HS}_\mathcal{H^\vee}=\frac{(1 + t^2 + t^4)^4 (1 - 2 t^2 + 3 t^4 + 3 t^{12} - 2 t^{14} + 
   t^{16})}{(1 - t^6)^2 (1 - t^{12})^2}.
\end{align}

\paragraph{Hasse diagram}

The Hasse diagrams are shown in Figure \ref{fig_224303Hasse}, using the method discussed in the previous section.

\begin{figure}[H]
    \centering
    \begin{subfigure}[t]{0.4\textwidth}
      \centering
    \begin{tikzpicture}
        \node[hasse] (0) at (0,0) {};
        \node[hasse] (1) at (0,1.5) {};
        \node at (-1.3,1.5) {$\urm(1)\times\Z_6$};
        \node at (-1.3,0) {$\urm(1)\times\urm(1)$};
        \draw (0)--(1);
        \node at (0.7,0.7) {$A_2$};
    \end{tikzpicture}
      \caption{}
      \label{224303left}
      \end{subfigure}
     \begin{subfigure}[t]{0.4\textwidth}
    \centering
    \begin{tikzpicture}
        \node[hasse] (0) at (0,0) {};
        \node[hasse] (1) at (0,1.5) {};
        \node[hasse] (2) at (0,3) {};
        \draw (0)--(1)--(2);
        \node at (-1,0) {$\urm(1)\times\Z_6$};
        \node at (-0.7,1.5) {$\Z_2$};
        \node at (-0.6,3) {$\mathrm{id}$};
        \node at (0.8,0.8) {$A_5$};
        \node at (0.8,2.3) {$A_1$};
    \end{tikzpicture}
     \caption{}
     \label{224303right}
     \end{subfigure}
    \caption{Hasse diagrams of the mirror pair defined in Section \ref{ex_224303}. \subref{224303left}: The Hasse diagram of the Higgs branch $\Hcal$ and the mirror Coulomb branch $\Ccal^\vee$; \subref{224303right}: The Hasse diagram of the mirror Higgs branch $\Hcal^\vee$ and the Coulomb branch $\Ccal$.}
    \label{fig_224303Hasse}
\end{figure}

\subsubsection{$\Z_{k_1}\times\dots\times\Z_{k_r}$ gauge theory}
\label{sec_wiki}

Consider a $\Z_{k_1}\times\dots\times\Z_{k_r}$ gauge theory with $r$ hypermultiplets and a full rank embedding matrix $\mathbf{b}$ of size $r\times r$. The mirror theory is a $\urm(1)^r$ gauge theory with charge matrix $\mathbf{b}^{-1,\mathsf{T}}$.
This mirror pair can be represented by the tuples:
\begin{align}
   [\lambda,(\mathbf{q}_0,\mathbf{b}),\gamma]& = [\varnothing,(\varnothing,\mathbf{b}_{r\times r}),\text{diag}(k_1,\dots,k_r)],\\
   [\lambda^\vee,(\mathbf{q}_0^\vee,\mathbf{b}^\vee),\gamma^\vee]& = [\text{diag}(k_1,\dots,k_r),(\mathbf{b}^{-1,\mathsf{T}}_{r\times r},\varnothing),\varnothing].
\end{align}
A nice property is that the Coulomb branch $\Ccal$ and the mirror Higgs branch $\Hcal^\vee$ is empty. In addition, the Higgs branch $\Hcal$ and the mirror Coulomb branch $\Ccal^\vee$ is the quotient singularity $\C^{2r}/\Z_{k_1}\times\dots\times\Z_{k_r}$ with the embedding specified by $\mathbf{b}$. 

An interesting example can be constructed by taking the matrix from \cite{Wikipedia_2024} as the charge matrix $\mathbf{q}$ of a $\urm(1)^3$ gauge theory:
\begin{equation}
    \mathbf{q}=\begin{pmatrix}
        2&4&4\\
        -6&6&12\\
        10&4&16
    \end{pmatrix}.
\end{equation}
The RSNF $\lambda=\text{diag}(156,2,2)$. After SL$(3,\Z)$ transformation, one can find $\mathbf{q}\to\mathbf{q}_0\cdot\lambda$ where:
\begin{equation}
    \mathbf{q}_0=\begin{pmatrix}
        -1&2&1\\
        2&3&-3\\
        -4&2&5
    \end{pmatrix}.
\end{equation}
The theory can be represented by the tuple:
\begin{equation}
   [\lambda,(\mathbf{q}_0,\mathbf{b}),\gamma] = [\begin{pmatrix}156&0&0\\0&2&0\\0&0&2\end{pmatrix},\left(\begin{pmatrix}
        -1&2&1\\
        2&3&-3\\
        -4&2&5
    \end{pmatrix},\varnothing\right),\varnothing].
\end{equation}

Applying formula (\ref{eq_generalmirror}), one finds the mirror theory is a $\Z_{156}\times\Z_2\times\Z_2$ gauge theory defined by the tuple:
\begin{equation}
    [\lambda^\vee,(\mathbf{q}_0^\vee,\mathbf{b}^\vee),\gamma^\vee] = [\varnothing,\left(\varnothing,\begin{pmatrix}
        -21&-2&-16\\
        8&1&6\\
        9&1&7
    \end{pmatrix}\right),\begin{pmatrix}156&0&0\\0&2&0\\0&0&2\end{pmatrix}].
\end{equation}

\begin{figure}[H]
    \centering
    \begin{tikzpicture}
        \node[hasse] (0) at (0,0) {};
        \node[hasse] (11) at (-2,2) {};
        \node[hasse] (12) at (0,2) {};
        \node[hasse] (13) at (2,2) {};
        \node[hasse] (21) at (-2,4) {};
        \node[hasse] (22) at (0,4) {};
        \node[hasse] (23) at (2,4) {};
        \node[hasse] (3) at (0,6) {};
        \draw (0)--(11) (0)--(12) (0)--(13) (11)--(21) (11)--(22) (12)--(21) (12)--(23) (13)--(23) (13)--(22) (21)--(3) (22)--(3) (23)--(3);
        \node at (0,-0.5) {$\Z_{156}\times\Z_2\times\Z_2$};
        \node at (0,6.5) {$\mathrm{id}$};
        \node at (-3,2) {$\Z_6\times\Z_2$};
        \node at (-0.8,2) {$\Z_4\times\Z_2$};
        \node at (3,2) {$\Z_6\times\Z_2$};
        \node at (-2.5,4) {$\Z_2$};
        \node at (2.5,4) {$\Z_2$};
        \node at (-0.6,4) {$\Z_6$};
        \node at (0.4,1) {$A_{77}$};
        \node at (1.6,1) {$A_{51}$};
        \node at (-1.5,1) {$A_{51}$};
        \node at (-1.5,2.6) {$A_1$};
        \node at (-2.5,3) {$A_5$};
        \node at (-1.5,3.7) {$A_3$};
        \node at (1.5,3.7) {$A_3$};
        \node at (2.5,3) {$A_5$};
        \node at (1.5,2.6) {$A_1$};
        \node at (-1.5,5) {$A_1$};
        \node at (1.5,5) {$A_1$};
        \node at (0.3,5) {$A_5$};
    \end{tikzpicture}
    \caption{Hasse diagram of The Coulomb branch and the mirror Higgs branch of the mirror pair defined in Section \ref{sec_wiki}.}
    \label{fig_wikiHasse}
\end{figure}
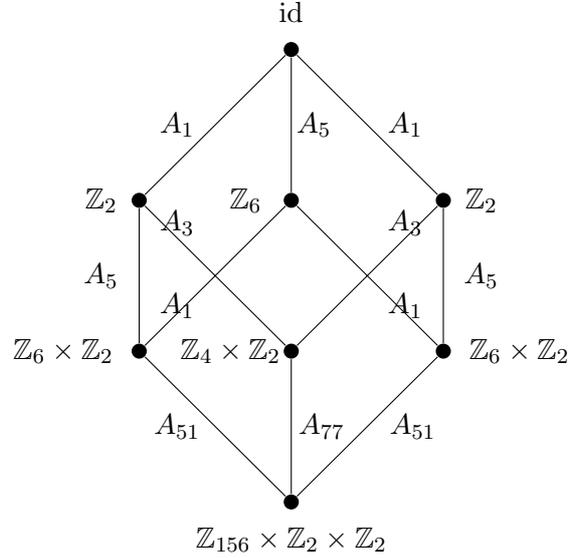

\paragraph{Hasse diagram} The Hasse diagram is shown in Figure \ref{fig_wikiHasse}.

\section{Outlook}


It is interesting to generalise the discussion in this paper to the family of $3d$ $\mathcal{N}=4$ non-Abelian theories with gauge group $G\subset\sprm(n)$ and hypermultiplets in any possible representation. 

\acknowledgments
We thank Chris Beem, Gwyn Bellamy, Federico Carta, Yoshinori Namikawa, Nicholas Proudfoot, Travis Schedler, Yinan Wang, and Michele Del Zotto for discussions.
JFG is supported by the EPSRC Open Fellowship (Schafer-Nameki) EP/X01276X/1 and the ``Simons Collaboration on Special Holonomy in Geometry, Analysis and Physics''. AH and DL are partially supported by STFC Consolidated Grants ST/T000791/1 and ST/X000575/1.

\appendix

\bibliographystyle{JHEP}
\bibliography{bibli.bib}

\end{document}